\documentclass[useAMS,usenatbib]{mn2e}
\usepackage{times}  % to get nice fonts for PDF
\usepackage{listings}
\usepackage{graphics}
\usepackage{subfigure}
\usepackage{graphicx}
\usepackage{amssymb}
\usepackage{hyperref}
\usepackage{aas_macros}

\def\pc{{\rm pc}}

\def\crate{\dot M_{\rm cool}}

\def\simprop{ \lower .75ex \hbox{$\sim$} \llap{\raise .27ex \hbox{$\propto$}} }

\title[Dynamical modelling of SNe feedback]{
A dynamical model of supernova feedback: gas outflows from the interstellar medium}
\author[Claudia del P. Lagos et al.]{
\parbox[t]{\textwidth}{
\vspace{-1.0cm}
Claudia del P. Lagos$^{1,2}$,
Cedric G. Lacey$^{1}$,
Carlton M. Baugh$^{1}$
}
\vspace*{6pt} \\
$^{1}$Institute for Computational Cosmology, Department of Physics,
University of Durham, South Road, Durham, DH1 3LE, UK. \\
$^{2}$European Southern Observatory, Karl-Schwarzschild-Strasse 2, 85748, Garching, Germany.
\vspace*{-0.5cm}}

\begin{document}
%\date{Accepted ???. Received ???; in original form ???}

\pagerange{\pageref{firstpage}--\pageref{lastpage}} \pubyear{2012}

\maketitle

\label{firstpage}

\begin{abstract}
We present a dynamical model of supernova feedback 
which follows the evolution of pressurised bubbles driven by supernovae
in a multi-phase interstellar medium (ISM). The bubbles are followed until the point of break-out into the halo, 
starting from an 
initial adiabatic phase to a radiative phase. 
We show that a key property which sets the fate of bubbles in the ISM 
is the gas surface density, through the work done by the expansion of bubbles and its role 
in setting the gas scaleheight.
The multi-phase description of the ISM is essential, and neglecting it leads to order of magnitude 
 differences in the predicted outflow rates.  
We compare our predicted mass loading and outflow velocities to observations 
 of local and high-redshift galaxies and find good agreement {over a wide range of stellar masses 
and velocities}. 
With the aim of analysing the
dependence of the mass loading of the outflow, $\beta$ (i.e. the ratio between the outflow and star
formation rates), on galaxy properties, we 
 embed our model in the galaxy formation simulation, {\texttt{GALFORM}}, set in the $\Lambda$CDM framework.
We find that a dependence of $\beta$ solely on the circular velocity, as is widely assumed in the literature, 
is actually a poor description of the outflow rate, as large variations with redshift 
and galaxy properties are obtained. Moreover, we find that below a circular velocity of  
$\approx 80\, \rm km\, s^{-1}$ the mass loading saturates. 
A more fundamental relation is that 
 between $\beta$ and the gas scaleheight of the disk, $h_{\rm g}$, and the gas fraction, $f_{\rm gas}$, 
as $\beta\propto h^{1.1}_{\rm g} f^{0.4}_{\rm gas}$, or the gas surface density, $\Sigma_{\rm g}$, and the gas fraction, as 
$\beta 
 \propto \Sigma^{-0.6}_{\rm g}  f^{0.8}_{\rm gas}$.    
We find that using the new mass loading 
model leads to a shallower faint-end slope in the predicted optical and near-IR galaxy luminosity functions. 
\end{abstract}


\section{Introduction}\label{Sec:IntroSNef}

An outstanding problem in astrophysics is to understand how galaxies form 
 in dark matter halos. The problem is highly non-linear: 
the stellar mass function of galaxies differs substantially from the 
dark matter halo mass function, with the stellar mass function being shallower at the low-mass end and 
steeper at the high-mass end than the halo mass function (see \citealt{Baugh06}). 
The main physical driver of these differences is thought to be 
gas cooling and feedback (\citealt{Larson74}; \citealt{Rees77}; \citealt{White78}; \citealt{Dekel86}; 
\citealt{White91}; \citealt{Cole00}; 
\citealt{Bower06}; \citealt{Croton06}). Feedback from supernovae (SNe) and active galactic 
nuclei (AGN) is thought to suppress star formation in low and high stellar mass 
galaxies, respectively, lowering the cold baryon fraction in these galaxies (e.g. 
\citealt{Fukugita98}; \citealt{Mandelbaum06}; \citealt{Liu10}). 

Observations suggest that SN-driven outflows are common in galaxies  
(e.g. \citealt{Martin99}; \citealt{Heckman00}; \citealt{Shapley03}; \citealt{Rupke05};
\citealt{Schwartz06}; \citealt{Weiner09}; \citealt{Sato09}; \citealt{Chen10};
 \citealt{Rubin10}; \citealt{Banerji11}; see \citealt{Veilleux05} for a review). 
In many cases 
the inferred outflow rate  exceeds the star formation rate (SFR; \citealt{Martin99}; \citealt{Martin05}; 
\citealt{Bouche12}), suggesting that SN feedback potentially has a large impact on galaxy evolution.
 The outflow rates inferred from absorption line studies correlate with galaxy properties 
such as SFRs and near-ultraviolet to optical colours, indicating that the influence 
of SN feedback might be differential with 
SFR and stellar mass (e.g. \citealt{Martin05}; \citealt{Kornei12}). 
Photometric and kinematic observations 
of atomic hydrogen shells and holes in the interstellar medium (ISM) of local galaxies, in addition to 
SN remnants observed in $X$-rays and radio, imply 
 that SNe lead to the formation of bubbles within the ISM 
and that the mass carried away 
is large and able to substantially change the gas reservoirs of galaxies (e.g. \citealt{Heiles79}; 
\citealt{Maciejewski96}; \citealt{Pidopryhora07}).
SN feedback is also thought to be responsible for the metal enrichment of the intergalactic medium 
 (e.g. see \citealt{Putman12} for a recent review).

Although the importance of SN feedback is clear from observations, 
the wide range of phenomenological models of SN feedback 
found in the literature reflect the uncertainty in how this process affects the ISM 
of galaxies and the intergalactic medium (IGM). 
The key questions are how does the mass loading of winds driven by SNe, $\beta=\dot{M}_{\rm out}/{\rm SFR}$ (the ratio between the 
outflow rate, $\dot{M}_{\rm out}$,  and the SFR), 
depend on galaxy properties and what is the effect of winds on the evolution of galaxies? 

A common assumption made in galaxy formation modelling is that the mass loading (sometimes called the 
`mass entrainment' of the wind) 
depends exclusively 
on the energy input by SNe and the circular velocity of the galaxy, which is taken as a proxy for the depth of the 
gravitational potential well (e.g. \citealt{White78}; \citealt{White91}). 
The specific form of the dependence  
contains adjustable parameters which are set by requiring that the model fits 
observations, such as the stellar mass function or luminosity function, etc 
(e.g. \citealt{Cole00}; \citealt{Springel01}; \citealt{Benson03}; \citealt{Croton06}).
Simple, physically motivated forms for the 
explicit dependence of $\beta$ on $v_{\rm circ}$ are based on 
 arguments which invoke momentum-driven or energy-driven winds, corresponding to dependences 
of $\beta \propto v^{-1}_{\rm circ}$ and $\beta \propto v^{-2}_{\rm circ}$, respectively 
(e.g. \citealt{Silk97}; \citealt{Silk03}; 
\citealt{Hatton03}; \citealt{Murray05}; \citealt{Stringer12}; see \citealt{Benson10b} for a review). 

Hydrodynamic simulations  
 commonly assume constant wind velocities, adopting a kinetic feedback scheme in which SNe inject momentum to 
neighbouring particles, which are assumed to become dynamically decoupled from the 
other particles for a period of time  
(\citealt{Springel03}; \citealt{Scannapieco06}; \citealt{DallaVecchia08}; \citealt{Narayanan08}; \citealt{Schaye10}).
 Alternatively, simple scaling relations between the outflow velocity and the halo circular velocity may be  
assumed (e.g. \citealt{Dave11}).
These calculations can qualitatively reproduce 
the properties of disk galaxies (\citealt{Scannapieco12}). 
The wind speed is a free parameter in these simulations with values 
of $v_{\rm w}\approx 300-1000 \,\rm km/s$ typically used 
(see \citealt{Schaye10} for an analysis of the impact of changing 
$v_{\rm w}$ on the predicted evolution of the global density of SFR in a hydrodynamical simulation, and 
\citealt{Scannapieco12} for a comparison between different simulations). 

However, 
such a scheme where the wind speed, $v_{\rm w}$, is constant fails to reproduce the 
stellar mass function, suggesting that this parametrisation is too effective in 
intermediate stellar mass galaxies, but not efficient enough in low stellar mass galaxies 
(\citealt{Crain09}; \citealt{Dave11}; \citealt{Bower12}). 
In addition to these problems, \citet{Bower12}, \citet{Guo12} and \citet{Weinmann12} show that 
simple phenomenological recipes for SN feedback are not able to 
 explain the observed shallow low-mass end of the stellar mass function  
 (\citealt{Drory05}; \citealt{Marchesini09}; 
\citealt{Li09}; \citealt{Caputi11}; \citealt{Bielby11}). This problem can be alleviated by introducing 
an ad-hoc dependence of the time it takes for  
the outflowing gas to fall back onto the galaxy on redshift \citep{Henriques13}. 
A possible explanation for this 
is that such parametrisations do not accurately describe the complex process 
of outflows driven by SNe in the interstellar medium and their subsequent propagation through the 
hot halo gas around galaxies. 

\citet{Creasey12} analysed the effect of a single SN in the ISM by simulating a
column through the disk of a galaxy with very high mass and spatial resolution. Creasey et al. varied the 
initial conditions in the disk with the aim of covering different gas surface densities
and gas-to-stellar mass ratios, and found that the mass outflow rate depends strongly on the local properties of the ISM, such as 
the gas surface density.
Similar conclusions were reached by \citet{Hopkins12} in $4$ simulations of individual galaxies including 
different types of feedback in addition to SN feedback. 
The SN feedback scheme used in Hopkins et al. was not fully resolved and hence depends on subgrid 
modelling of the momentum deposition of the different types of feedback. Regardless of the details of each simulation, 
both studies point to a breakdown of the classical parametrisations  
used for $\beta$. However, since the simulations of both Creasey et al. and Hopkins et al. 
cover a narrow range of environments, the generality of their results is not clear.

In this paper we implement a fully numerical
treatment of SN feedback due to bubbles inflated by SNe which expand into the ISM. 
We follow the bubbles during the adiabatic and radiative phases assuming spherical symmetry, 
starting in the star-forming regions in the ISM and continuing until the bubble breaks out of the galactic disk or 
is confined. The aims of this paper are (i) to study the effect of different physical processes on the expansion of bubbles, such as 
the multi-phase ISM, the gravity from stars and dark matter (DM), the temporal changes in the ambient pressure, etc., 
and (ii) to extend previous theoretical work by using the new dynamical SN feedback model in the cosmological
semi-analytic model of galaxy formation, {\tt GALFORM}.
Semi-analytic models have the advantage 
of being able to simulate large cosmological volumes containing
millions of galaxies over cosmic epochs and making multiwavelength predictions
(\citealt{Baugh06}). This approach makes it possible to study a wide enough 
range of properties and epochs to reach robust conclusions about 
the dependence of $\beta$ on galaxy properties
and to characterise the combination of properties that best 
quantifies the mass outflow rates in galaxies. 

Previous dynamical models of SN feedback 
in the context of cosmological galaxy formation 
have focused on the evolution of bubbles either in the ISM or the hot halo. 
For instance, \citet{Larson74} (see also \citealt{Monaco04b} and \citealt{Shu05}) 
implemented analytic solutions for the evolution of bubbles 
until their break-out from the ISM by neglecting gravity, external pressure and temporal changes in the 
ambient gas. Bertone et al. (2005, 2007) and \citet{Samui08} 
followed the evolution of bubbles in the hot halo assuming an ad-hoc
mass outflow rate and wind velocity from the disk into the halo.
\citet{Dekel86} implemented a simpler model which aimed to estimate 
the mass ejection rate from both the ISM and the halo, using analytic solutions for the evolution of bubbles
in the ISM 
to calculate an average rate of mass injection from the ISM into the halo.
 \citet{Efstathiou00} went a step further, implementing bubble evolution in a multi-phase ISM 
with the hot phase dominating the filling factor, 
using analytic solutions for the evolution of adiabatic bubbles.  
We improve upon previous calculations by 
including the effects of gravity, radiative losses,
external pressure from the diffuse medium and temporal changes in the ambient gas on the 
expansion of bubbles, all embedded in a multi-phase medium. 
We use the information about the radial profiles of 
galaxies to calculate mass outflow rates locally.  
In addition to the sophistication of our calculation, another 
key difference in our work is that bubbles expand into the warm 
component of the ISM instead of the hot component, as is assumed in some previous work. 
This is motivated by the results from detailed simulations 
and observations in our Galaxy which point to a 
rather small volume filling factor of hot gas, $\lesssim 20$\%, with  
little mass contained in this gas phase 
(e.g. \citealt{MacLow89}; 
\citealt{Ferriere01}; \citealt{Avillez04}; see \citealt{Haffner09} for a review on the warm phase of the ISM). 

In this paper we focus on the ejection of gas from the disk and do not attempt to model the expansion of 
bubbles in the hot halo or the rate of gas ejection  
from the halo into the IGM. 
In paper II, we will implement a full model of the expansion of bubbles in the hot halo, 
following a similar approach to that adopted in this paper, 
and analyse the rate at which mass and metals 
escape the halo and go into the IGM, 
and how this depends on galaxy and halo properties (Lagos, Baugh \& Lacey, 2013, in prep.).

This paper is organised as follows. 
$\S$\ref{Sec:DynModel_SN} describes the dynamical model of SN feedback and the evolution of individual bubbles in 
the ISM. 
$\S$\ref{DiffISMandClumps} describes the calculation the properties of the diffuse medium and how we locate 
giant molecular clouds (GMCs) in the disk. 
In $\S$\ref{Sec:SFeqs_sn} we describe how we include the full dynamical model of SN feedback in the galaxy formation  
simulation {\tt GALFORM}.
In $\S$\ref{PhysicalChar} we analyse the properties of bubbles and the 
 mass and metal outflow rate, and their dependence on galaxy properties. We also present 
 analytic derivations of some of the relations found in this work, giving insight into the physics which sets the outflow rate. 
We study the physical regimes of SN feedback and compare with 
observations of mass outflow rates and velocities 
in galaxies. In $\S$\ref{NewBetasMod} we present a new parametrisation of the outflow rate that accurately describes 
the full dynamical calculation of SN feedback and compare this to parametrisations that are 
widely used in the literature.
In $\S$\ref{Sec:Galaxies} we show how the new SN feedback model affects the galaxy luminosity function 
and the SFR density evolution. 
We discuss our results and present our
conclusions in $\S$\ref{Sec:Conclus_SNepaper}. In Appendix~\ref{App:YieldRecycle} we describe how we calculate the 
recycled fraction and yield from supernovae, in Appendix~\ref{Profiles} we explain how we calculate the stellar 
and DM mass enclosed by bubbles, and in Appendix~\ref{App:MassRates} we describe how we calculate the overall 
rates of break-out and confinement 
 of bubbles in the ISM.

\section{Modelling superbubble expansion driven by supernovae}\label{Sec:DynModel_SN}

In this section we describe the physical treatment we apply to bubbles and their expansion 
in the ISM. We consider that galaxies have an ISM which is initially characterised by two gas phases: 
the diffuse, atomic phase and the dense, molecular phase. The molecular gas 
is assumed to be locked up in GMCs and stars are allowed to form 
only in these regions. We use the empirical relation proposed by \citet{Blitz06} which connects the 
atomic-to-molecular surface density ratio to the hydrostatic gas pressure (see $\S$~\ref{Sec:modeldetails} for details). 
We use the observed molecular star formation (SF) rate coefficient, $\nu_{\rm SF}$, 
to calculate the rate at which stars form from molecular gas (e.g. Bigiel et al. 2008, 2010).
 
The onset of star formation 
in GMCs results in SNe. SNe inject mechanical energy and momentum into the surrounding 
medium, which pressurises the immediate region inflating a cavity of hot gas, called a SNe driven bubble. 
We follow the evolution of the bubbles from an initial adiabatic phase to a possible radiative phase. 
The interiors of bubbles correspond to a third phase in the ISM of galaxies: a hot, low density gas phase. 
Bubbles start their expansion conserving energy, but soon after 
the expansion starts (after a cooling time), the interiors of bubbles become radiative. Bubbles then enter into a 
pressure-driven phase, in which the interior gas is still hot and highly pressurised. Once this interior gas cools radiatively, 
bubbles continue their evolution conserving momentum.

The main considerations 
we take into account when following the evolution of bubbles are:
\begin{itemize}
\item The injection of energy by SNe lasts for a finite period of time, which corresponds to 
the lifetime of a GMC.
\item The gravity of stars and dark matter is included and can decelerate the expansion of bubbles.
\item Temporal changes are followed in the atomic, molecular, stellar and dark matter contents, with 
bubbles evolving in this dynamical environment.
\item We allow bubbles to be offset from the centre of the galaxy but they are centered on the midplane 
of the disk. We therefore consider local 
properties when calculating the expansion of bubbles.
\item Metal enrichment in the ISM due to massive stars takes place through bubbles.
\item We follow the radiative cooling in the interior of bubbles to make an accurate estimate 
of the transition between the adiabatic and radiative stages of bubble evolution.  
\end{itemize}

\noindent We solve the equations describing the evolution of bubbles numerically to prevent having to apply 
restrictive 
assumptions to features we would like to test, such as the effect of ambient pressure and  gravity on the expansion of bubbles.
We make three key assumptions when solving for the evolution of bubbles:
\begin{itemize}
\item Star formation taking place in a single GMC gives rise to a new generation of SNe. 
We assume that the group of SNe in a single GMC inflate a single bubble. 
Thus, each bubble is accelerated by a number of SNe, the value of which depends on the SFR in the GMC and 
the initial mass function of stars (IMF).  
\item We assume bubbles are spherically symmetric. 
Observations of SNe remnants show that the geometry of bubbles is 
close to spherical in most cases (e.g. \citealt{Green09}).
This assumption does not restrict the level of accuracy that can
be added into the equations of momentum and energy describing the evolution of bubbles.
\item We assume that bubbles expand only
through the diffuse atomic medium and that the gas in GMCs is not
affected by these expanding bubbles. This is motivated by the fact that GMCs
are characterised by large gas densities
which tend to reflect the energy carried out by bubbles rather than
absorbing it (e.g. \citealt{McKee75}; \citealt{Elmegreen99}). 
In addition, \citet{Dale12}  
 and \citet{Walch12} show that
 at the moment of explosion of massive stars, the surrounding gas
has already been photo-ionised by the radiation emitted by those stars. \citet{Hopkins12} 
show that this effect is also present in their simulations of individual galaxies.
This implies that SNe can efficiently accelerate the surrounding diffuse gas, 
causing the adiabatic expansion of a bubble to last for longer.
\end{itemize}

\begin{figure}
\begin{center}
\includegraphics[width=0.45\textwidth]{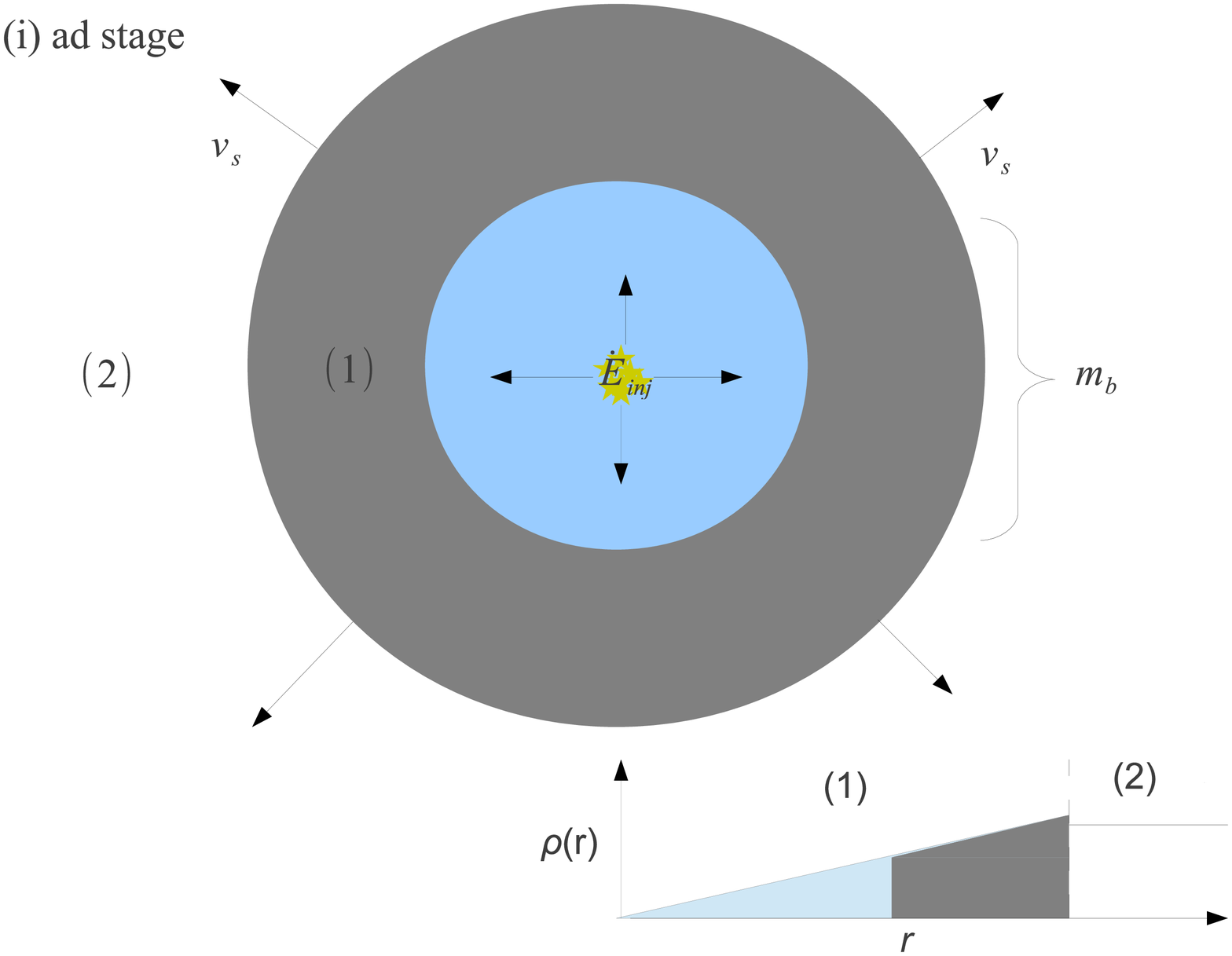}
\includegraphics[width=0.45\textwidth]{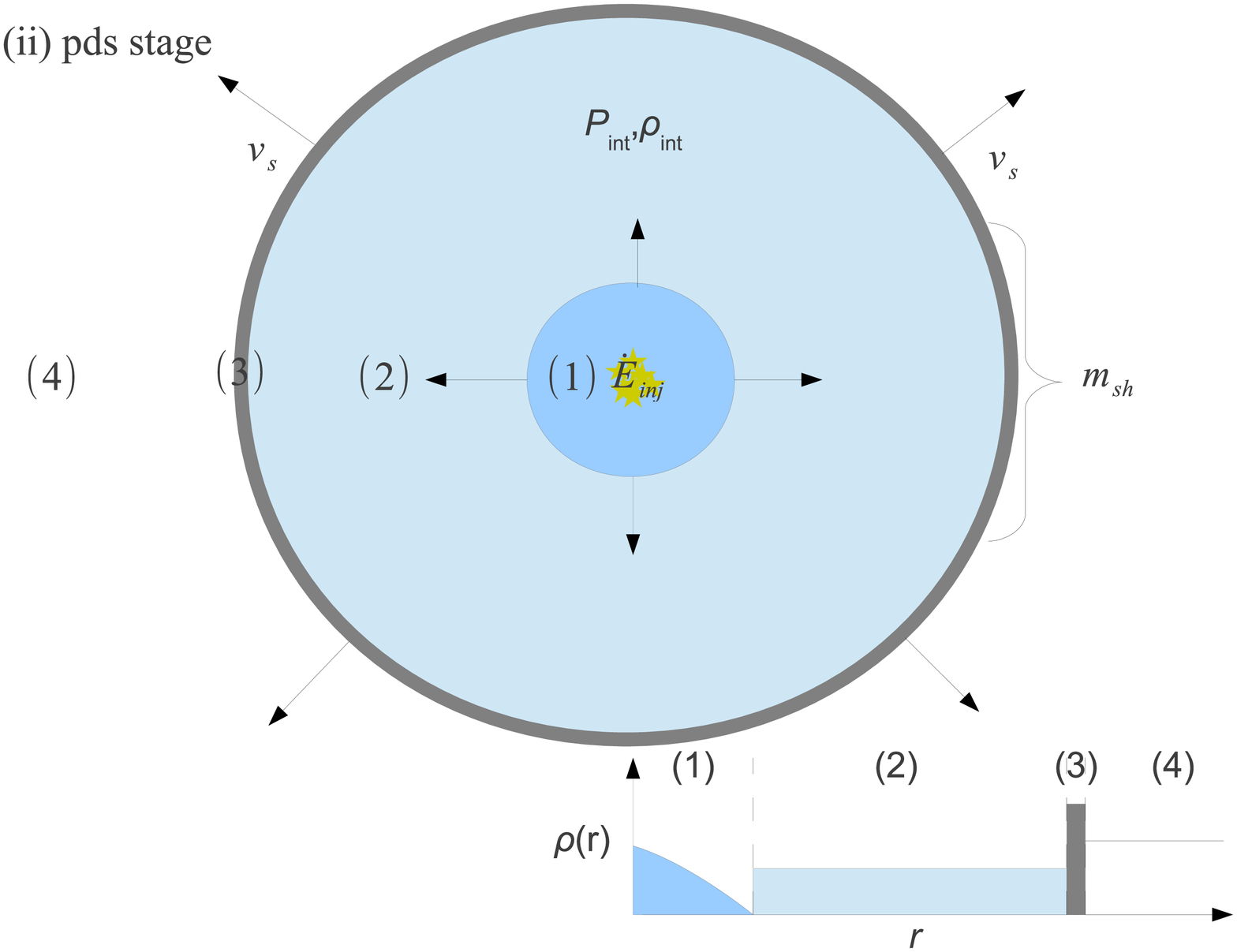}
\includegraphics[width=0.45\textwidth]{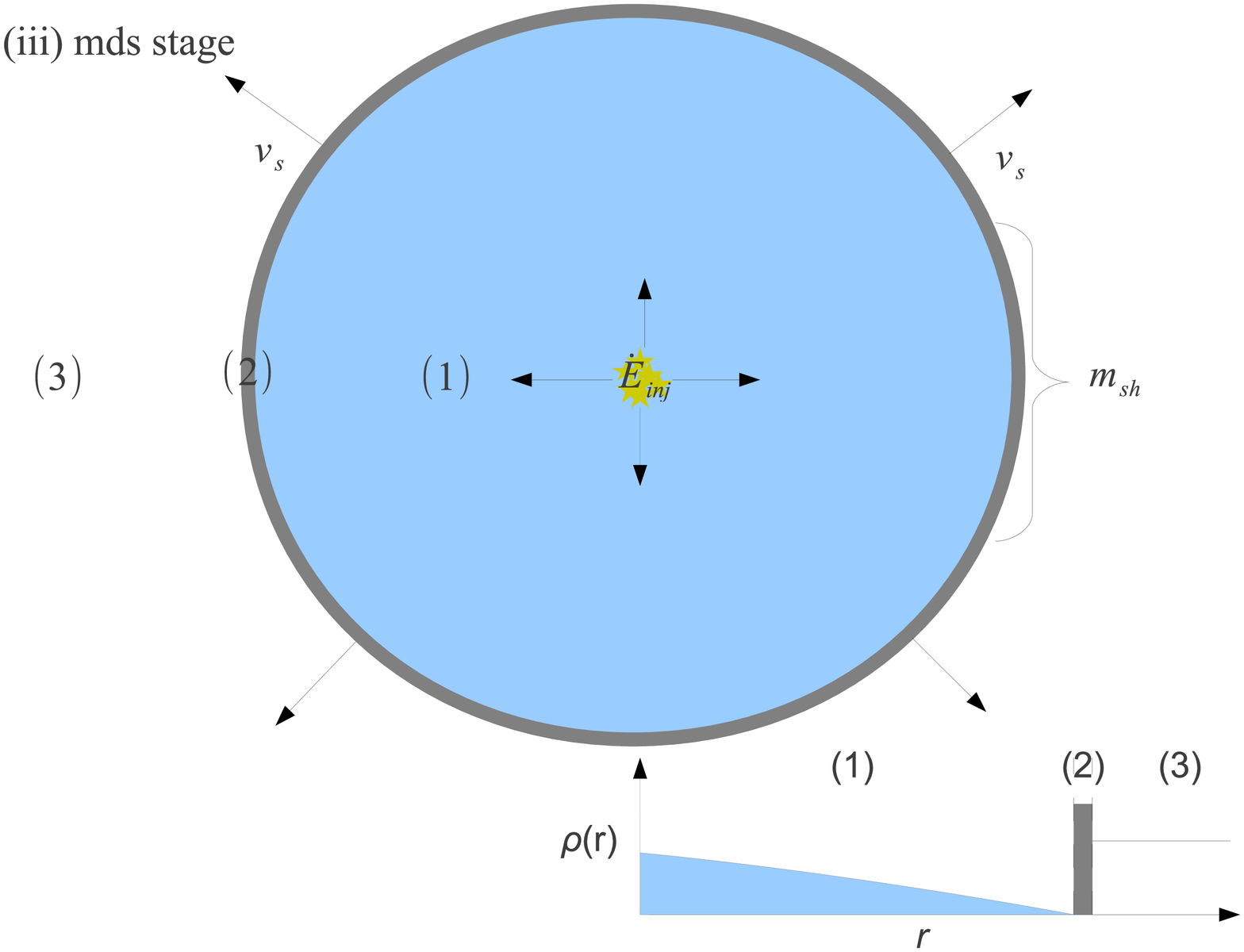}
\caption[Schematic view of the inner structure of bubbles in three of the expansion stages 
considered in our dynamical model of SNe feedback.]{Schematic 
 of the inner structure of bubbles in three of the expansion stages considered in our 
dynamical model of SNe (see $\S$~\ref{Sec:DynModel_SN}).  
SNe inject energy at a rate $\dot{E}_{\rm inj}$, at the centre of the bubble 
and the ambient medium surrounds the bubble. 
A schematic of the gas densities as a function of radius depicting the inner structure of 
the bubble is shown in the bottom right of each panel.
{\it Top panel:} The adiabatic (`ad') stage. The overpressurised region 
initially expands adiabatically, with the density increasing towards the edge of the bubble due to the swept-up gas, 
producing a thick shell. {\it Middle panel:} The pressure-driven snowplough (`pds') stage. 
Once the cooling time becomes shorter than the expansion time, 
the internal mass collapses to a shell. The interior mass fueled by the injected mass from SNe 
remains adiabatic.
The interior accelerates the outer shell through 
pressure. {\it Bottom panel:} The momentum-driven snowplough (`mds') stage. 
Once the cooling time in the interior becomes shorter than the expansion time in the 
`pds' stage, the interior mass collapses to the shell and forms a bubble with a cooled, low density interior. The mass and 
energy injected by SNe modify directly the motion of the outer shell through momentum injection.} 
\label{GeomBubble2}
\end{center}
\end{figure}

In $\S$~\ref{EvoSingleBubble} we describe 
the three evolutionary stages for a single bubble 
outlined above and give the equations we use to determine the mass, radius, velocity 
and temperature of the expanding bubbles. In $\S$~\ref{DiffISMandClumps} we describe how we estimate 
the properties of GMCs and the diffuse medium, and how we connect these to the global properties 
of galaxies.

\subsection{Expansion of a single bubble}\label{EvoSingleBubble}

Let us consider a bubble located at a distance $d$ from the galactic centre and expanding in 
a diffuse medium characterised by density $\rho_{\rm d}$, velocity dispersion 
$\sigma_{\rm d}$, pressure $P_{\rm d}$, internal energy density $u_{\rm d}$ and metallicity $Z_{\rm g}$.  

A single GMC has a SFR of $\psi_{\rm GMC}$ and lasts for a time 
$\tau_{\rm life,GMC}$. Within the cloud, the rate of SNe events is 
$\eta_{\rm SN}\psi_{\rm GMC}$, where $\eta_{\rm SN}$ is the number of SNe 
per solar mass of stars formed. The latter depends on the IMF adopted.  
Individual SNe release $E_{\rm SN}=10^{51}\,{\rm erg}$ 
(\citealt{Arnett89}; \citealt{Woosley95}). With these definitions in mind 
we set out the equations we use to follow the expansion of bubbles in the following 
three subsections. 

\subsubsection{The adiabatic expansion}\label{Sec:Ener}

The pressure generated by SNe can significantly exceed that of the ISM, producing 
a hot cavity. When radiative losses are negligible, the hot cavity evolves 
like a stellar wind bubble which cools adiabatically.
 The interior of the bubble is thermalised and its motion drives a shock 
into the ISM and starts to sweep up the surrounding gas (\citealt{Ostriker88}).
The inner structure of the bubble corresponds to a thick shell of gas swept-up from the 
ambient interstellar medium. 
The top-panel of Fig.~\ref{GeomBubble2} shows a schematic of the inner structure of bubbles in this stage, 
which we refer to with the label ``ad''. The internal gas density profile is illustrated 
in the bottom-right corner. 

The bubble at this stage is characterised by kinetic and thermal 
energies $E_{\rm K}$ and $E_{\rm th}$, respectively, a radius $R$ and an expansion speed 
$v_{\rm s}={\rm d}R/{\rm d}t$, which evolve with time. The total mass of the bubble, $m_{\rm b}$, 
 corresponds to the sum of the mass injected by SNe, $m_{\rm inj}$, and 
the swept-up from the diffuse ISM, $m_{\rm sw}$. The rate of mass injection depends 
on $\psi_{\rm GMC}$ and the fraction of the total mass that is returned to the medium by massive stars,
$R_{\rm SN}$, through $\dot{m}_{\rm inj}=R_{\rm SN}\,\psi_{\rm GMC}$.
 Explicit expressions for $\eta_{\rm SN}$ and $R_{\rm SN}$ are given in 
Appendix~\ref{App:YieldRecycle}.

The expansion of the inflated bubble is described by the equations of 
energy and mass conservation,

\begin{eqnarray}
E&=& E_{\rm th}+E_{\rm K}=\kappa_{\rm E}\, m_{\rm b}\, v^2_{\rm s}\label{ener1}\\ 
\frac{\rm d\it E}{{\rm d}t} & = & \dot{E}_{\rm inj}+4\pi\, R^2\, v_{\rm s}\cdot\\
                          & &  \left(u_{\rm d}-\rho_{\rm d}\frac{G\,M_{\rm t}(R,d)}{R}-\rho_{\rm t}\frac{G\,m_{\rm b}}{R}\right)\nonumber\label{ener1.5} \\
\frac{\rm d\it m_{\rm b}}{{\rm d}t} & = & \dot{m}_{\rm inj} + 
4\pi\,R^2\,\rho_{\rm d} \, v_{\rm s}.\label{ener2}
\end{eqnarray}

\noindent Here, $E$ is the total energy of the bubble in the 
adiabatic stage and $\dot{E}_{\rm inj}$ is the energy injection rate from SNe.

The total stellar plus DM mass enclosed by a bubble is $M_{\rm t}(R,d)$ 
and the average density of stars and DM within the bubble is $\rho_{\rm t}$. 
Both terms act to decelerate the expansion of the bubble and come from the gravitational term 
 $\int^{V_{\rm b}}_0 \rho(r)\,v(r)\,g(r)\,{\rm d}V$ in the energy conservation equation, where $V_{\rm b}$ 
is the volume enclosed by the bubble.
The term $G\rho_{\rm t}\,m_{\rm b}/R$ represents the increase of gravitational energy internal to the bubble due to the expanding shell 
(see Appendix~\ref{Profiles} for a description of the calculation 
of the stellar and DM profiles and the mass enclosed in $R$).
Note that here we neglect the self-gravity of the bubble, given that $m_{\rm b} \ll M_{\rm t }(R,d)$. 

The ratio $E/(m_{\rm b}\, v^2_{\rm s})=\kappa_{\rm E}$ is calculated using a single power-law dependence 
of the velocity and density on the radius inside the bubble ($\rho\propto r$ and $v\propto r$), which gives $\kappa_{\rm E}=3/4$, for 
a ratio of specific heats of $\gamma=5/3$ (corresponding to a monoatomic gas; \citealt{Ostriker88}). 
 The energy injection rate is calculated from the SNe rate, $\eta_{\rm SN}\, \psi_{\rm GMC}$, 
and the mechanical energy produced by an individual SN, $E_{\rm SN}$,  

\begin{equation}
\dot{E}_{\rm inj}=E_{\rm SN}\, \eta_{\rm SN}\, \psi_{\rm GMC}.
\label{Eq:Einj}
\end{equation}

Note that the pressure of the diffuse medium 
does not affect the energy of the bubbles, given that the diffuse ISM is static with respect to the bubbles. 
This means that there is no coherent motion in the ISM, only random motions characterised by a velocity dispersion 
$\sigma_{\rm d}$.

For the rate of change in the mass internal to the bubble in Eq.~\ref{ener2}, the right-hand side of the equation 
corresponds to the rate at which mass is incorporated from the diffuse medium into the bubble. 
We also keep track of the swept-up mass, $m_{\rm sw}$, in order to subtract it from the diffuse ISM component 
when solving the SF equations (see $\S$~\ref{Sec:SFeqs_sn}), 

\begin{equation}
\frac{\rm d\it m_{\rm sw}}{{\rm d}t}  = 4\pi\,R^2\,\rho_{\rm d}\, v_{\rm s}.
\label{mdrag}
\end{equation}

Metals produced by nucleosynthesis in stars and ejected by SNe are added 
to the hot cavities. The rate of metal injection by SNe into the hot cavity depends on 
the SFR, $\psi_{\rm GMC}$, the SNe metal yield, $p_{\rm SN}$, and the metallicity of the gas from which the stars were formed, 
$Z_{\rm g}$, and is given by 
$\dot{m}^{Z}_{\rm inj}=(p_{\rm SN}+R_{\rm SN}Z_{\rm g})\psi_{\rm GMC}$.
 The term $p_{\rm SN}\psi_{\rm GMC}$ corresponds to the      
newly synthesized metals and $R_{\rm SN} Z_{\rm g}\psi_{\rm GMC}$ to the metals 
present in the gas from which stars were made 
(see Appendix~\ref{App:YieldRecycle} for a description of how the recycled fraction and yield are calculated). 
 
The rate of change in the mass of metals in the interior of bubbles and 
in the swept-up gas component are given by: 

\begin{eqnarray}
\frac{{\rm d}  m^{\rm Z}_{\rm b}}{{\rm d}t}&=& \dot{m}^{Z}_{\rm inj}+\frac{{\rm d}  m^{\rm Z}_{\rm sw}}{\rm dt}, \\
\frac{{\rm d}  m^{\rm Z}_{\rm sw}}{{\rm d}t} &=& 4\pi R^2 \rho_{\rm d}\, v_{\rm s}\, Z_{\rm g}.
\end{eqnarray}

\noindent Similarly to Eq.~\ref{mdrag}, 
it is possible to isolate the metals that have been incorporated into bubbles from the ISM, 
$m^{Z}_{\rm sw}$.  
The internal metallicity of a bubble is therefore 
$Z_{\rm b}=m^{\rm Z}_{\rm b}/m_{\rm b}$. This way, the enrichment of
 the ISM will depend on the rate of bubble confinement and break-out. 

The high temperature of the interior of bubbles results in a large sound speed, $c_{\rm s}\gg v_{\rm s}$,
which makes the time for a sound wave to cross the interior much shorter than the expansion time.
This causes the interior to be isobaric, characterised by a mean pressure $P_{\rm b}$.
We calculate the internal bubble pressure, temperature ($T_{\rm b}$)  
and cooling time ($t_{\rm cool}$), with the latter two properties defined just behind the shock at $R$ 
(see top panel of Fig.~\ref{GeomBubble2}), using 

\begin{eqnarray}
P_{\rm b} &=& \frac{2}{3}\, u=\frac{E_{\rm th}}{2\pi R^3},\label{Pint1}\\
T_{\rm b}(R) &=& \frac{\mu\, m_{\rm H}\, P_{\rm b}}{\kappa_{B} \rho_{\rm b}(R)},\label{Teq1}\\
t_{\rm cool}(R) &=& \frac{3\,k_{\rm B}\, T_{\rm b}(R)}{n_{\rm b}\,\Lambda(T_{\rm b}(R),Z_{\rm g})}\label{tcooleq1}.
\end{eqnarray}

\noindent Here, the internal pressure of a bubble is calculated from its  
internal energy, $u$, $k_{\rm B}$ is Boltzmann's constant, the mean molecular 
weight of a fully ionised gas (i.e. internal to the bubble) is $\mu=0.62$, $m_{\rm H}$ is 
the mass of a hydrogen atom, $\Lambda(T_{\rm b},Z_{\rm b})$ corresponds to 
the cooling function and 
 $n_{\rm b}=\rho_{\rm b}(R)/(\mu\, m_{\rm H})$, is the volume number density behind the shock. 
We adopt the cooling function tables of \citet{Sutherland93}.

In order to set the correct initial conditions for the expansion in the adiabatic phase, 
we use the analytic solutions to the set of Eqs.~\ref{ener1}-\ref{ener2} 
given by \citet{Weaver77}. These analytic solutions are obtained by neglecting the pressure and internal energy of the ambient medium, and 
the gravity of the stellar plus dark matter component and by assuming that the injected mass
is small compared to the swept-up mass.  
We do this for an initial short period of time, $t^{\prime}$, which we quantify in terms of the 
cooling time, $t^{\prime}\le 0.1\,t_{\rm cool}$. At $t>t^{\prime}$, 
we follow the solution in the adiabatic stage numerically to accurately track
the transition to the radiative phase.
Our results are insensitive to the precise values of $t^{\prime}$, provided that $t^{\prime}<0.3\,t_{\rm cool}$. 
The properties of bubbles during this early adiabatic period are: 

\begin{eqnarray}
R_{\rm b}(t) & = & \alpha \, \left(\frac{\dot{E}_{\rm inj}}{\rho_{\rm d}}\right)^{1/5}\,t^{3/5}, \label{ener:analytic1}\\
v_{\rm s}(t) & = & \frac{3}{5}\,\alpha\,\left(\frac{\dot{E}_{\rm inj}}{\rho_{\rm d}}\right)^{1/5}\,t^{-2/5},\\
m_{\rm sw}(t) & = &  \frac{4 \pi}{3}\,\alpha^3\, \dot{E}_{\rm inj}^{3/5}\,\rho^{2/5}_{\rm d} \, t^{9/5}, \label{ener:analytic1.5}\\
m^{\rm Z}_{\rm sw}(t) &=& m_{\rm sw}(t)\, Z_{\rm g},\\ 
m_{\rm b}(t) & = & m_{\rm sw}(t)+R_{\rm SN}\psi_{\rm GMC}\, t, \label{ener:analytic1.7}\\
m^{\rm Z}_{\rm b}(t) & = & m^{\rm Z}_{\rm sw}(t)+(p_{\rm SN}+R_{\rm SN} Z_{\rm g})\psi_{\rm GMC}\, t, 
\label{ener:analytic2}
\end{eqnarray}

\noindent where $\alpha=0.86$. 
 Eqs.~\ref{ener:analytic1.7}-\ref{ener:analytic2} account for the injected metals and mass from the dying stars.

\subsubsection{Pressure-driven snowplough expansion}\label{Sec:Mom}

As the temperature of the bubble decreases with time, 
the cooling time becomes sufficiently short so as to 
be comparable with the expansion time of the bubble. At this stage, radiative losses from  the 
expanding thick shell 
can no longer be neglected and the shocked swept-up material quickly becomes thermally unstable and 
collapses into a thin, dense shell.  
The shocked mass ejected by SNe in the interior of the thin shell still conserves its energy and the 
bubble enters a pressure-driven phase. The energy injected by SNe modifies the thermal energy 
of the shocked interior.  
We refer to properties of bubbles in this stage with the label ``pds'', denoting pressure-driven snowplough 
(see middle panel of Fig.~\ref{GeomBubble2}). 

In this phase bubbles are characterised by the swept-up mass 
accumulated in a thin shell, $m_{\rm sh}$, and an interior mass, $m_{\rm int}$.
The interior of the bubble is still isobaric, characterised by a mean pressure, $P_{\rm int}$. 
We consider that 
the density of the shocked SNe injected material is constant and is calculated as 
$\rho_{\rm int}=m_{\rm int}/(4/3\pi R^3)$.

We calculate $P_{\rm int}$ using Eq.~\ref{Pint1}, 
$P_{\rm int}=E_{\rm int}/2\pi R^3$, where $E_{\rm int}$ is the interior energy of the bubble and 
is calculated from the 
energy gained from SNe ($\dot{E}_{\rm inj}$) and the energy loss due to the work done by the interior gas 
on the expanding shell,

\begin{equation}
\frac{{\rm d}E_{\rm int}}{{\rm d}t}=\dot{E}_{\rm inj}-4\pi\, R^2\, v_{\rm s}\, P_{\rm int}.
\label{Eint}
\end{equation}

\noindent The rate of change of mass and metals in the interior of bubbles are set by the mass and metals
 injection rates by SNe, $\dot{m}_{\rm int}=\dot{m}_{\rm inj}$ and 
$\dot{m}^Z_{\rm int}=\dot{m}^Z_{\rm inj}$.

The temperature and cooling time in the interior of the bubble are calculated following Eqs.~\ref{Teq1} and 
\ref{tcooleq1}, but replacing $\rho(R)$ by $\rho_{\rm int}=m_{\rm int}/(\frac{4}{3}\pi R^3)$,  
$P_{\rm b}$ by $P_{\rm int}$ and $Z_{\rm b}$ by $Z_{\rm int}=m^{Z}_{\rm int}/m_{\rm int}$.

The equations of motion and 
 of the conservation of the total mass and mass in metals for the shell in the pressure-driven stage are 

\begin{eqnarray}
\frac{\rm d\rm (\it m_{\rm sh}\,v_{\rm s})}{{\rm d}t} & = &4 \pi\, R^2\,(P_{\rm int}-P_{\rm d})-\frac{G\,M_{\rm t}(R,d)}{R^2}\,m_{\rm sh}\label{press1}\\
\frac{{\rm d}m_{\rm sh}}{{\rm d}t}\it  & = & 4\pi\,R^2\,\rho_{\rm d}\, v_{\rm s},  \\
\frac{{\rm d} m^{\rm Z}_{\rm sh}}{{\rm d}t} &=& 4\pi\,R^2\,\rho_{\rm d}\, v_{\rm s}\, Z_{\rm g}. \label{press2}
\end{eqnarray}

\noindent Note that the expansion of the bubbles is driven by the pressure difference $(P_{\rm int}-P_{\rm d})$.
The gravitational term $G\,M_{\rm t}m_{\rm sh}/R^2$ comes from integrating $g\delta M$ over all the mass elements inside a radius that is 
comoving with the diffuse medium in the equation of motion for an element of fluid of mass $\delta M$. We 
neglect the shell self-gravity, given that $m_{\rm s} \ll M_{\rm t}(R,d)$.
 
\subsubsection{Momentum-driven snowplough expansion}\label{Sec:Mom}

When the expansion time in the pds stage becomes longer than the cooling time of the interior, the bubble 
enters to the momentum-driven phase. 
The cavity interior to the bubble is composed of low-density cooled gas 
of total mass $m_{\rm int}$. This interior mass corresponds to the 
ejected mass from SNe that has not yet had enough time to encounter the shell. 
The explosions at the centre inject mass and momentum into the shell.
The interior density is calculated from the continuity equation,

\begin{equation}
\rho_{\rm int}=\frac{\dot{m}_{\rm inj}}{4\pi R^2 v_{\rm inj}}.
\label{rhoint}
\end{equation}

\noindent The density of the ejected material 
drops with radius and by the time the ejected gas encounters the shell, most of the energy input by SNe  has become 
kinetic energy. Therefore, SNe ejected material acts on the shell by increasing the momentum of the shell 
(see schematic in the bottom panel of Fig.~\ref{GeomBubble2}).
We therefore 
consider that $v_{\rm inj}=\sqrt{2\,\dot{E}_{\rm inj}/\dot{m}_{\rm inj}}$.
The equations describing the change of mass 
and mass in metals of the bubble interior are: 

\begin{eqnarray}
\frac{{\rm d}m_{\rm int}}{{\rm d}t} & = & \dot{m}_{\rm inj}\, \frac{v_{\rm s}}{v_{\rm inj}},  \label{mint_mds}\\
\frac{{\rm d}m^{\rm Z}_{\rm int}}{{\rm d}t} & = & \dot{m}^Z_{\rm inj}\, \,\frac{v_{\rm s}}{v_{\rm inj}}.\label{int1}
\end{eqnarray}

\noindent Here, the amount of injected mass that remains in the interior of the bubble depends on the velocity ratio 
$v_{\rm s}/v_{\rm inj}$, which means that if the shell expands slowly, most of the mass injected by SNe quickly reaches 
the shell. Note that gravity is neglected in the motion of the interior material. 

The equations describing the conservation of momentum, 
total mass and mass in metals for the mds stage are, 

\begin{eqnarray}
\frac{\rm d\rm (\it m_{\rm sh}\,v_{\rm s})}{{\rm d}t} & = &  \dot{m}_{\rm inj}\,(v_{\rm inj}-v_{\rm s})-\frac{G\,M_{\rm t}(R,d)}{R^2}\,m_{\rm sh}\nonumber \\
 & & -4\pi\, R^2\,P_{\rm d}, \label{mom1}\\
\frac{{\rm d} m_{\rm sh}}{{\rm d}t} & = & \dot{m}_{\rm inj}\,\left(1-\frac{v_{\rm s}}{v_{\rm inj}}\right)+4\pi\,R^2\,\rho_{\rm d}\, v_{\rm s},  \\
\frac{{\rm d}m^{\rm Z}_{\rm sh}}{{\rm d}t}&=&\dot{m}^Z_{\rm inj}\,\left(1-\frac{v_{\rm s}}{v_{\rm inj}}\right) \nonumber, \\
 & & + 4\pi\,R^2\,\rho_{\rm d}\, v_{\rm s}\, Z_{\rm d}. \label{mom2}
\end{eqnarray}

\noindent Note that the expansion of the bubbles is driven by the velocity gradient $(v_{\rm inj}-v_{\rm s})$.

If the bubble has a radius which exceeds the scale height of the galaxy, part of the bubble 
would be expanding in a lower density medium (see bottom panel of Fig.~\ref{GeomBubble}). We account for this 
by including a correction factor 
in the density of the diffuse medium when $R>h_{\rm g}$, $\rho^{\prime}_{\rm d}=\rho_{\rm d}\, (1-h_{\rm g}/R)$, 
 which accounts for the 
fraction of the surface of the bubble outside the disk. We replace 
$\rho_{\rm d}$ by $\rho^{\prime}_{\rm d}$ in the set of equations describing the evolution of bubbles.

\subsection{Properties of molecular clouds and the diffuse medium in galaxies}\label{DiffISMandClumps}

In this section, we describe how we calculate the properties of GMCs and the diffuse 
medium, and explain the techniques used to follow their evolution throughout the ISM.

\subsubsection{Molecular cloud properties}\label{Sec:GMCs}

The dynamical evolution described above corresponds to a single bubble driven by the 
SF taking place in one GMC. 
In order to incorporate this evolution into the galaxy formation context, we 
 consider GMC formation in the ISM of galaxies and subsequent SF 
in GMCs. For this, it is necessary to define the GMC mass, SF efficiency and the 
timescales for the formation and destruction of 
GMCs. We first define individual GMC properties 
and then connect them to galaxy properties to estimate their number and radial distribution in $\S$~\ref{SubSubSec:ConnectingGMCs}. 

{\it GMC mass.} Motivated by observations of the Milky Way and nearby galaxies, we consider GMCs to have typical masses of 
$m_{\rm GMC}\approx 10^5-10^6\, M_{\odot}$ (e.g. \citealt{Solomon87}; \citealt{Williams97}; \citealt{Oka01}; 
\citealt{Rosolowsky05}). We assume that GMCs are fully molecular and that all the molecular gas in galaxies is 
locked up in GMCs. This is a good approximation for most local galaxies, in which more than 
$90$\% of the molecular gas is in gravitationally bound clouds (\citealt{Ferriere01}). However, 
it is important to note that in the densest nearby starburst galaxies, some molecular gas is also found 
in the diffuse component (e.g. M64; \citealt{Rosolowsky05}). 

{\it The SFR per GMC}. $\psi_{\rm GMC}$ depends on the GMC mass and 
the molecular SF coefficient rate, $\nu_{\rm SF}$, 
as $\psi_{\rm GMC}=\nu_{\rm SF}\, m_{\rm GMC}$. 
To ensure consistency with the global SF law, 
we use the same SF rate coefficient defined in $\S$~\ref{Sec:DynModel_SN}. This implies that, as we incorporate the 
dynamical SNe feedback model in the galaxy formation simulation,  
GMCs forming stars in the disk have different depletion timescales than those forming stars in the bulge 
(see $S$~\ref{Sec:modeldetails} for details).
This difference in the SF timescales of GMCs in normal star-forming galaxies and starbursts (SBs) 
has been proposed theoretically by \citet{Krumholz09}. Krumholz et al. 
 argue that in normal galaxies the ambient pressure is negligible compared to 
the internal pressure of GMCs, and therefore, the properties setting the SF are close to universal. However, 
in high gas density environments appropriate to SBs, the ambient pressure becomes equal to 
the typical GMC pressure, and therefore, in order to maintain GMCs as bound objects, their properties need to 
change according to the ambient pressure. This naturally produces a dichotomy between normal star-forming 
galaxies and starburst galaxies. 

{\it GMC lifetime.} The formation and destruction timescales of GMCs 
depend on the properties of the ISM: gas density, convergence flow velocities, magnetic fields, 
turbulence, etc. (\citealt{McKee07}).
GMCs can form through large-scale self-gravitating instabilities, which can include
Parker, Jeans, magneto-Jeans and/or magneto-rotational
instabilities (e.g. \citealt{Chieze87}; \citealt{Maloney88};
\citealt{Elmegreen89}; \citealt{McKee99};
\citealt{Krumholz05}), or 
through collisions of large-scale gas flows 
(e.g. \citealt{Ballesteros-Paredes99}; \citealt{Heitsch05}; \citealt{Vazquez-Semadeni06}).
 GMCs in these formation scenarios tend to last $\sim 1-3$ crossing times before being destroyed by stellar
feedback (i.e. proto-stellar and stellar winds, and HII regions).
Observationally, the lifetime of GMCs is inferred from 
 statistical relations between the location of GMCs and young star clusters and is 
in the range $10-30$~Myr (e.g. \citealt{Blitz80}; \citealt{Engargiola03}; \citealt{Blitz07}). 
We therefore restrict the range of the lifetimes of GMCs to $\tau_{\rm life,GMC}=10-30$~Myrs.

\subsubsection{Properties of the pervasive interstellar medium}\label{SubSec:DiffMedProps}

We assume that the diffuse pervasive medium in the ISM is fully atomic.  
We define the relevant properties of the diffuse medium (see Eqs.\ref{ener1}-\ref{mom2}) 
as a function of radius for the disk and bulge. 

For the gas surface density profiles of the disk and bulge, we assume 
that both are well described by exponential profiles with half-mass radii, 
$r_{\rm 50,d}$ and $r_{\rm 50, b}$, respectively. 
This is done for simplicity. However, it has been shown that the neutral gas 
(atomic plus molecular) in nearby spiral galaxies 
 follows an exponential radial profile \citep{Bigiel12}. 
\citet{Davis12} found that this is also the case in a large percentage 
of early-type galaxies in the local Universe.
In interacting galaxies and galaxy mergers, Davis et al. show 
 that the gas can have very disturbed kinematics, and in these cases our approximation 
is no longer valid.  

To calculate the HI surface density we follow \citet{Lagos10} and use 
the \citet{Blitz06} pressure law ($\S$~\ref{Sec:SFeqs_sn}). 
We assume this pressure-law also holds in higher gas density media, typical of SBs. 
Hydrodynamic simulations including 
the formation of H$_2$ have shown that, for extreme gas densities, the relation between 
hydrostatic pressure and the $\Sigma_{\rm H_2}/\Sigma_{\rm HI}$ ratio deviates from the 
empirical pressure law resulting in more 
 H$_2$ \citep{Pelupessy09}. If the conclusions of Pelupessy et al. are correct, 
 our assumption that the Blitz \& Rosolowsky law holds 
for SBs would represent an upper limit for the HI mass. 
The effect of this systematic on the final result of SNe feedback is highly non-linear given that having more HI mass makes 
 the expansion of bubbles more difficult, but in the case of escape, more outflow mass is released 
from the galaxy.

We assume that gas motions in the diffuse medium are dominated by a random component 
and we choose the vertical velocity dispersion to be $\sigma_{\rm d}=10\, \rm km\, s^{-1}$ 
\citep{Leroy08}. 
The source of the motion of the diffuse ISM is not relevant so long as it gives rise 
to gas dominated by random motions. 
The assumption of random motions is consistent with turbulence and thermally driven motions   
(e.g. \citealt{Wada02}; \citealt{Schaye04}; \citealt{Dobbs11}).
We estimate the gaseous disk scale height, the volume density and thermal pressure as a function 
of radius, $h_{\rm g}(r_{\rm i})$, $\rho_{\rm d}(r_{\rm i})$ and $P_{\rm d}(r_{\rm i})$, 
respectively. The set of equations defining these properties is  

\begin{eqnarray}
h_{\rm g}(r_{\rm i})&=& \frac{\sigma^{2}_{\rm d}}{\pi\, G\, \left[\Sigma_{\rm g}(r_{\rm i})+\frac{\sigma_{\rm d}}{\sigma_{\star}(r_{\rm i})}\, \Sigma_{\star}(r_{\rm i})\right]},\label{SetEqsDiff1}\\
\rho_{\rm d}(r_{\rm i})&=&\frac{\Sigma_{\rm atom}(r_{\rm i})}{2 \,h_{\rm g}(r_{\rm i})},\\
P_{\rm d}(r_{\rm i})&=&\rho_{\rm d}(r_{\rm i})\sigma^2_{\rm d}.
\label{SetEqsDiff2}
\end{eqnarray}

\noindent Here $\sigma_{\star}$ is the 
velocity dispersion of the stars, and 
$\Sigma_{\rm atom}(r_{\rm i})$, $\Sigma_{\rm g}(r_{\rm i})$ and $\Sigma_{\star}(r_{\rm i})$
are the atomic, total gas (molecular plus atomic) and stellar surface densities, respectively, at $r_{\rm i}$.
In Appendix~\ref{App:BR} we describe the calculation of $\sigma_{\star}$ and the origin of the expression for $h_{\rm g}$. 
The choice of $\sigma_{\rm d}$ fixes the internal energy of the diffuse medium 
throughout the disk and bulge, so that $u=3/2\,P_{\rm d}$. 

Note that we include the contribution of helium in 
$\rho_{\rm d}(r_{\rm i})$. The filling factor of molecular clouds in the ISM is very small, 
typically $F_{\rm GMC}\approx 0.01$ \citep{McKee07}, so we assume that the filling factor of the diffuse gas 
is $F_{\rm d}=1$ and therefore we do not include it in Eq.~\ref{SetEqsDiff1}-\ref{SetEqsDiff2}.

The gas scaleheight includes the gravitational effect of stars through $\Sigma_{\star}(r_{\rm i})$.
The underlying assumption in Eq.~\ref{SetEqsDiff1} is that the galaxy is in vertical equilibrium and that the 
diffuse medium is characterised by a
uniform pressure\footnote{\citet{Shetty12} use a set of vertically resolved hydrodynamic 
simulations to show that 
vertical equilibrium is reached within a vertical crossing time and \citet{Koyama09} show that 
variations in pressure vertically are within a factor of $2$.}. {Using Eq.~\ref{SetEqsDiff1} and for 
$\sigma_{\rm d}=10\,\rm km\,s^{-1}$, we find that
 the mean 
scaleheight of starburst galaxies at $z=0$ is $\approx 50$~pc for galaxies 
with stellar mass in the range $10^8\,M_{\odot}<M_{\rm stellar}<10^9\,M_{\odot}$, and 
 $\approx 10$~pc for galaxies with $10^{10}\,M_{\odot}<M_{\rm stellar}<10^{11}\,M_{\odot}$. 
At $z=7$, these numbers decrease to $\approx 5$~\pc and $\approx 1$~pc, respectively. 
In the case of quiescent galaxies at $z=0$, the mean $h_{\rm g}$ is $\approx 450$~pc for galaxies with 
$10^8\,M_{\odot}<M_{\rm stellar}<10^9\,M_{\odot}$, and $\approx 100$~pc for galaxies 
with $10^{10}\,M_{\odot}<M_{\rm stellar}<10^{11}\,M_{\odot}$.
 At $z=7$, these numbers decrease to $\approx 60$~pc and $\approx 5$~pc, respectively.
Note that $h_{\rm g}$ is very sensitive to 
the velocity dispersion of the gas, and therefore if we assume higher values for $\sigma_{\rm d}$ 
 (see Sec.~\ref{Sec:xtmISM}), we would find scaleheights larger 
by factors of $20$ to $100$.

We warn the reader that observations have shown that local starburst galaxies have 
gas velocity dispersions systematically 
larger compared to spiral and dwarf galaxies 
 (e.g. \citealt{Solomon97}; \citealt{Downes98}), with values that range 
between $\sigma_{\rm d}=20-100\rm \, km\, s^{-1}$, with a median of $\sigma_{\rm d}\approx 60\rm \, km\, s^{-1}$. 
 These values of $\sigma_{\rm d}$ may drive the typical GMC mass to increase too, as the Jeans mass in a disk  
scales with the gas velocity dispersion as $M_{\rm J}\propto \sigma^4_{\rm d}/\Sigma_{g}$.  
In this paper we analyse the general effect of 
 increasing $\sigma_{\rm d}$ and $M_{\rm GMC}$ in the mass loading and velocity of the outflow in Sec.~\ref{Sec:xtmISM}. However, 
 we assume the same velocity dispersion and GMC mass in starbursts as quiescent galaxies for simplicity. 
In a future paper we investigate the effect of assuming different 
 $\sigma_{\rm d}$ and $M_{\rm GMC}$ for starbursts.}

\subsubsection{Connecting GMCs and galaxy properties}\label{SubSubSec:ConnectingGMCs}

We follow the evolution of bubbles in rings within the disk and the bulge, and assume 
cylindrical symmetry: all bubbles at a given radius $r_{i}$ from the centre are 
identical, where $i=1..N_{\rm r}$. 
We estimate the number of molecular clouds in the ISM at a given timestep that give rise to 
a new generation of bubbles. If at a timestep $t=t_{\rm j}$ the radial profile of molecular mass 
is $\Sigma_{\rm mol}(r,t_{\rm j})$, the 
total number of GMCs in an annulus of radius $r_{\rm i}$ and width $\delta r$ is,

\begin{equation}
N_{\rm GMCs,i,j}= \frac{2\pi \int_{r_{\rm i}-\delta r/2}^{r_{\rm i}+\delta r/2} \Sigma_{\rm mol}(r,t_{\rm j})\, r\, {\rm d}r }{m_{\rm GMC}}. 
\label{Ngmcs_annuli}
\end{equation}

\noindent The rate of GMC formation in the annulus $i$  
in a given time $t_{\rm j}$ is therefore estimated as, 

\begin{eqnarray}
\dot{N}_{\rm GMC,new,i,j}&=&\frac{N_{\rm GMCs,i,j}}{\tau_{\rm life,GMC}}.
\label{GMClife}
\end{eqnarray}

\noindent Note that by fixing the SF rate coefficient, $\nu_{\rm SF}$, and the properties of GMCs,  
we are implicitly assuming that all GMCs at a given timestep are forming stars.\\ 

We performed tests to choose the 
value of $N_{\rm r}$ to ensure convergence in the results presented in this work. These tests suggests 
$N_{\rm r}=10$. The spatial extent of each ring $i$ depends on the total extent of the disk we choose to 
resolve. We model out to $5r_{\rm 50}$ in disk radius, 
so the molecular mass enclosed is $>99.999$\% of the total. This defines the extent of the individual annuli,  
$\delta r=5r_{\rm 50}/N_{\rm r}$. 

\begin{figure}
\begin{center}
\includegraphics[trim = 0.0mm 60mm 0mm 30mm,clip,width=0.5\textwidth]{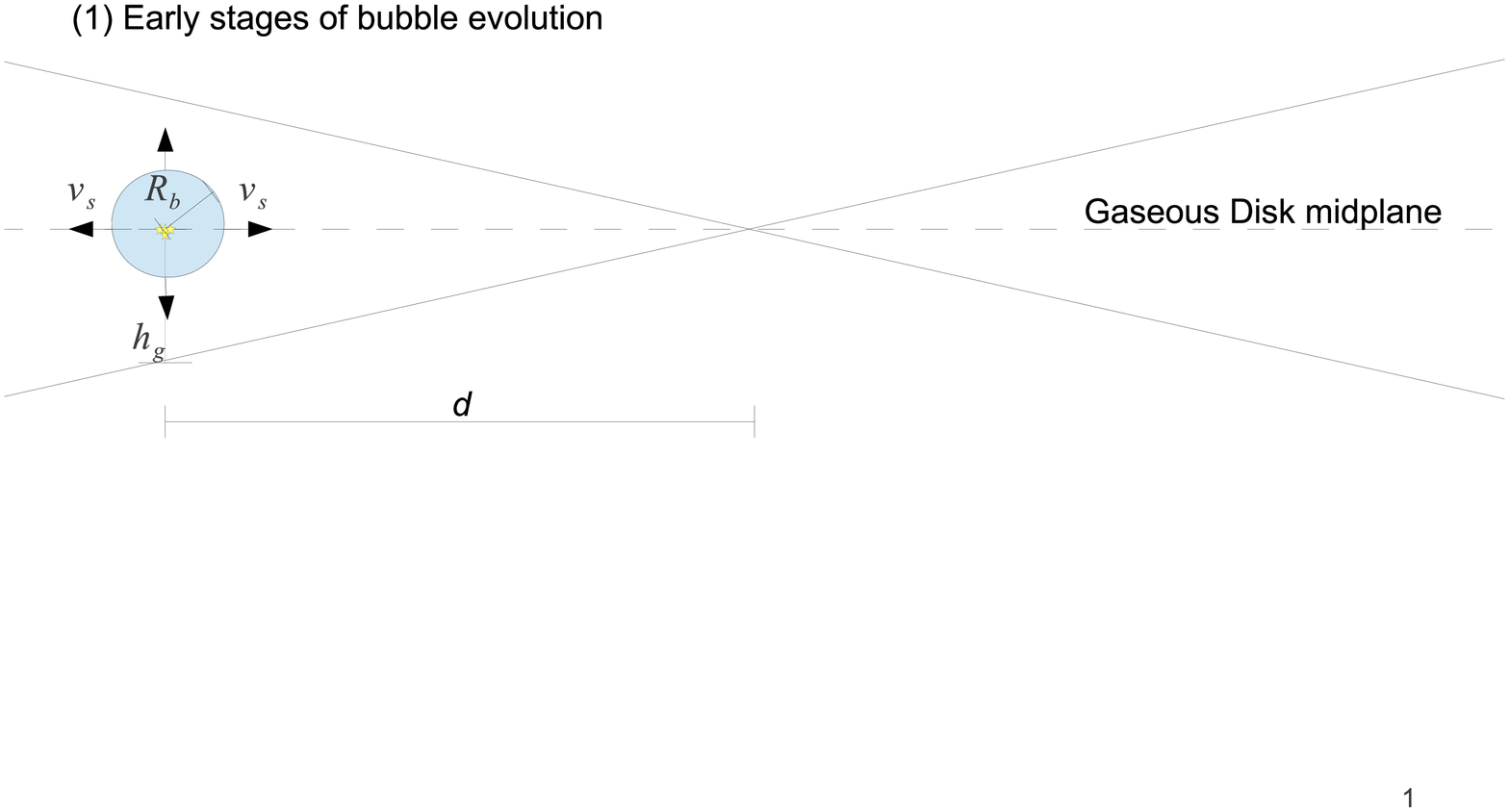}
\includegraphics[trim = 0.0mm 60mm 0mm 30mm,clip,width=0.5\textwidth]{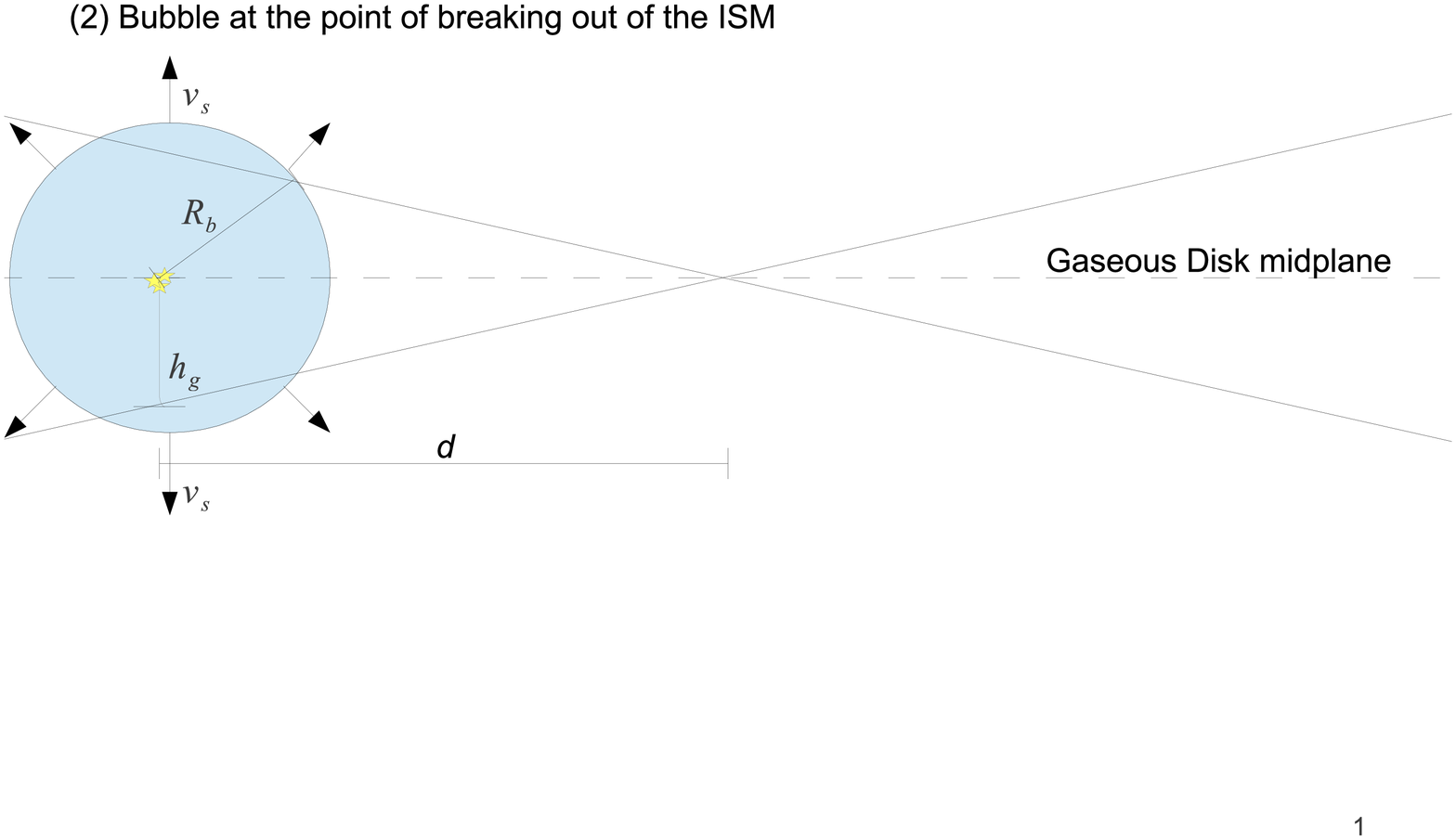}
\caption[Geometry of the dynamical model for supernovae feedback.]{Geometry of the dynamical model for supernovae feedback. 
{\it Top panel:} the early stages of pressurised bubble growth due to SNe, where 
the bubble is fully embedded in the ISM, at a distance $d$ from the galaxy centre, where 
the disk has a gas scale height of $h_{\rm g}$. 
The bubble radius and expansion velocity are $R_{\rm b}$ and $v_{\rm s}$, respectively.  
{\it Bottom panel:} Schematic showing the stage of bubble evolution just before breaking out from the 
ISM. At this stage the bubble has just exceeded the gas scale height.}
\label{GeomBubble}
\end{center}
\end{figure}

{Note that, at high-redshift, galaxies can have large fractions of molecular gas \citep{Lagos11}.
Due to our assumptions, namely, that the molecular gas is locked up in GMCs and that bubbles do 
work against the diffuse medium,  
this large molecular gas content has an effect on the dynamics of bubbles only through its
gravitational effect on the midplane of the disk and the higher SFRs, which result in more SNe. 
 Although our model can be improved to include other physical effects that are enhanced at the 
 contact surface between the supperbubble and high density media, 
we show in $\S$~\ref{ParameterEffects} and $\S$~\ref{Sec:xtmISM} 
that our predictions for the mass loading and velocity of the winds 
are currently limited by our choice of parameters describing the ISM and GMCs.}

\subsubsection{Bubble confinement and break-out}

\begin{table*}
\begin{center}
\caption{List of parameters in the dynamical SNe feedback model. 
In the right-hand column, theoretical and observational constraints on these parameters 
are described. The values adopted in our preferred model (referred to as the standard model in the text) 
are indicated in parentheses.}\label{freepars}
\begin{tabular}{c c c l}
\\[3pt]
\hline
\small{symbol} & \small{parameter} & \small{range and value adopted} & \small{constraints from obs. and theory}\\
\hline
GMC parameters & & & \\
\hline
$M_{\rm GMC}$ & typical mass of a GMC& $M_{\rm GMC}=10^5-{10^8\, M_{\odot}}$ & Estimated by \citet{Solomon87}, \\
 & & (std. model $M_{\rm GMC}=10^6\, M_{\odot}$) & \citet{Williams97}.\\
$\tau_{\rm life,GMC}$ & Lifetime of a GMC & $t_{\rm life,GMC}=10-30$~Myr   & Observations and theoretical arguments\\
             &  & (std. model $t_{\rm life,GMC}=10$~Myr)& favour values in the range given here\\
 & & &  (e.g. \citealt{Blitz80}; Dobbs et al. 2011).\\
\hline
Diffuse medium parameters & & & \\
\hline 
$\sigma_{\rm d}$ & velocity dispersion of & $\sigma_{\rm d}\approx 5-{70}\, \rm km\,s^{-1}$ & \citet{VanDerKruit11}. Used to \\
 &  the gas in disks & (std. model $\sigma_{\rm d}=10\, \rm km\,s^{-1}$) & calculate $P_{\rm d}$, $u_{\rm d}$ and $h_{\rm g}$.\\
\hline
Disk parameters & & & \\
\hline
$f_{\star}$ & ratio of the scale radius to& $f_{\star}=r_{\rm s}/h_{\rm star}\approx 7.3$ & \citet{Kregel02}. Used to calculate\\
 & the scale height of the stellar disk& & $P_{\rm ext}$ and $h_{\rm g}$.\\
$f_{\rm r}$ & Defines the radius at which & $f_{\rm r}=1.1-2$ & In principle $f_{\rm r}$ is a free parameter.\\
 & bubbles are assumed to have & (std. model $f_{\rm r}=1.5$) & However, we set a range within which\\
 & escaped the galaxy. & & we vary $f_{\rm r}$ as to get a break-out\\
 & & & mass fraction consistent with previous\\ 
 & & &  theoretical estimates \\
 & & & (e.g. \citealt{MacLow88}; \\
 & & & \citealt{Fujita09}).\\
\hline
SF parameters & & & \\
\hline
$\nu_{\rm SF}$ & SFR coefficient & $\nu_{\rm SF}=0.25-1\, {\rm Gyr}^{-1}$ & Determines the SFR per unit\\
               &                 & (std. model $\nu_{\rm SF}=0.5\, {\rm Gyr}^{-1}$) & molecular mass $\Sigma_{\rm SFR}=\nu_{\rm SF}\, \Sigma_{\rm mol}$.\\
               &                 & & Measured by e.g. \citet{Leroy08}.\\
$P_{0}$        & Pressure normalisation & $\rm log(P_{0}/k_{\rm B} [\rm cm^{-3} K])=4.19-4.54$ & $\Sigma_{\rm H_2}/\Sigma_{\rm HI}=(P_{\rm ext}/P_{0})^{\alpha_{\rm P}}$. Measured\\ 
               &                         &  (std. model & by e.g. \citet{Wong02}, Blitz \&\\
               &                         &  $\rm log(P_{0}/k_{\rm B} [\rm cm^{-3} K])=4.54$)& Rosolowsky (2006), \citet{Leroy08}.\\
$\alpha_{\rm P}$ & Power-law index in & $\alpha_{\rm P}=0.73-0.92$ & Measured (see authors above).\\
               & pressure law       & (std. model $\alpha_{\rm P}=0.92$) & \\
\hline
\end{tabular}
\end{center}
\end{table*}

{\it Confinement.} If bubbles are slowed down sufficiently, they are assumed to 
mix with the surrounding medium. The condition for mixing to take place is obtained 
by comparing the bubble expansion velocity 
to the velocity dispersion of the diffuse component of the ISM.
Confinement takes place if $v_{\rm s}\le \sigma_{\rm d}$. If this happens,
we assume instantaneous mixing and add the mass and metals of the bubble to the diffuse medium of 
the ISM. 

\noindent {\it Break-out from the ISM.} If a bubble reaches the edge of the disk or the bulge with an 
expansion velocity exceeding the sound speed of the diffuse ISM, 
it is assumed to break out from the ISM. The edge is defined as a fixed fraction of the 
gas scale height, $f_{\rm r}\, h_{\rm g}$ (see $\S$~\ref{SubSec:DiffMedProps} for the definition of gas scale height). 
 The opening angle of the wind 
at the moment it escapes from the galaxy is given by $\theta\equiv 2\, \arccos(1/\it f_{\rm r})$, assuming that 
bubbles are centered at the midplane of the disk. 
A fraction $f_{\rm bo}$ of the mass and metals carried away by bubbles will escape from the galaxy. 
This depends on the choice of $f_{\rm r}=R/h_{\rm g}$ is given by 

\begin{eqnarray}
f_{\rm bo} = \left(1-\frac{h_{\rm g}}{R} \right)=1-f^{-1}_{\rm r}.
\label{fo}
\end{eqnarray}

\noindent A fraction $(1-f_{\rm bo})$ of the mass and metals carried away by bubbles 
is assumed to be confined in the ISM. The physical motivation for this 
choice is that the gas expanding along the major axis of the disk does not escape and 
that, in the case of the gas expanding perpendicular to the midplane of the disk, 
Rayleigh-Taylor instabilities grow at the edge
of the ambient gas due to the drastic change of density.
These instabilities produce fragmentation in the swept-up 
mass and some of this material is reincorporated into the 
galaxy. \citet{MacLow88} and \citet{MacLow89}, by means of hydrodynamical simulations, estimated 
$f_{\rm r}\approx 1-2$ for a Milky Way-like galaxy. \citet{MacLow89} show that approximately $10$\% of the mass 
contained in shells at the point of break-out accelerates upwards and $\approx 90$\% stays in the ISM. 
Similar values have been obtained by 
 more sophisticated hydrodynamical simulations (e.g. \citealt{deAvillez01}; \citealt{Fujita09}).
In detail, the break-out radius and the mass in shells escaping the galaxy disk 
is thought to mainly depend on the 
density contrast between the disk and halo gas which sets the development of instabilities which 
fragments the bubble shells. 
Other hydrodynamical effects, such as 
weak magnetic fields in the ISM, can inhibit the generation of Rayleigh-Taylor instabilities 
and/or help accelerate the cool shell gas even further away through magnetic pressure 
(e.g. \citealt{Fujita09}). 
These effects influence the cold dense gas of bubbles, while the hotter, interior material is 
shown to escape to the hot halo in all of the simulations.
Taking into account these results, 
we restrict the range of values of $f_{\rm r}$ to $f_{\rm r}\approx 1.1-2$, implying that 
a significant fraction of the swept-up mass in bubbles stays in the ISM. The hot gas 
contained in the interior of bubbles is assumed to fully escape into the hot halo. 
In our standard model, we adopt $f_{\rm r}=1.5$. In $\S$~\ref{ParameterEffects} we show how 
the mass outflow rate varies when $f_{\rm r}$ takes the lowest and highest values in the range above.

Fig.~\ref{GeomBubble} shows a schematic of the evolution of bubbles 
in the ISM. 
We summarise all the parameters needed to characterise GMCs and the ISM of galaxies in Table~\ref{freepars}. 
We give there the reference value used for our standard SNe feedback model but also give the ranges 
motivated by observations and theory, which we also test in $\S$~\ref{ParameterEffects} { and 
 $\S$~\ref{Sec:xtmISM}}.

\section{Incorporating dynamical supernova feedback into a galaxy formation simulation}\label{Sec:SFeqs_sn}

One of the aims of this paper is to study how the outflow rate depends on galaxy properties in a galaxy population 
which has a representative set of star formation histories and which resembles observed 
galaxy properties. We achieve this by incorporating the full dynamical model described in 
$\S$~\ref{Sec:DynModel_SN} into the semi-analytic galaxy formation model {\tt GALFORM}, 
which is set in the $\Lambda$ cold dark 
matter framework.
 
In $\S$~\ref{Sec:modeldetails}, we briefly describe the {\tt GALFORM} model and in 
$\S$~\ref{Sec:SFequations} we give details on how we modify the model to include 
the dynamical model of SNe presented in $\S$~\ref{Sec:DynModel_SN} and \ref{Sec:GMCs}.
 
\subsection{The {\texttt{GALFORM}} model}\label{Sec:modeldetails}

The {\texttt{GALFORM}} model 
takes into account the main physical processes
that shape the formation and evolution of galaxies  \citep{Cole00}. These are: (i) the collapse
and
merging of DM halos, (ii) the shock-heating and radiative cooling
of gas inside
DM halos, leading to the formation of galactic disks, (iii) quiescent star
formation (SF) in galaxy disks, (iv) feedback
from supernovae (SNe), from AGN and from photo-ionization of the
IGM, (v) chemical
enrichment of stars and gas, and (vi) galaxy mergers driven by
dynamical friction within common DM halos, which can trigger bursts of SF 
and lead to the formation of spheroids (for a review of these
ingredients see \citealt{Baugh06} and  \citealt{Benson10b}).
Galaxy luminosities are computed from the predicted star formation and
chemical enrichment histories using a stellar population synthesis model.
Dust extinction at different wavelengths is calculated
self-consistently from the gas and metal contents of
each galaxy and the predicted scale lengths of the disk and bulge components
using a radiative transfer model
(see \citealt{Lacey11} and \citealt{Gonzalez-Perez12}). 

{\texttt{GALFORM}} uses 
the formation histories of DM halos as a starting point 
to model galaxy formation (see \citealt{Cole00}). 
In this paper we use halo merger trees extracted from the Millennium N-body
simulation \citep{Springel05}, which assumes the following  cosmological parameters: 
$\Omega_{\rm m}=\Omega_{\rm DM}+\Omega_{\rm baryons}=0.25$ (with a
baryon fraction of $0.18$), $\Omega_{\Lambda}=0.75$, $\sigma_{8}=0.9$
and $h=0.73$. The resolution of the $N$-body
simulation corresponds to a minimum halo mass of $1.72 \times 10^{10} h^{-1} M_{\odot}$, which in the \citet{Lagos12} model 
corresponds to a stellar mass limit of $7\times 10^{7} h^{-1} M_{\odot}$. 
This is sufficient to resolve the halos that contain most of the 
H$_2$ in the universe at $z<8$ \citep{Lagos11}. The construction of the merger trees 
used by {\texttt{GALFORM}} is described in \citet{Merson12}.

In this paper we focuses on the Lagos et al. (2012; hereafter Lagos12) model, which includes a two-phase description of the ISM, 
i.e. composed of the atomic and molecular contents of galaxies, and 
adopt the empirical SF law of \citet{Blitz06}. The physical treatment of the ISM in the 
Lagos et al. model is a key feature affecting the predicted outflow rate of 
galaxies, as we show in $\S$~\ref{PhysicalChar}, which justifies our choice of exploring the full dynamical model 
of SNe in this model.

The \citet{Blitz06} empirical SF law has the form 

\begin{equation}
\Sigma_{\rm SFR} = \nu_{\rm SF} \,\rm f_{\rm mol} \, \Sigma_{\rm g},
\label{Eq.SFR}
\end{equation}
\noindent where $\Sigma_{\rm SFR}$ and $\Sigma_{\rm g}$ are the surface
densities of the  SFR and the total cold gas mass, respectively,
$\nu_{\rm SF}$ is the inverse of the SF
timescale for the molecular gas, $\nu_{\rm SF}=\tau^{-1}_{\rm SF}$, 
and $\rm f_{\rm mol}=\Sigma_{\rm mol}/\Sigma_{\rm g}$ is the
molecular to total gas mass surface density ratio. The molecular and total gas
contents include the contribution from helium, while the HI and H$_2$ masses only include 
hydrogen (helium accounts for $26$\% of the overall cold gas mass). 
The integral of $\Sigma_{\rm SFR}$ over the disk corresponds to the instantaneous SFR, 
$\psi$. The ratio $\rm f_{\rm mol}$ is assumed to depend on 
the internal hydrostatic pressure of the disk 
as \citep{Blitz06}

\begin{equation}
\frac{\Sigma_{\rm mol}}{\Sigma_{\rm atom}}=\rm f_{\rm mol}/(f_{\rm mol}-1)=\left(\frac{P_{\rm ext}}{P_{0}}\right)^{\alpha_{\rm P}}.
\label{Eq.SFR2}
\end{equation}

\noindent For a description of how we calculate 
$P_{\rm ext}$ see Appendix~\ref{App:BR}. 
The parameter values we use for $\nu_{\rm SF}$, $\rm P_{0}$ and $\alpha_{\rm P}$ 
are the best fits to observations of nearby spiral and dwarf galaxies, $\nu_{\rm SF}=0.5\, \rm Gyr^{-1}$,  
$\alpha_{\rm P}=0.92$ and $\rm log(P_{0}/k_{\rm B} [\rm cm^{-3} K])=4.54$
(\citealt{Blitz06}; \citealt{Leroy08}; \citealt{Bigiel11}; \citealt{Rahman11}).

For SBs the situation is less clear. Observational uncertainties,
such as the conversion factor between CO and H$_2$
in SBs, and the intrinsic compactness of star-forming regions,
have not allowed a clear characterisation of the SF law in this case (e.g.
\citealt{Kennicutt98}; \citealt{Genzel10}; \citealt{Combes11};  see \citealt{Ballantyne13} for 
an analysis of how such uncertainties can bias the inferred SF law). Theoretically,
it has been suggested that the SF law in SBs is different
from that in normal star-forming galaxies (\citealt{Pelupessy09}). 
The ISM of SBs is predicted to always be dominated by H$_2$ independently
of the exact gas pressure.
 For these reasons we choose to apply Eq.~\ref{Eq.SFR}
only during quiescent SF (i.e. SF fuelled by the accretion of cooled gas 
onto galactic disks) and retain the original SF prescription
for SBs, which are driven either by galaxy mergers or disk instabilities 
(see \citealt{Cole00} and L11 for details). In the SBs,
the SF timescale is taken to be proportional to the bulge dynamical timescale
above a minimum floor value (which is a model parameter) and involves the whole ISM gas content
in the SB, giving $\rm SFR = \it M_{\rm gas}/\tau_{\rm SF,SB}$ (see
\citealt{Granato00} and \citealt{Lacey08} for details), with  

\begin{equation}
\tau_{\rm SF,SB}=\rm max(\tau_{min},f_{\rm dyn}\tau_{\rm dyn}). 
\label{SFlawSB}
\end{equation}

\noindent Here we adopt 
$\tau_{\rm min}=100\, \rm Myr$ and $f_{\rm dyn}=50$ following \citet{Lagos12}. 

Throughout the paper we will refer to galaxies 
as `starburst galaxies' if their total SFR is dominated by the starburst mode, 
$\rm SFR_{\rm starburst}>SFR_{\rm quiescent}$, 
while the remainder of the model galaxies will be referred to as `quiescent galaxies'.

\subsection{Predicting the star formation history of galaxies}\label{Sec:SFequations}

\begin{figure*}
\begin{center}
\includegraphics[width=0.7\textwidth]{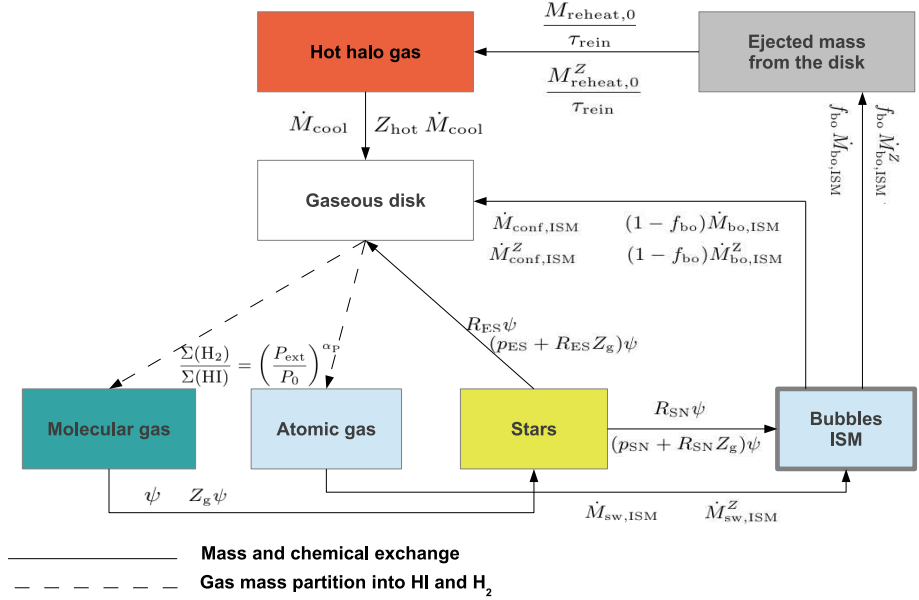}
\caption[Schematic showing the flow of mass and metals in the dynamical model of SNe feedback
{\texttt{GALFORM}}.]{Schematic of the flow of mass and metals in the dynamical model of SNe feedback. 
The scheme shows the exchange of mass and metals (solid lines) between the hot halo, 
stars and the three gas phases in the ISM, and the partition of the ISM gas into the 
atomic and molecular gas components (dashed lines), 
corresponding to Eqs.~\ref{Eqs:SFset1}-\ref{Eqs:SFset3} in the text.  
Note that the same 
scenario would apply to SBs without the inflow of cooled gas from the hot halo.} 
\label{fig:scheme}
\end{center}
\end{figure*}

The {\tt GALFORM} model includes two gas phases in the ISM of galaxies, an atomic and a molecular phase, 
which correspond to the warm and cold phases, respectively. By including 
 dynamical modelling of SNe feedback, we introduce 
a new phase into the ISM of galaxies corresponding to the interiors of 
expanding bubbles (see $\S$~\ref{Sec:DynModel_SN}). 

The equations of SF need to be modified accordingly to 
include the contribution from the mass and metals in bubbles.
The chemical enrichment is also assumed to proceed through the expansion of 
SNe inflated bubbles: stellar winds and SNe feedback shock the surrounding 
medium and inflate bubbles through thermal energy, so the new metals produced 
by recently made intermediate and high mass stars will be contained in the interiors of bubbles. 
In the case of low mass stars, recycling of mass and newly synthesised metals feed the ISM directly.
In the case of confinement, metals contained in the thin, dense shell of swept-up 
gas and the interior of bubbles are mixed instantaneously with the cold and warm ISM. 
Note that we do not apply any delay to the mixing of metals given that the cooling time 
for the hotter phases is typically small ($t_{\rm cool}=5\times 10^2-10^5$~yr).

The five mass components of the system are: the stellar mass of the disk, $M_{\star}$, 
the total gas mass in the ISM (molecular plus atomic), $M_{\rm g,ISM}$, the mass in bubbles (interior plus shell) in the ISM, $M_{\rm b,ISM}$,
the mass of the hot gaseous halo of the galaxy, $M_{\rm hot}$, 
and the mass escaping the galaxy disk through bubbles,
$M_{\rm eject}$. The latter represents all gas that has not yet mixed with the hot 
halo gas; i.e. that is thermally/kinematically decoupled from the hot halo gas. 
The underlying assumption is that all gas ejected from the disk ends up 
in reheated gas reservoir.  
The reincorporation time, $\tau_{\rm rein}$, 
of the ejected component into the halo is always larger than the timestep 
over which we perform the integration. We therefore calculate the rate of reincorporation of gas into the hot halo component 
only with the ejected mass available at the beginning of the timestep, $M_{\rm eject}$. 
We remind the reader that in this paper we 
we use the standard approach of {\tt GALFORM} to calculate $\tau_{\rm rein}$. This consists 
of parametrising $\tau_{\rm rein}$ as depending linearly on the dynamical timescale of the halo regulated by an efficiency, 
which is a free parameter of the model, $\tau_{\rm rein}=\tau_{\rm dyn}/\alpha_{\rm reheat}$ 
(we retain the value of $\alpha_{\rm reheat}=1.2$ used in Lagos12).
In paper II 
we introduce a physical modelling of the reincorporated gas and the timescale for this process.

Fig.~\ref{fig:scheme} depicts the exchange of mass and metals between the different components 
of galaxies: the hot halo, ISM, stars and bubbles expanding in the ISM. 
As in the original model of \citet{Cole00}, 
we assume that during SF, the inflow rate from the hot halo, $\dot M_{\rm cool}$, 
is constant, implicitly assuming that SNe heating plays no role in the inflow rate until the ejected mass and metals 
are incorporated into the hot halo after timescale $\tau_{\rm rein}$.
The gas mass in the ISM is affected by $\dot M_{\rm cool}$, the rate at which 
mass is recycled from evolved stars (assumed to go straight to the ISM),   
the rate at which bubbles sweep up mass from the ISM, $\dot{M}_{\rm sw,ISM}$, 
and the rate of bubble confinement, $\dot{M}_{\rm conf,ISM}$ and 
break-out, $\dot{M}_{\rm bo,ISM}$ (the calculation of each of these are described in 
detail in Appendix~\ref{App:MassRates}). 
At each substep in the numerical solution scheme, we update the values of each of the mass variables. 
It is therefore possible to replenish the atomic/molecular gas contents and also modify the 
H$_2$/HI ratio, as the gas and stellar surface densities change. 

The set of equations describing the flow of mass and metals between the different 
phases are
 
\begin{eqnarray}
&&{\rm Mass\, exchange:}\nonumber\\
\dot{M}_{\star}&=&(1-R_{\rm ES}-R_{\rm SN})\psi,\label{Eqs:SFset1}\\
\dot{M}_{\rm g,ISM} &=& \dot{M}_{\rm cool}+(R_{\rm ES}-1)\psi-\dot{M}_{\rm sw,ISM}+\dot{M}_{\rm conf,ISM}\nonumber\\
                     & &+(1-f_{\rm bo})\dot{M}_{\rm bo,ISM},\\
\dot{M}_{\rm b,ISM} &=& R_{\rm SN}\psi+\dot{M}_{\rm sw,ISM}-\dot{M}_{\rm conf,ISM}-\dot{M}_{\rm bo,ISM}\\
\dot{M}_{\rm eject} &=& f_{\rm bo}\, \dot{M}_{\rm bo,ISM}-\frac{M_{\rm eject}}{\tau_{\rm rein}},\label{Eqs:SFsetMeje}\\
\dot{M}_{\rm hot} &=& -\dot{M}_{\rm cool}+\frac{M_{\rm eject}}{\tau_{\rm rein}}.\\
\nonumber \\
&&{\rm Metallicity\, exchange:}\nonumber\\
\dot{M}^{Z}_{\star}&=&(1-R_{\rm ES}-R_{\rm SN})Z_{\rm g}\psi,\label{Eqs:SFMZ}\\
\dot{M}^{Z}_{\rm g,ISM} &=&\dot{M}_{\rm cool}Z_{\rm hot}+(p_{\rm ES}+R_{\rm ES}Z_{\rm g})\psi-\dot{M}^{Z}_{\rm sw,ISM}\nonumber\\
                        & &+\dot{M}^{Z}_{\rm conf,ISM}+(1-f_{\rm bo})\dot{M}^{\rm Z}_{\rm bo,ISM},\\
\dot{M}^{Z}_{\rm b,ISM} &=& (p_{\rm SN}+R_{\rm SN}Z_{\rm g})\psi+\dot{M}^{Z}_{\rm sw,ISM}-\dot{M}^{Z}_{\rm conf,ISM} \label{Eqs:SFset3}\\
  &&-\dot{M}^{Z}_{\rm bo,ISM}\nonumber\\
\dot{M}^{Z}_{\rm eject} &=& f_{\rm bo}\, \dot{M}^{Z}_{\rm bo,ISM}-\frac{M^{Z}_{\rm eject}}{\tau_{\rm rein}},\label{Eqs:SFset4}\\
\dot{M}^{Z}_{\rm hot} &=& -\dot{M}_{\rm cool}Z_{\rm hot}+\frac{M^{Z}_{\rm eject}}{\tau_{\rm rein}}.\label{Eqs:SFset2}
\end{eqnarray}

The recycled mass from newly formed stars is specified separately 
for SNe, $R_{\rm SN}$, and intermediate and 
low mass stars, $R_{\rm ES}$ (namely, evolved stars). We calculate the recycled 
fractions of each stellar mass range following Eq.~\ref{Eq:ejec}. SNe are considered 
to be all stars with $m>8\, M_{\odot}$, and less massive stars 
in the range $1<m/M_{\odot}<8$ are considered as evolved stars (intermediate and low mass stars). 
Stars less massive than $1 M_{\odot}$ have lifetimes larger than the age of the Universe and therefore 
do not recycle mass into the ISM.
The yield is also defined 
separately for SNe and evolved stars in order to inject the metals from SNe 
into the bubbles, whilst metals from evolved stars go directly into the ISM. 
We adopt the instantaneous mixing approximations for the metals in the ISM. This implies that the metallicities  
of the molecular and atomic phases in the ISM are equivalent and equal to $Z_{\rm g}=M^{Z}_{\rm g,disk}/M_{\rm g,disk}$. 
The metallicity of the hot gas in the halo is $Z_{\rm hot}=M^{Z}_{\rm hot}/M_{\rm hot}$. 

The system of SF Eqs.~\ref{Eqs:SFset1}-\ref{Eqs:SFset3} applies for quiescent SF and SBs. In 
 the latter case $\crate=0$.
During a SB, we assume that all bubbles expanding in galaxy disks are 
destroyed, as well as bubbles expanding in the satellite galaxy in the case of a galaxy merger. The new 
generation of stars made in the SB creates a new generation of 
inflated bubbles expanding over the bulge.

\section{Physical characterisation of bubbles in the ISM}\label{PhysicalChar}

In this section we explore the physical properties of bubbles and the main drivers of their evolution in the ISM of galaxies. 
In $\S$~\ref{SubSec:PropsSingle}, we focus on individual examples of bubbles in ad-hoc galaxies. We explore how the bubble mass 
depends on different global galaxy properties, such as the gas fraction, gas metallicity and scaleheight, and local properties, such as gas density and surface density. 
In $\S$~\ref{Subsec:RadialBetas}, $\S$~\ref{SubSec:statisprops} and $\S$~\ref{Sec:ComparisonObsHydro} 
we focus on the outflow properties of {\tt GALFORM} galaxies 
when the full dynamical model
for SNe feedback is included (see $\S$~\ref{SubSec:DiffMedProps}). 
Comparisons with observations and previous theoretical work 
are presented and discussed in $\S$~\ref{Sec:ComparisonObsHydro}.

\subsection{Properties of individual bubbles}\label{SubSec:PropsSingle}
\begin{table}
\begin{center}
\caption{Properties of the three example galaxies used to study the effect of the 
different physical parameters on the evolution of bubbles in the ISM.
We list the $10$ properties we need to characterise the radial profiles 
of the stellar, gaseous and DM components, disk and bulge half-mass radii, $r_{\rm d}$ and $r_{\rm b}$, 
stellar mass in the disk and the bulge, $M_{\star,\rm d}$ and $M_{\star,\rm b}$, cold gas mass, $M_{\rm gas,ISM}$, 
gas metallicity, $Z_{\rm g}$, halo virial mass, $M_{\rm halo}$, radius, $r_{\rm vir}$, and halo concentration, $c$. 
 We also fix the distance to the galaxy centre at which the
example bubble is located, $d$. The properties listed define the local properties 
of the ISM (see Appendix~\ref{Profiles}). For those parameters which we vary, we give the range chosen 
to study their effect on the bubble expansion, and in the line below this we give the reference value.}\label{ModelsSingle}
\begin{tabular}{l c c c}
\\[3pt]
\hline
\small{Model} & Dwarf & Spiral & Giant \\
\hline
\hline
Varying parameters  & & & \\
\hline
\hline
$M_{\rm gas,ISM}/M_{\odot}$    & $10^7$-$10^{9.5}$ & $10^8$-$10^{11}$ & $10^9$-$10^{12}$ \\
 ref. value    & $5\times 10^9$ & $8\times 10^{10}$ & $1\times 10^{11}$\\
\hline
$M_{\star,\rm d}/M_{\odot}$ &  $10^7$-$10^{9.5}$ & $10^8$-$10^{11}$ & $10^9$-$10^{12}$ \\
  ref. value      & $10^9$& $5\times 10^{10}$ & $10^{11}$\\
\hline
 $Z_{\rm g}/Z_{\odot}$    & $10^{-3}-2$ &$10^{-3}-2$ &$10^{-3}-2$\\
   ref. value             & 0.1& 1 & 2\\
\hline
 $d/r_{\rm d}$    & 0-6 & 0-6 & 0-6 \\
   ref. value                  & 0.5 & 0.5  & 0.5 \\
\hline
\hline
Fixed parameters & & & \\
\hline
\hline
$r_{\rm d}/{\rm kpc}$              & 2.5 & 6 & 10\\
$M_{\star,\rm b}/M_{\odot}$        & 0 & $8\times 10^9$ & $2\times 10^{11}$\\ 
$r_{\rm b}/{\rm kpc}$              & 0 & 0.5 & 3 \\
$M_{\rm halo}/M_{\odot}$           & $5\times 10^{10}$ & $10^{12}$  & $10^{14}$ \\
$r_{\rm vir}/{\rm Mpc}$            & 0.08 & 0.2 & 1 \\
$c$                                & 5 & 5 & 5 \\
\hline
\end{tabular}
\end{center}
\end{table}
\begin{figure}
\begin{center}
\includegraphics[trim = 5mm 3.3mm 4mm 0mm,clip,width=0.45\textwidth]{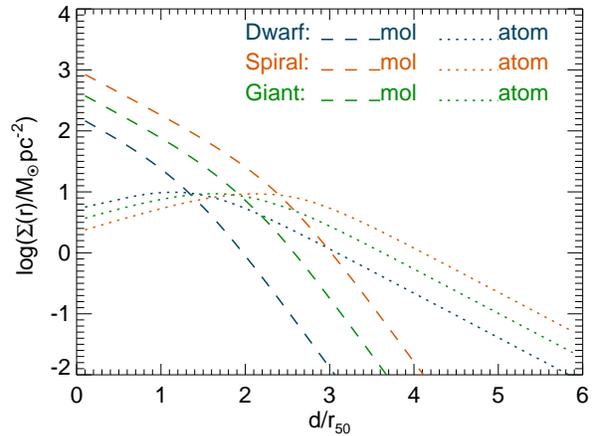}
\caption{Surface density of molecular and atomic gas as a function of the distance from the galactic centre in units of 
the half-mass radius 
of the three example galaxies listed in  
Table~\ref{ModelsSingle}. Line styles and colours show different components of the gas content in the 
different galaxies as labelled.}
\label{singlebs0}
\end{center}
\end{figure}
\begin{figure*}
\begin{center}
\includegraphics[trim = 3mm 0mm 4mm 0mm,clip,width=0.99\textwidth]{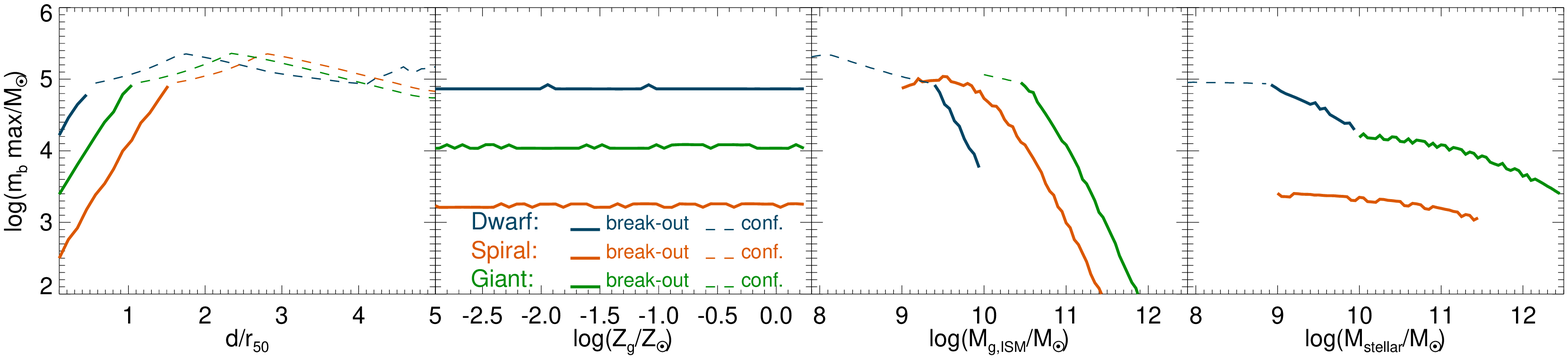}
\includegraphics[trim = 3mm 0mm 4mm 0mm,clip,width=0.99\textwidth]{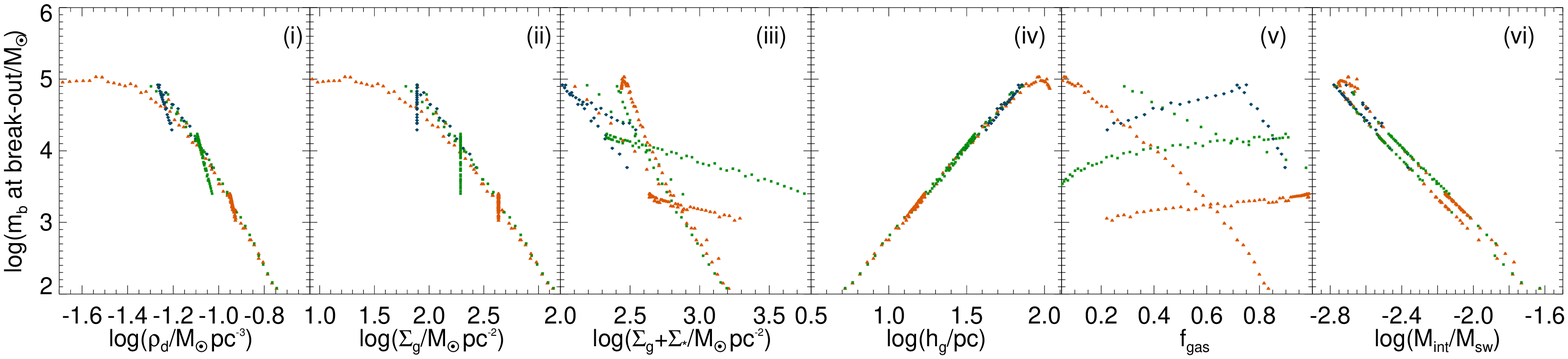}
\caption[Schematic showing the flow of mass and metals in the dynamical model of SNe feedback
{\texttt{GALFORM}}.]{{\it Top panel:} Bubble mass at the point of break-out or confinement, maximum $m_{\rm b}$, 
as function of the distance to the galaxy centre in units of the 
half-mass radius, $d/r_{50}$ (left panel), the gas metallicity in units of the solar metallicity, $Z_{\rm g}/Z_{\odot}$ (middle-left panel), the ISM mass 
(molecular plus atomic), $M_{\rm g,ISM}$ (middle-right panel) and the stellar mass, $M_{\rm stellar}$ (right panel).  
The segments of the curves shown with solid lines correspond to those regions of the planes where 
bubbles end up breaking out from the galactic disk. 
Those segments shown with dashed lines correspond to regions where bubbles end up confined in the ISM of the galaxy.
{\it Bottom panel:} Bubble mass at the point of break-out  
as a function of the local properties (i) atomic gas density, (ii) total (molecular plus atomic) gas surface density, 
(iii) surface density of total gas plus stars, (iv) gas scaleheight, (v) gas fraction and (vi) the ratio 
between the interior and the swept-up mass of bubbles (the interior mass corresponds to the 
fraction of the total mass injected 
by SNe that has not yet cooled down or hit the shell). 
Individual realizations for each galaxy are shown as points in the colours labelled.} 
\label{singlebs}
\end{center}
\end{figure*}

We study the dependence of the mass in a single bubble (interior plus shell) on the properties of the 
diffuse medium with the aim of determining which local properties are the more relevant in setting the mass of bubbles at the point of 
break-out or confinement (i.e. their maximum mass). 

In order to fully characterise a single bubble in the ISM of a galaxy,  
we need to choose values for the galaxy properties which are required in the dynamical SN feedback model, 
namely the gas and stellar mass in the disk and the bulge, the half-mass radii of both stellar components, 
the halo virial mass, radius and concentration, the gas metallicity and the 
location of the bubble in the galaxy disk.
We focus on three example galaxies with properties within a representative range 
which are listed in Table~\ref{ModelsSingle}.

To calculate the expansion of a single bubble in the ISM of these galaxies, we use the standard set of 
parameters in Table~\ref{freepars} to describe GMCs and the ISM.
In Fig.~\ref{singlebs0}, we show the radial profiles of the atomic and molecular gas 
for the three galaxies of Table~\ref{ModelsSingle}. 
We construct these profiles using the \citet{Blitz06} 
relation (Eq.~\ref{Eq.SFR2}).
The three galaxies plotted in 
Fig.~\ref{singlebs0} show central regions dominated by molecular gas, and atomic gas surface densities which saturate at  
$\approx 10\, M_{\odot}\, {\rm pc^{-2}}$, above which the gas is mainly molecular.

In order to study the dependence of the maximum mass of bubbles on galaxy properties, 
we vary the mass of gas and stars, the gas metallicity and the distance 
of the bubbles from the galaxy centre for the three 
galaxies in Table~\ref{ModelsSingle}. These parameters
 are expected to have an effect on the expansion of bubbles 
by varying the gas density, scale height, cooling timescale, gravitational field, etc. The strategy is to vary one property at a time 
leaving the other ones unchanged, to see how the predictions change. 
We evolve bubbles until they become confined or break 
out from the galaxy disk. When we fix $d$, we arbitrarily choose $d=0.5\, r_{50}$ for 
illustration. This value of $d$ typically corresponds to a region where bubbles break 
out. The $4$ experiments (i.e. changing $d$,  $Z_{\rm d}$, $M_{\rm gas,ISM}$ and  $M_{\star,\rm d}$)  
are performed for each of the galaxies of Table~\ref{ModelsSingle} and the results are shown in the top panel of 
Fig.~\ref{singlebs}. The maximum mass of a single bubble shown in Fig.~\ref{singlebs} 
corresponds to the mass at the point of break-out or confinement. 

In the central regions of galaxies, bubbles break-out from the galaxy disk, while in the outskirts bubbles tend to 
be confined. In the case of the `dwarf' galaxy, the break-out region 
is restricted to $d\lesssim 0.5r_{50}$, while in the case of the 
`spiral' and `giant' galaxies, the region of break-out extends out to $d> r_{50}$. In the break-out regions, 
there is a strong relation between the bubble mass and the distance from the galactic centre. This is 
driven by an underlying relation between $m_{\rm b}$ and the gas scaleheight or gas surface density. 

Variations in the gas metallicity have very little effect
on the resulting bubble mass. When the gas surface density is high, the metallicity
plays only a minor role because the cooling time is already very short and bubbles become radiative very quickly.
In the case of low gas surface densities, the cooling time becomes long even for high metallicities, which preserves
the energy of the bubbles. In the case that metallicity does have an effect on the bubble mass, the differences
found are always less than a factor of $\sim 2$.

Strong variations in the maximum mass of the bubble are obtained when varying $M_{\rm gas,ISM}$.
In the regime of break-out from the galaxy disk, the bubble mass quickly decreases when increasing $M_{\rm gas,ISM}$. 
As $M_{\rm gas,ISM}$ increases, the surface density of gas also increases. This 
 reduces the gas scaleheight, which reduces the bubble mass. The reason for this is that 
the radius the bubble needs to reach to escape the galaxy decreases, and therefore 
 also the total mass that it is able to sweep-up also decreases, as this is proportional to the bubble volume.
The higher $M_{\rm gas,ISM}$ results in an overall decrease of the bubble mass by a factor of 
$100-500$.

Variations in stellar mass have a non-negligible effect on the bubble mass, 
particularly at the massive end of the range tested (see second row of Table~\ref{ModelsSingle}). 
There is a trend of decreasing bubble mass with increasing stellar mass in the region of break-out. This happens due to the increasing 
gravitational field driven by the higher stellar surface densities, which decreases the gas scaleheight of the disk and 
the radius the bubble needs to reach to break-out. 
 The bubble mass obtained when increasing the 
stellar content of galaxies can be lower by up to a factor of $3$. 
The effect of the more efficient deceleration of bubbles due to the larger gravitational field 
when the stellar mass increases is secondary to the 
effect of the stellar surface density on the gas scaleheight, and represents only $\approx 0.1-5$\% of the total effect observed 
when increasing $M_{\star,\rm d}$. 

The distance to the galactic centre and the gas content of the galaxies shown in 
Fig.~\ref{singlebs} drive the strongest variations in 
 bubble mass. This is due to the dependence of $m_{\rm b}$ 
on the gas density (atomic plus molecular) and the gas scaleheight, which is shown 
in the bottom-right panel of Fig.~\ref{singlebs}.
We include only those examples in which the bubble breaks out from the galaxy disk. Bubble masses 
in the cases tested here are always dominated by the swept-up mass (see bottom-right panel of Fig.~\ref{singlebs}). 
However, there is an increasing contribution from $m_{\rm int}$ to $m_{\rm b}$ for decreasing $m_{\rm b}$.   
We give physical insight into the relations between $m_{\rm b}$, $h_{\rm g}$ and $\Sigma_{\rm g}$ 
in the next subsection. 

In the case of the gas fraction, we find that there is a complex dependence of $m_{\rm b}$ on $f_{\rm gas}$. 
The gas fraction acts to modify the 
normalisation of the relation between the total outflow rate and $h_{\rm g}$ 
and the power-law index of the relation between the total outflow rate and $\Sigma_{\rm g}$. {The gas fraction 
is also responsible for the dispersion at fixed $\Sigma_{\rm g}$ 
in panel (ii) in the bottom of Fig.~\ref{singlebs}.}

\subsubsection{Analytic derivation of the scaling relations of single bubbles}\label{Sub:Analytic}

At the point of break-out, the volume of the gas disk 
occupied by a single bubble is $V=2\pi\, h^3_{\rm g} (f^2_{\rm r}-1/3)$. In the regime 
where $m_{\rm inj}\ll m_{\rm sw}$, which is a representative limit for most bubbles 
(see the bottom-right panel of Fig.~\ref{singlebs}), and neglecting temporal changes in the 
gas density of the diffuse medium during the evolution of bubbles in the ISM, one can write the bubble mass as 

\begin{equation}
m_{\rm b}=\rho_{\rm d}\, V=(1-f_{\rm mol}) \pi (f^2_{\rm r}-1/3)\, \Sigma_{\rm g}\, h^2_{\rm g}. 
\label{Eq:mb}
\end{equation}

\noindent In order to find an expression for $m_{\rm b}$ in terms of $\Sigma_{\rm g}$ and $h_{\rm g}$ alone, we need to 
express $f_{\rm mol}$ as a function of the same variables. 

We can write $f_{\rm mol}$ in terms of the gas (atomic and molecular) density  

\begin{equation}
1-f_{\rm mol}=\frac{1}{1+(P_{\rm ext}/P_0)^{\alpha_{\rm P}}}=\frac{1}{1+\left(\frac{\Sigma_{\rm g}}{2 h_{\rm g}}\, \sigma^2_{\rm d}/P_0 \right)^{\alpha_{\rm P}}}. 
\end{equation}

\noindent By introducing the expression for $f_{\rm mol}$ into Eq.~\ref{Eq:mb}, we find that 

\begin{eqnarray}
m_{\rm b} \approx \left\{
  \begin{array}{l l} 
   \pi \left(f^2_{\rm r}-\frac{1}{3}\right) \, \Sigma_{\rm g}\, h^{2}_{\rm g} &\quad \left(\frac{\Sigma_{\rm g}}{2 h_{\rm g}}\, \sigma^2_{\rm d}/P_0 \right) \ll 1\\
   \pi \left(f^2_{\rm r}-\frac{1}{3}\right) \, \left(\frac{2 P_{0}}{\sigma^2_{\rm d}}\right)^{\alpha_{\rm P}} & \quad \left(\frac{\Sigma_{\rm g}}{2 h_{\rm g}}\, \sigma^2_{\rm d}/P_0 \right) \gg 1\\
   \,\,\,\,\,\,\,\,\,\,\,\,\,\,\,\,\, \cdot \Sigma^{1-\alpha_{\rm P}}_{\rm g}\, h^{2+\alpha_{\rm P}}_{\rm g.}
  \end{array} \right.
\label{bubblemassform}
\end{eqnarray}

\noindent If we now apply the limit 
 $\Sigma_{\rm g}\gg (\sigma_{\rm g}/\sigma_{\star})\Sigma_{\star}$, where gas dominates over stars in the gravity acting on 
the gas layer, we find that $h_{\rm g}\propto \sigma^{2}_{\rm d}/\Sigma_{\rm g}$ and  

\begin{eqnarray}
m_{\rm b} \propto \left\{
  \begin{array}{l l}
     h_{\rm g} \propto \Sigma^{-1}_{\rm g} & \quad f_{\rm mol}\ll 1 \\
     h_{\rm g}^{1+2\alpha_{\rm P}}\propto \Sigma^{-(1+2\alpha_{\rm P})}_{\rm g} & \quad  f_{\rm mol}\approx 1
  \end{array} \right.
\end{eqnarray}

\noindent These expressions describe the relations shown in the bottom panel of Fig~\ref{singlebs}, where 
we obtain, in the high-density regime, $\Sigma_{\rm g} \gtrsim 70 M_{\odot}\, {\rm pc}^{-2}$, the power-law relations  
$m_{\rm b}\propto h^{2.5}_{\rm g}$ and $m_{\rm b}\propto \Sigma^{-2.3}_{\rm g}$, 
and in the lower density regime, we find $m_{\rm b}\propto h^{0.7}_{\rm g}$ and $m_{\rm b}\propto \Sigma^{-0.8}_{\rm g}$. 
 These power-law relations are approximate as the exact value of the power-law index changes 
slightly from case to case. From this analytic derivation of the scaling relations it is fair to say that the transition 
from the atomic- to molecule-dominated media has a large impact on the mass of a bubble at the point of break-out. 

If we assume a steady state (i.e. the SFR is constant), we can write the outflow rate per annulus as a function of 
each individual bubble mass as, 

\begin{equation}
\dot{M}_{\rm eject}=\frac{f_{\rm bo}\, m_{\rm b}\, M_{\rm mol}}{\tau_{\rm life,GMC}\, M_{\rm GMC}}.
\end{equation}

\noindent Considering $\psi=\nu_{\rm SF}\, M_{\rm mol}$, we can directly write $\beta$ per annulus in terms of a single bubble mass

\begin{equation}
\beta = \frac{\dot{M}_{\rm eject}}{\psi}=\frac{f_{\rm bo}}{\nu_{\rm SF}\, \tau_{\rm life,GMC}\, M_{\rm GMC}}\, m_{\rm b}.
\label{analyticbeta}
\end{equation}

\noindent There is a direct relation between $\beta$ and $m_{\rm b}$ in the case of a steady state. 
We therefore expect to see a similar transition in the relation between the outflow rate and the gas surface density to 
the one obtained for $m_{\rm b}$: 
from a steeper relation in galaxies with molecule-dominated ISM to a shallower relation in galaxies with atomic-dominated ISM. 
From Eqs.~\ref{bubblemassform} and ~\ref{analyticbeta} we also see how each of the parameters describing the ISM and GMCs affect 
individual bubble masses and the global outflow rate.

\subsection{Radial profile of the mass loading factor and outflow velocity}\label{Subsec:RadialBetas}

\begin{figure}
\begin{center}
\includegraphics[trim = 5mm 2mm 7mm 0mm,clip,width=0.43\textwidth]{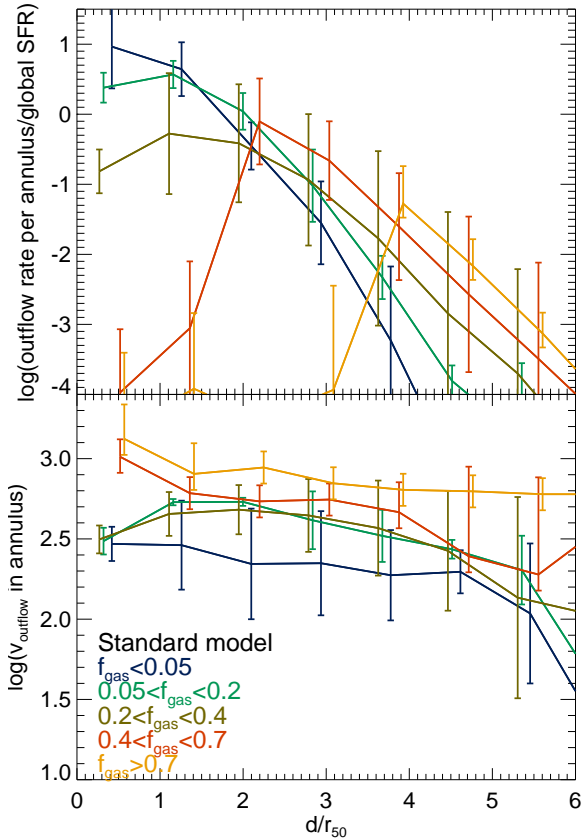}
\caption[The outflow rate contribution by each annulus in units of the global SFR, as a function of the radius to 
scale length ratio.]{{\it Top panel:} The outflow rate contributed by each annulus in units of the global SFR, as a function of the 
distance from the galactic centre in units of the half-mass radius, $d/r_{\rm 50}$, 
for the dynamical model with the standard
choice of parameters (see Table~\ref{freepars}) and for galaxies at $z<0.1$ and with 
$M_{\star}>10^{10}\, h^{-1} M_{\odot}$. 
For quiescent SF we use $r_{\rm 50}$ of the disk, and for starbursts, $r_{\rm 50}$ of the bulge.
Solid lines and errorbars represent the median and 
$10$ to $90$\% range of the distributions. The predictions are plotted for galaxies with different gas fractions, as labelled.
{\it Bottom panel:} As in the top panel, but here we show the outflow velocity of the gas at the point of break-out 
 as a function of distance from the galactic centre in units of the half-mass radius.}
\label{fig:beta_annuli1}
\end{center}
\end{figure}

In order to physically characterise the outflow rate 
in a galaxy population which resembles the observed one, 
we use the {\tt GALFORM} semi-analytic model, 
into which we incorporate the dynamical feedback described in $\S$~\ref{Sec:DynModel_SN}. 
The key difference with the analysis of $\S$~\ref{SubSec:PropsSingle} is that here we explore the whole 
 galaxy population and 
the outflow rate with the aim of
characterising: (i) a preferred radius from which most of the material escapes and 
the outflow velocity, and
(ii) the scaling relations between the mass loading factor, $\beta$, and local properties of the disk, computed in an annulus 
which is at a distance $d$ from the galactic centre. 
The galaxies used in the analysis in this section are selected so that 
they are close to the break of the stellar mass function at low-redshift, $M_{\star}>10^{10}\, M_{\odot}\, h^{-1}$, 
 and have $z<0.1$. This selection makes the galaxy properties comparable to those simulated by \citet{Creasey12}.

In order to gain insight into (i), we show in 
the top panel of Fig.~\ref{fig:beta_annuli1} the outflow rate in each radial annulus in units of the global SFR  
as function of the 
distance from the galactic centre.
We distinguish between galaxies 
with different gas fractions, $f_{\rm gas}=M_{\rm g,ISM}/(M_{\rm g,ISM}+M_{\star})$.
There is a tendency for gas-rich galaxies to have most of the mass breaking-out from the disk 
 at $d \approx r_{\rm 50}$, while in gas-poor galaxies most of the 
mass escapes from close to the galactic centre. We calculate the radius inside which half of the global outflow mass 
escapes, $\dot{M}_{\rm out}(d<r_{\rm out})=\dot{M}_{\rm eject}/2$, where $\dot{M}_{\rm eject}$ is the global outflow rate. 
Galaxies in Fig.~\ref{fig:beta_annuli1} with $f_{\rm gas}>0.8$ have $r_{\rm out}=0.8\, r_{\rm 50}$ and 
 those with $f_{\rm gas}<0.1$ have $r_{\rm out}=0.4\, r_{\rm 50}$.
This is consistent with the picture presented 
in $\S$~\ref{SubSec:PropsSingle}, where the gas-poor dwarf galaxy has a more centrally concentrated 
outflow than galaxies that are gas rich.
%Given that high-redshift galaxies are predicted to be more gas-rich than their low-redshift counterparts 
%(see \citealt{Geach11}; \citealt{Lagos11}), it is therefore expected from this model 
%that high-redshift galaxies exhibit more extended outflows relative to the size of the galaxy than 
%low-redshift galaxies.

In the bottom panel of Fig.~\ref{fig:beta_annuli1} 
we show the mass-weighted velocity of the 
gas escaping the galaxy disk as a function of the distance from the galactic centre, $d$, for galaxies 
with different gas fractions. 
There is a trend of increasing outflow velocity with increasing $f_{\rm gas}$. Gas rich galaxies
typically have a molecule-dominated ISM. In these galaxies the density of atomic, diffuse gas is lower, 
resulting in a more inefficient deceleration of bubbles.
The predicted values of the outflow velocity are comparable with the observed values. We directly compare 
with observations of the outflow velocity in $\S$~\ref{Sec:ComparisonObsHydro}.  

\begin{figure}
\begin{center}
\includegraphics[trim = 5mm 2mm 7mm 0mm,clip,width=0.43\textwidth]{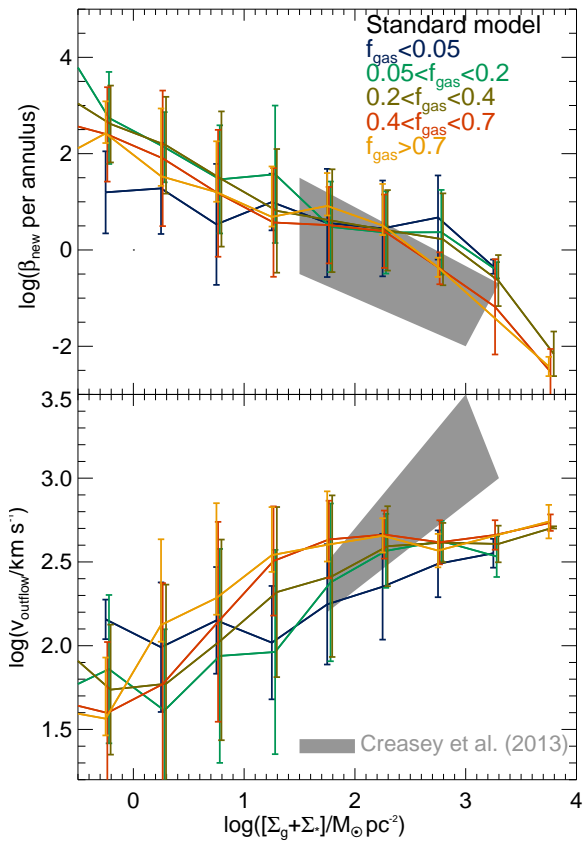}
\caption[The outflow rate to SFR ratio per annulus as a function of the surface density of gas plus stars for different gas fractions.]{
{\it Top panel:} The ratio of the outflow rate to SFR per annulus as a function of the surface density of gas plus stars 
for galaxies at $z<0.1$ and with $M_{\star}>10^{10}\, h^{-1} M_{\odot}$ and for different 
gas fractions, as labelled, in the model with the standard set of parameters (see Table~\ref{freepars}).
Solid lines and errorbars correspond to the median and $10$ and $90$ percentiles of the distributions.
The shaded region corresponds to the predictions of \citet{Creasey12}, and
is plotted over the range of surface density of gas plus stars probed by the simulations. 
{\it Bottom panel:} As in the top panel but for the outflow velocity per annulus as a function of $(\Sigma_{\rm g}+\Sigma_{\star})$.}
\label{fig:beta_annuli4}
\end{center}
\end{figure}

Concerning the scaling relations of the outflow (listed as (ii) above), 
we calculate the ratio between the mass outflow rate and the SFR in each annulus, 
$\beta_{\rm annulus}$, and investigate its dependence 
on the local properties of the disk, as estimated at the 
mean radius of each annulus. 
The top panel of Fig.~\ref{fig:beta_annuli4} shows the relation between
$\beta_{\rm annulus}$ and $(\Sigma_{\rm g}+\Sigma_{\star})$, evaluated at $r_{\rm annulus}$, 
for galaxies with different
gas fractions. There is a tight correlation between
the two quantities, with only a modest dependence on other galaxy properties, such as the gas fraction.  
This is expected from the correlation between $m_{\rm b}$ and $(\Sigma_{\rm g}+\Sigma_{\star})$ (\S~\ref{SubSec:PropsSingle}). 
The results of \citet{Creasey12} (see $\S$~\ref{Sec:IntroSNef} for details) are 
 also shown in Fig.~\ref{fig:beta_annuli4} by the shaded region, plotted over the range of surface densities probed by their simulations. 
Our predicted relation is similar to what Creasey et al. found using a completely 
different approach (see $\S$~\ref{Sec:IntroSNef}).

The best fit to the relation in Fig.~\ref{fig:beta_annuli4} is 

\begin{eqnarray}
\beta_{\rm annulus} &=& \left[\frac{\Sigma_{\rm g}+\Sigma_{\star}}{69\, M_{\odot}\, {\rm pc}^{-2}} \right]^{-1.3}.
\label{betashell2}
\end{eqnarray}

The bottom panel of  Fig.~\ref{fig:beta_annuli4} shows the outflow velocity, $v_{\rm outflow}$, 
as a function of $(\Sigma_{\rm g}+\Sigma_{\star})$, 
evaluated at $r_{\rm annulus}$. There is a trend of increasing $v_{\rm outflow}$ for increasing 
$(\Sigma_{\rm g}+\Sigma_{\star})$. Our predictions for $v_{\rm outflow}$ also overlap with those of 
Creasey et al., although we find that outflow velocities $>1000\, \rm km\, s^{-1}$ are  
statistically unlikely. These velocities can occur for starbursts in our model (see $\S$~\ref{Sec:TestingEffects}).  
Note that for a given $(\Sigma_{\rm g}+\Sigma_{\star})$ there is a trend of $\beta$ decreasing 
 with and $v_{\rm outflow}$ increasing with increasing gas fraction. This prediction is also in agreement 
with the findings of Creasey et al.. 

Note that changes in the SNe feedback model parameters,
which are summarised in Table~\ref{freepars}, produce similar deviations to those found for 
the galaxy-wide $\beta$ and mass-weighted $v_{\rm outflow}$ 
in $\S$~\ref{ParameterEffects}. We find that the surface density normalisation and power-law index in 
Eq.~\ref{betashell2} increase with increasing redshift, in a similar way that 
the global $\beta$ does (Fig.~\ref{fig:beta1}). 
Therefore, the similarity between our predictions and 
those of Creasey et al. is confined to our low-redshift galaxy sample. 
Note that the results of Fig.~\ref{fig:beta_annuli4} for a fixed gas fraction do not depend on stellar mass or redshift, but the  
global normalisation and power-law index of Eq.~\ref{betashell2} do due to the predominance of gas poor galaxies at low redshift and 
of gas-rich galaxies at high redshift.

\subsection{Statistical properties of the outflow rate and velocity}\label{SubSec:statisprops}

In this section, we attempt to answer three questions: 
What is the effect of the multiphase treatment of the ISM on $\beta$?
What is the overall effect of varying the physical parameters of the ISM and GMCs on the outflow rate? 
Is the outflow rate dominated by adiabatic or radiative bubbles? 

Here we analyse galaxies from {\tt GALFORM}, after the full dynamical model of 
SNe feedback is included in the calculation. 
 At each redshift we focus on galaxies with $M_{\star}>10^8\, h^{-1} M_{\odot}$, 
to be safely above the resolution limit of the Millennium simulation ($\S$~\ref{Sec:modeldetails}).  
We consider the total mass loading rate of the outflow, $\beta$, 
which we define as $\beta=\dot{M}_{\rm eject}/\psi$, where
$\dot{M}_{\rm eject}$ corresponds to the total mass breaking out from the ISM
(given by $f_{\rm bo}\dot{M}_{\rm bo,ISM}$ in Eqs.\ref{Eqs:SFset1}-\ref{Eqs:SFset3}) and
$\psi$ is the
instantaneous SFR. In $\S$~\ref{MetalsRate} we analyse the metal loading of the wind, 
which we define as $\beta^{Z}=\dot{M}^{Z}_{\rm eject}/Z_{\rm g} \psi$.
This $\beta$ differs from the $\beta_{\rm annulus}$ of $\S$~\ref{Subsec:RadialBetas} in two respects; the former is integrated 
over the galaxy and over longer timesteps.

In $\S$~\ref{Sec:TestingEffects}, \ref{ParameterEffects}, \ref{Sec:xtmISM} and \ref{Sec:PhysicalRegime}, 
we show the total mass loading $\beta$ as a function of the gas scaleheight 
 at the half-mass radii of galaxies, $h_{\rm g}$.
This can be understood from the strong dependence of 
$m_{\rm b}$ on $h_{\rm g}$ and the small dispersion in this relation  
(see $\S$~\ref{SubSec:PropsSingle}). In $\S$~\ref{MetalsRate} we show how and where 
$\beta^{Z}$ differs from $\beta$ and the reasons for such differences.

\subsubsection{Testing the effect of the multiphase medium and gravity on the outflow properties}\label{Sec:TestingEffects}

\begin{figure}
\begin{center}
\includegraphics[trim = 5mm 0mm 7.0mm 1.0mm,clip,width=0.44\textwidth]{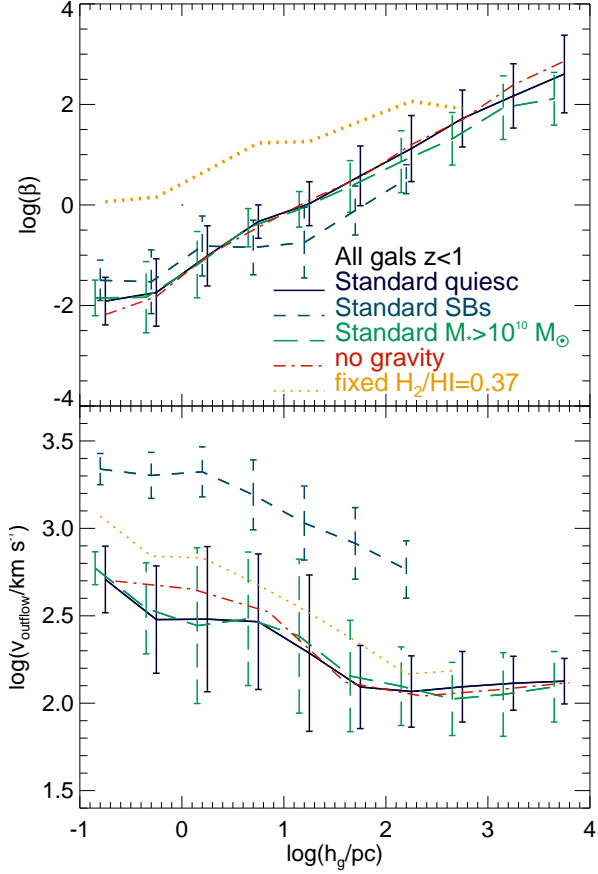}
\caption[The outflow rate to SFR ratio, as a function of
 the surface density of gas plus stars for different model 
considerations.]{{Top panel:} The
mass loading, $\beta$, as a function of
 the gas scaleheight at the half-mass radius for: quiescent (solid line), starburst (dashed line) and 
a subsample of massive galaxies, $M_{\star}>10^{10}\, h^{-1} M_{\odot}$ (long-dashed line), in the 
model with the standard set of parameters (Table~\ref{freepars}). 
In the case of quiescent SF, $h_{\rm g}$ is evaluated at 
$r_{\rm 50}$ of the disk, and for starbursts, at $r_{\rm 50}$ of the bulge.
We include in the plot all galaxies in {\tt GALFORM} at $z<1$ and with $M_{\star}>10^{8}\, h^{-1} M_{\odot}$.
We also show the effect of suppressing gravity on the expansion of bubbles (dot-dashed line),  
and of assuming a constant H$_2$/HI ratio instead of that derived from the 
Blitz \& Rosolowsky pressure law (dotted line). Solid lines and errorbars indicate the median and 10 and 90\% ranges of 
the predictions. For clarity, errorbars are shown only for selected cases. {\it Bottom panel} As in the top panel, but here we show 
the mass-weighted outflow velocity as a function of the gas scaleheight.}
\label{fig:quiesc_burst}
\end{center}
\end{figure}

The top panel of Fig.~\ref{fig:quiesc_burst} shows the correlation between
$\beta$ and $h_{\rm g}$ at the half-mass radius obtained
with and without considering gravity from stars and DM in Eqs.~\ref{ener1}-\ref{ener2},
\ref{press1}-\ref{press2} and \ref{mom1}-\ref{mom2}, and using the standard set of parameters 
to describe GMCs and the ISM of galaxies (see Table~\ref{freepars}). 
{We plot the gas scaleheight at the half-mass radius in the range from $0.1$~pc to $10^4$~pc, but galaxies with such extreme 
half-mass radius are very rare. In fact, the median 
$h_{\rm g}$ for starbursts ranges from $50$~pc in low mass galaxies to $10$~pc in 
high-mass galaxies, and for quiescent galaxies it ranges from $450$~pc in low mass galaxies to $80$~pc in 
high-mass galaxies.}

We find that $\beta$ is only slightly affected when
gravity is not included. This agrees with what we find for individual bubbles, in which 
gravity has an effect of at most $5$\% on the final bubble mass.  
The effect of including the H$_2$/HI ratio calculated from the Blitz \& Rosolowsky pressure law 
in the modelling of the ISM is much larger than the direct gravitational
effect, as the dotted line in Fig.~\ref{fig:quiesc_burst} shows.
The omission of self-consistent multiphase modelling is represented by the results obtained with 
a fixed H$_2$/HI$=0.37$ ratio, which is
the value used in previous work to estimate HI from the total cold gas content (e.g. \citealt{Power10}; \citealt{Kim10}).
With a fixed H$_2$/HI ratio, the mass loading increases by factors of up to
$100$ for galaxies with the smallest gas scaleheights (i.e. highest density regimes). This is due to the anticorrelation
between H$_2$/HI and $h_{\rm g}$ \citep{Lagos11}. Galaxies with
very high gas and/or stellar surface densities have smaller $h_{\rm g}$ and 
larger H$_2$/HI, driving a lower overall content of HI and
therefore providing less material for bubbles to sweep up, reducing the outflow mass. This effect is very large in
more extreme cases, where the pressure law predicts little HI. This is also clear from the single bubble examples of 
$\S$~\ref{SubSec:PropsSingle}, in which the bubble mass is greatly reduced in molecule-dominated media.
 This demonstrates the importance of the
ISM modelling introduced in \citet{Lagos10} and \citet{Lagos11}, and also included in
some other recent models (e.g. \citealt{Fu10}).

In the top panel of  Fig.~\ref{fig:quiesc_burst} we show the relations for starburst and massive galaxies 
separately. This stresses the similarity between the relations displayed by quiescent and starburst galaxies in the 
$\beta$-$h_{\rm g}$ plane and the fact that massive galaxies follow the same relation as the 
overall galaxy population, which is dominated in number by lower mass systems.
This is because the mass loading $\beta$ is primarily determined by the gas scaleheight 
 and the gas fraction, as we show later in $\S$~\ref{SubSec:NewBetaModel}.

In the bottom panel of Fig.~\ref{fig:quiesc_burst} we show the mass-weighted outflow velocity as a function of 
the gas scaleheight. 
There is a trend of decreasing velocity for increasing $h_{\rm g}$. Starburst galaxies 
exhibit a relation with a similar slope to that of quiescent galaxies but offset by $\approx 0.5$dex to larger velocities. 
This is due to the different star formation laws assumed in the model for the starburst and quiescent star formation modes
(see $\S$~\ref{Sec:modeldetails}). For a fixed $h_{\rm g}$, a starburst galaxy generally has a larger SFR than its 
quiescent counterpart. This drives 
larger energy and momentum injection, resulting in larger outflow velocities. 
The effect of gravity in the outflow velocity is only minor, as is also the case for $\beta$. 
The effect of including the Blitz \& Rosolowsky pressure law in the modelling of the ISM on 
the outflow velocity is more significant, and 
its omission results in velocities that are larger by a factor of $\approx 2$ at small $h_{\rm g}$. 
In \S~\ref{Sec:ComparisonObsHydro} we compare our predicted velocities with observations. 

\subsubsection{Assessing the impact of ISM and GMC parameters on the outflow properties}\label{ParameterEffects}

\begin{figure}
\begin{center}
\includegraphics[trim = 5mm 0mm 7.8mm 1mm,clip,clip,width=0.44\textwidth]{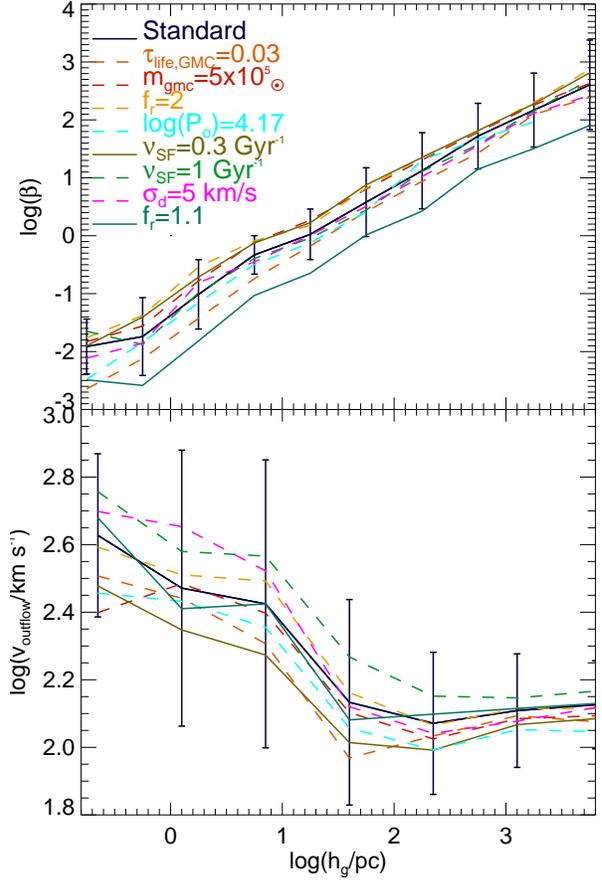}
\caption[The outflow rate to SFR ratio, as a function of
 surface density of gas plus stars for
different choices of GMC and ISM model parameters.]{{\it Top panel:} The
predicted mass loading, $\beta$, as a function of
the gas scaleheight.
In the case of quiescent SF, $h_{\rm g}$ is evaluated at
$r_{\rm 50}$ of the disk, and for starbursts, at $r_{\rm 50}$ of the bulge.
The predictions are shown for different choices of the model parameters, as labelled.
We include in the plot all galaxies in {\tt GALFORM} at $z<1$ with $M_{\star}>10^{8}\, h^{-1} M_{\odot}$.
 Lines and errorbars indicate the median and 10 and 90 percentile ranges of the relations. For clarity,
the percentile range is shown only for one model as they are all similar. Solid lines are used for the 
model with the standard set of parameters and those predicting the lowest and the highest $\beta$ for a given $h_{\rm g}$. 
Dashed lines are used for the rest of the models (see Table~\ref{freepars}).
 {\it Bottom panel} As in the top panel, but here we show
the mass-weighted outflow velocity as a function of the gas scaleheight. }
\label{fig:betaall}
\end{center}
\end{figure}

The top panel of Fig.~\ref{fig:betaall} shows the predicted mass loading as a function of the gas scaleheight 
when varying the parameters associated with the modelling of GMCs and the diffuse medium 
(see Table~\ref{freepars}).
Changes in the GMC and diffuse medium model parameters drive different normalisations in the
$\beta$-$h_{\rm g}$ relation but have a weak impact on the shape of the relation.
The variations between the models that produce the smallest and largest $\beta$ values,
which correspond to adopting $f_{\rm r}=1.1$ and $\nu_{\rm SF}=0.3\, \rm Gyr^{-1}$, respectively,
are at most a factor of $\approx 10$.
It is reasonable to argue that a better understanding of the multi-phase nature 
 of the ISM and the properties of GMCs is very important, even more so 
than including some of the physical mechanisms in the expansion of bubbles, such as gravity. 
This was also hinted at in Fig.~\ref{fig:quiesc_burst} from the effect of adopting a multi-phase ISM 
description of the outflow rate. 

The effect of each of the parameters in Table~\ref{freepars} on $\beta$ is summarised below. 
\begin{itemize}
\item Smaller values of $f_{\rm r}$
 result in smaller $\beta$ values by a factor $\approx 3-5$. This is expected from the
role $f_{\rm r}$ plays in determining the break-out radius of bubbles and therefore the bubble mass (Eq.~\ref{bubblemassform}). 
\item Adopting a smaller SF coefficient or a smaller GMC mass drives an increase in $\beta$ 
 {due to the lower SFR predicted by the former and the higher number of GMCs predicted by the latter.}
The effect of increasing $\nu_{\rm SF}$ or $M_{\rm GMC}$ is therefore a smaller $\beta$. 
Adopting a longer lifetime for GMCs also decreases $\beta$ due to the anticorrelation between 
$\beta$ and $\tau_{\rm life,GMC}$.
\item A smaller hydrostatic pressure normalisation in the Blitz \& Rosolowsky law (see $\S$~\ref{Sec:SFeqs_sn}) drives 
larger $\beta$ but only
 in galaxies which have a molecule-dominated ISM, as it only affects this regime (see Eq.~\ref{bubblemassform}). 
In these cases, the lower $P_{0}$ drives smaller individual bubble masses and therefore smaller $\beta$ (see Eq.~\ref{analyticbeta}).
Similarly, the effect of decreasing $\sigma_{\rm d}$ is to slightly decrease $\beta$, which is also expected from the analysis of 
$\S$~\ref{Sub:Analytic}. 
\end{itemize}

The effect of varying the parameters 
above on the mass-weighted outflow velocities, $v_{\rm outflow}$, is shown in the bottom panel of Fig.~\ref{fig:betaall}.
Variations in $v_{\rm outflow}$ due to different ISM parameter choices are smaller than in the case of $\beta$, with 
a difference between the minimum and maximum $v_{\rm outflow}$ of $\approx 0.5$dex. The models predicting the highest and lowest $\beta$ 
are not the same as those predicting the highest and lowest $v_{\rm outflow}$. This is because $v_{\rm outflow}$ is more affected by those 
parameters directly changing the energy injection into 
the ISM by SNe. Indeed, the parameter that is most important in setting 
$v_{\rm outflow}$ is the star formation coefficient, $\nu_{\rm SF}$. The more efficient the conversion from gas to stars, the higher the 
outflow velocity. This is consistent with what is shown for quiescent and starburst galaxies in Fig.~\ref{fig:beta_annuli4}.

\subsubsection{The outflow rate and velocities in galaxies with extreme ISM conditions}\label{Sec:xtmISM}

\begin{figure*}
\begin{center}
\includegraphics[trim = 1mm 0mm 7.8mm 1mm,clip,clip,width=0.4\textwidth]{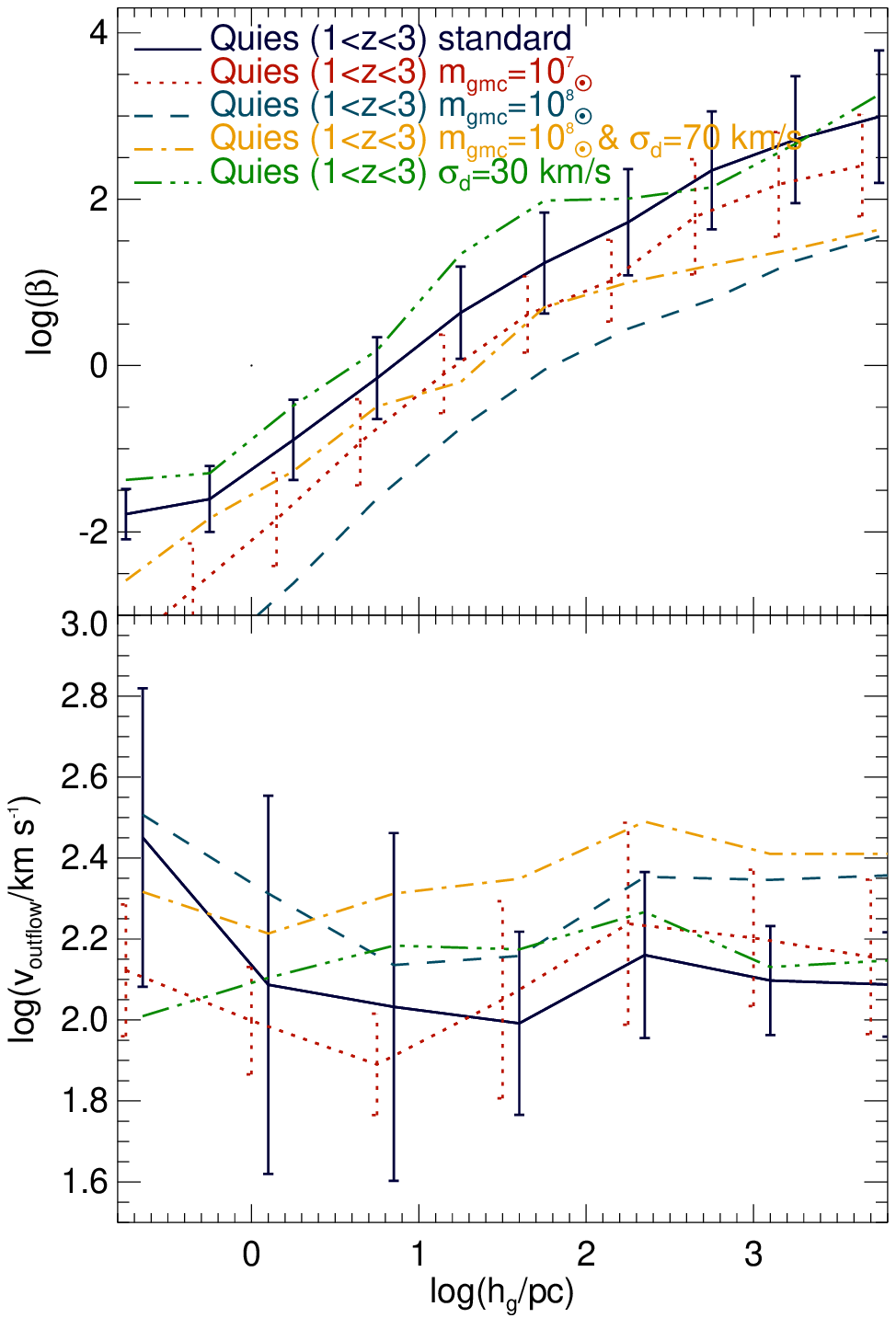}
\includegraphics[trim = 1mm 0mm 7.8mm 1mm,clip,clip,width=0.4\textwidth]{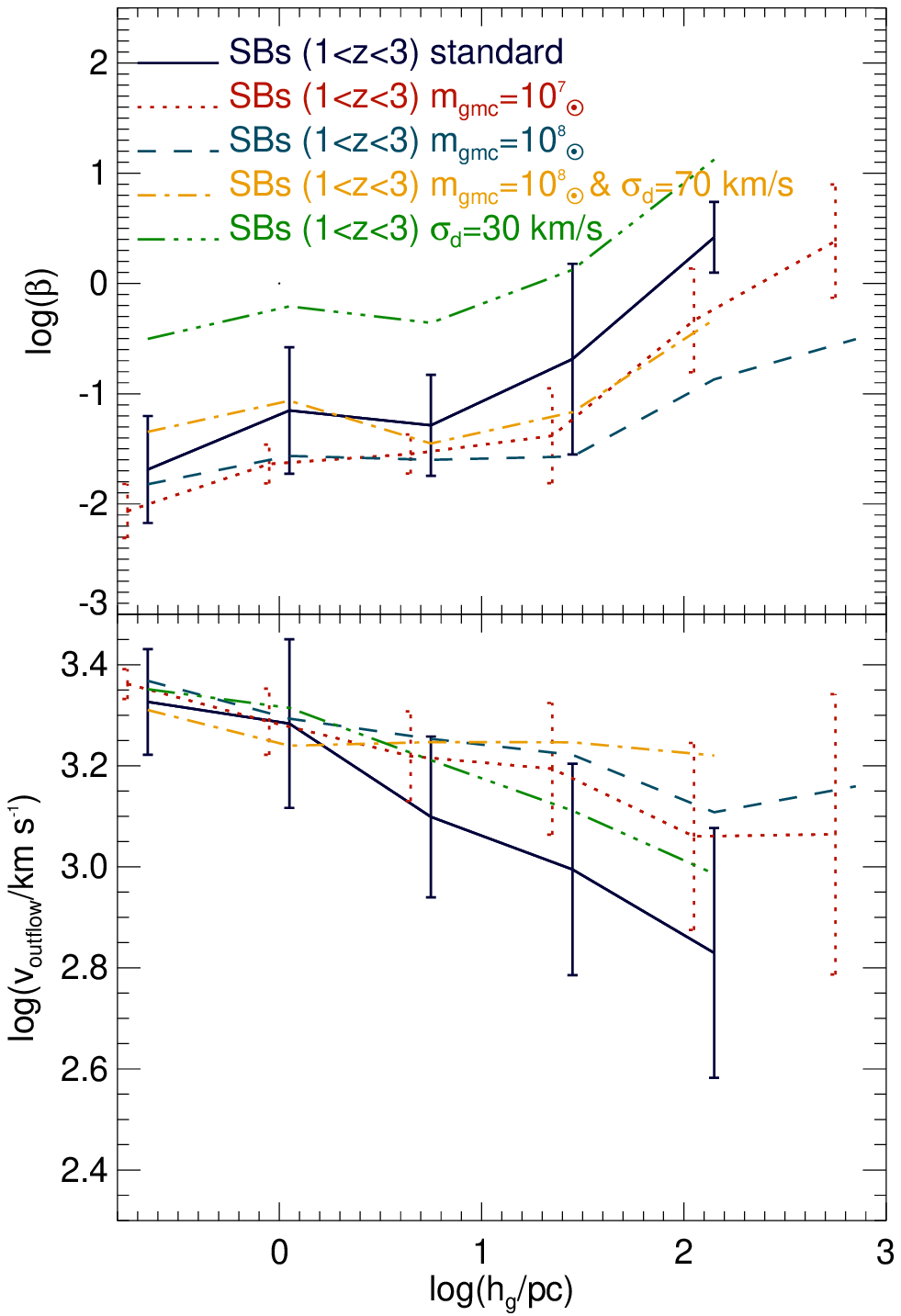}
\caption[The outflow rate to SFR ratio, as a function of
 surface density of gas plus stars for
galaxies with extreme ISM conditions.]{{ {\it Top panel:} The
predicted mass loading, $\beta$, as a function of
the gas scaleheight, $h_{\rm g}$, for quiescent SF (left panel) and starbursts (right panel). 
In the case of quiescent SF, $h_{\rm g}$ is evaluated at
$r_{\rm 50}$ of the disk, and for starbursts, at $r_{\rm 50}$ of the bulge.
The predictions are shown for the standard choice of parameters, and for extreme values of $M_{\rm GMC}$ and 
$\sigma_{\rm d}$, as labelled, which could be representative of the conditions of high-redshift star-forming galaxies.
Since we want to investigate these high-redshift galaxies, we include in the plot 
all galaxies in {\tt GALFORM} at $1<z<3$ with $M_{\star}>10^{8}\, h^{-1} M_{\odot}$.
 Lines and errorbars indicate the median and 10 and 90 percentile ranges of the relations. For clarity,
the percentile range is shown only for two models as they are all similar. 
 {\it Bottom panel} As in the top panel, but here we show
the mass-weighted outflow velocity as a function of $h_{\rm g}$.} }
\label{fig:betaxtmAll}
\end{center}
\end{figure*}

{Resolved observations of the ionised gas in star-forming galaxies at $1\lesssim z \lesssim 3$ have shown that they have 
velocity dispersions that are systematically 
larger than the ones measured for the neutral gas content of local spiral and dwarf galaxies, 
and that they host star-forming clumps which can be more extended and luminous in H$\alpha$ than local clumps 
e.g. \citealt{Law07}; \citealt{Puech07}; \citealt{Genzel08}; \citealt{Livermore12}; see \citealt{Glazebrook13} 
for a recent review), similarly 
to local starbursts (see $\S$~\ref{SubSec:DiffMedProps}). 
Galaxies more massive than $M_{\rm stellar}\gtrsim 10^{11}\,M_{\odot}$ built-up more than half of their stellar mass at $z>1$ 
(e.g. \citealt{Perez-Gonzalez08}), and therefore 
 may form most of it in a clumpy, turbulent ISM. 
However, it is important to bear in mind that the 
low number of galaxies on the observational samples does not allow to conclusively determine 
how representative these are of the overall galaxy population.
Another important warning is that the velocity dispersion measured at high-redshift correspond to the 
ionised component of the ISM, while the relevant quantity for our model is the atomic and molecular gas velocity dispersion. 
 Other systematics effects include 
the point-spread function and the limited spatial resolution that can bias the 
inferred values toward higher observed velocity dispersion and more extended clumps 
(e.g. see \citealt{Glazebrook13} for a discussion of systematics).

Given the important role an `extreme' ISM phase could play on galaxy evolution, we investigate in this section  
the effect on the mass loading and velocity of the outflow of increasing $\sigma_{\rm d}$ and $M_{\rm GMC}$.
We adopt $M_{\rm GMC}=10^8\,M_{\odot}$ and $\sigma_{\rm d}=70\rm \, km\, s^{-1}$ as representative values for 
clumpy galaxies. We also test intermediate values for the GMC mass, $M_{\rm GMC}=10^7\,M_{\odot}$, and 
for the gas velocity dispersion, $\sigma_{\rm d}=30\rm \, km\, s^{-1}$, to better test the effects of increasing 
$M_{\rm GMC}$ and $\sigma_{\rm d}$.}

{We ran $3$ simulations with increased $M_{\rm GMC}$ or $\sigma_{\rm d}$ and 
one with both quantities increased with respect to the standard choice of ISM and GMC parameters 
(see Table~\ref{freepars}). The results of those runs 
 are shown in Fig.~\ref{fig:betaxtmAll} for quiescent and starburst galaxies. We focus on galaxies in the redshift range 
$1<z<3$ to match the redshift range of the surveys described above. The increase in $M_{\rm GMC}$ by 
two orders of magnitude decreases $\beta$ by $\approx 1.5$~dex, while the increase in $\sigma_{\rm d}$ by a factor of 
$3$ increases 
$\beta$ by $\approx 1$~dex. This is consistent with the variations we expect from our simplified analytic solution 
for $\beta$ ($\S$~\ref{Sub:Analytic}).
When we increase both, $\sigma_{\rm d}$ and $M_{\rm GMC}$, the variations 
in $\beta$ compensate in a way that adopting 
$\sigma_{\rm d}=70\,\rm km \,s^{-1}$ and $M_{\rm GMC}=10^8\,M_{\odot}$ causes 
$\beta$ to decrease by at most $0.5$~dex with respect to the values obtained in our standard choices for these parameters.
From the Jeans mass in a disk, $M_{\rm J}\propto \sigma^4_{\rm d}/\Sigma_{g}$, we expect both quantities to increase together and 
thus we expect 
net variations in $\beta$ of at most a factor of $3$ in galaxies with more extreme ISM conditions, which could be representative of 
the high-redshift population.}

{In the case of the outflow velocity (lower panels in Fig.~\ref{fig:betaxtmAll}), we find that the increase 
in $\sigma_{\rm d}$ and $M_{\rm GMC}$ drive smaller variations than in $\beta$, in the range of $0.3-0.4$~dex. This is consistent 
with the picture presented in Sec.~\ref{ParameterEffects}, where $\nu_{\rm SF}$ drives the largest variations in the outflow velocity. 
Note that the effect of adopting different values of these parameters is different for 
quiescent galaxies than it is for starbursts. This is driven by the different 
star formation laws assumed in each SF mode (see Sec.~\ref{Sec:modeldetails}).}

\subsubsection{The physical regimes of the outflow}\label{Sec:PhysicalRegime}
\begin{figure}
\begin{center}
\includegraphics[trim = 4.3mm 2.5mm 4.8mm 0mm,clip,width=0.43\textwidth]{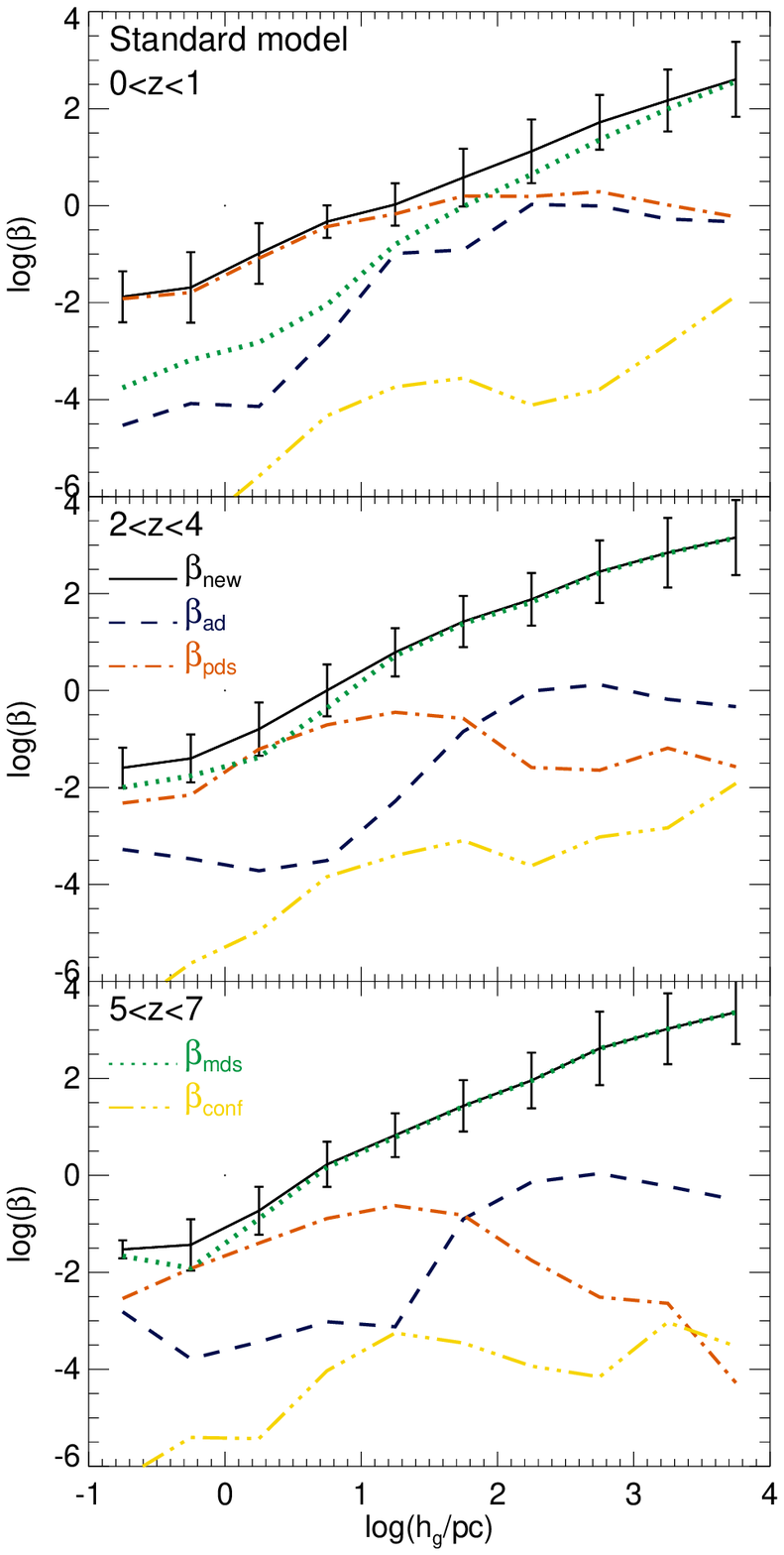}
\caption[Physical regimes of supernovae feedback: when and where each evolutionary stage
of bubbles dominate the outflow rate.]{The mass loading factor, 
$\beta$, as a function of
 the gas scaleheight at the half-mass radius 
for galaxies with $M_{\star}>10^{8}\, h^{-1} M_{\odot}$ in three redshift 
ranges, as labelled in each panel. 
In the case of quiescent SF, $h_{\rm g}$ is evaluated at
$r_{\rm 50}$ of the disk, and for starbursts, at $r_{\rm 50}$ of the bulge. 
The contribution to the total
$\beta$ (solid line) from bubbles escaping in the adiabatic, pressure-driven
and momentum-driven snowplough phases are shown as dashed, dot-dashed and dotted lines, respectively.
 The ratio between the rate of mass confinement and the SFR, $\beta_{\rm conf}$,
is shown as triple-dot-dashed line. Lines represent the medians and the errorbars, which are shown for clarity 
only for the total $\beta$, represent the $10$ to $90$ percentile range.}
\label{PhyscRegimes}
\end{center}
\end{figure}

Bubbles inflated by SNe feedback can escape the galaxy in any of the three evolutionary stages
described in $\S$~\ref{Sec:DynModel_SN}. We now quantify where and when
each of these stages dominates the outflow of material.

Fig.~\ref{PhyscRegimes} shows the mass loading, $\beta$, as a function of
the gas scaleheight, $h_{\rm g}$, evaluated at the half-mass radius 
for the model with the standard set of parameters. 
We find that at high redshift, most of the outflow in galaxies is 
produced by bubbles escaping in the momentum-driven stage, while 
 low-redshift galaxies with small gas scaleheights have mass outflow rates  
dominated by bubbles escaping in the pressure-driven stage.
High-redshift galaxies have a gas scaleheight set by the
gas surface density with a negligible contribution from the surface density of stars.
In the low-redshift regime, galaxies with small gas scaleheight have, by comparison, 
a more important contribution from the stellar component. In fact, the median gas fraction 
of the galaxy sample with $h_{\rm g}<10\, \rm pc$ in the high- and low-redshift samples is 
$0.98$ and $0.18$, respectively. Galaxies which have the gas scaleheight set mainly by 
the stellar surface density, have bubbles where the cooling time for the interior gas 
is large enough for bubbles to escape the disk in the pds stage.
In the case of the larger gas scaleheight galaxy population, the scaleheight set 
mainly by the gas surface density, so no significant difference with redshift is obtained. 

When bubbles escape the ISM in the radiative phase (i.e. pds or mds), 
this implies that most of the outflow mass is in 
a cold, dense phase (i.e. molecular or neutral atomic gas) and that the interior mass of the bubbles 
 is only a minor contributor. 
This qualitatively agrees with what is observed in local galaxies (e.g. Tsai et al. 2012a,b). 
A quantitative comparison will be presented in a forthcoming paper (Lagos, Baugh \& Lacey, in prep.). 

The adiabatic phase only rarely dominates the outflow rate, since 
 the transition from the ad to the pds stage takes place early on in the evolution of bubbles.
This transition almost always takes place on a timescale of $\approx 10^3-10^5$~years.
Full confinement due to deceleration of bubbles rarely takes place (i.e. the case in which 
no bubbles break-out from the galaxy disk), 
and happens mainly in places where the scaleheight
is large and the bubble has time to decelerate to the velocity dispersion of the diffuse gas 
(i.e. at low gas densities). Most of the gas which remains in the ISM therefore corresponds to gas expanding
in the direction close to the plane (i.e. the fraction $(1-f_{\rm bo})$ in Eqs.~\ref{Eqs:SFset1}-\ref{Eqs:SFset3}) rather than to 
bubbles which are fully confined in the ISM. 
The tendency we find for bubbles to break-out in the radiative phase
contrasts with what \citet{Monaco04} found, whose model predicts that 
most bubbles escape during the adiabatic phase. This difference may be due 
to the assumptions Monaco makes that bubbles expand against the hot phase.
In our model, bubbles expand against the warm phase, whose density is typically higher than the hot phase,
which results in larger cooling rates. We find that our approach gives answers more similar
to fully hydrodynamical simulations in the range where they overlap (see $\S$~\ref{Subsec:RadialBetas}).

\subsubsection{Outflow rates of mass and metals}\label{MetalsRate}

\begin{figure}
\begin{center}
\includegraphics[trim = 0.2mm 0mm 3.8mm 0mm,clip,width=0.43\textwidth]{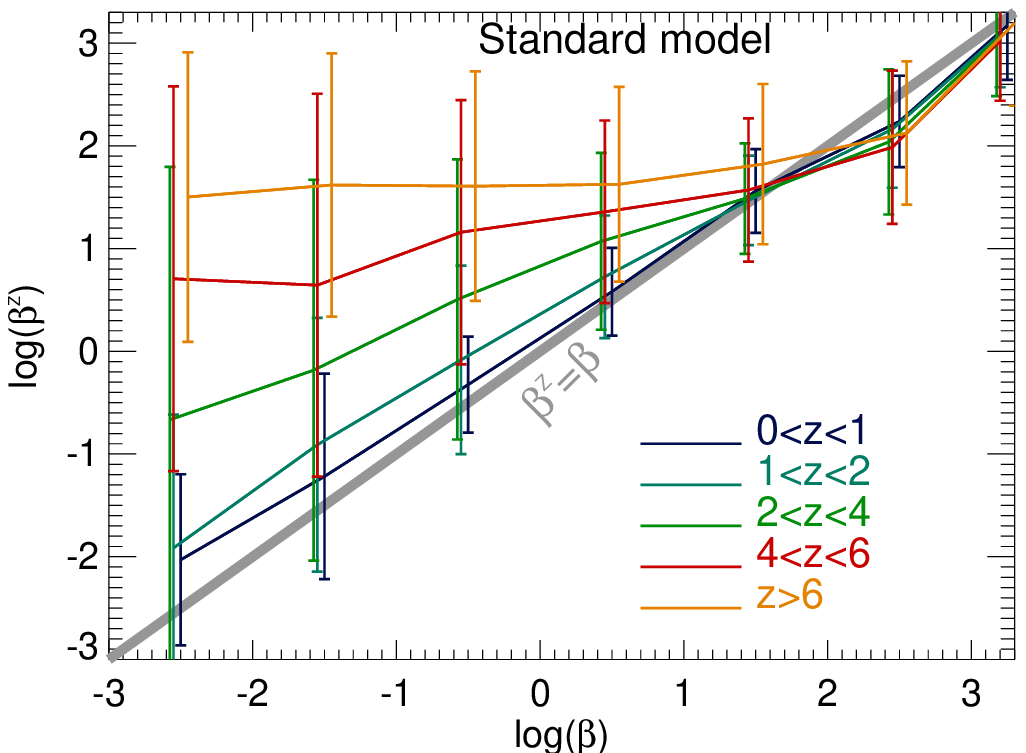}
\includegraphics[trim = 0.2mm 0mm 3.8mm 0mm,clip,width=0.43\textwidth]{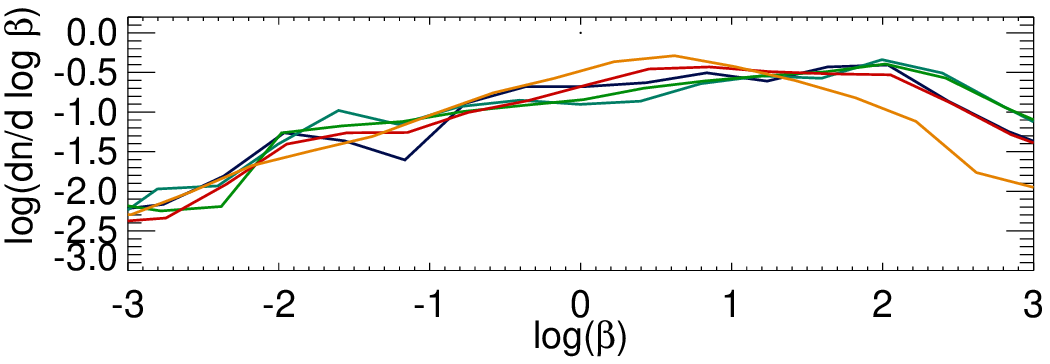}
\caption[Physical regimes of supernovae feedback: when and where each evolutionary stage
of bubbles dominate the outflow rate.]{{\it Top panel:} The metal loading factor, 
$\beta^Z=\dot{M}^Z_{\rm eject}/(Z_{\rm g}\psi)$, as a function of the 
mass loading factor, $\beta=\dot{M}_{\rm eject}/\psi$, for galaxies 
with $M_{\star}>10^{8}\, h^{-1} M_{\odot}$. 
Both quantities are integrated over the galaxy and in the same timesteps.
Lines with errorbars represent the median and the $10$ to $90$ percentile range, respectively, for galaxies 
at different redshifts, as labelled. The thick, straight line shows $\beta^Z=\beta$. 
{\it Bottom panel:} normalised distribution of $\beta$ for galaxies in the 
same redshift ranges as in the top panel.}
\label{MetalsOut}
\end{center}
\end{figure}

We have analysed the physics behind the dependence of $\beta$ on galaxy properties and gave 
analytic derivations for such relations. However, a key part of the impact of outflows on 
galaxy evolution is the fate of the metals carried away by bubbles. In the model, we assume that the metals 
which flow out from the galaxy accumulate in the ejected mass component, which is later reincorporated into the 
hot halo gas (see Eqs.~\ref{Eqs:SFset1}-\ref{Eqs:SFset2}). The amount of metals outflowing from 
the galaxy therefore has a direct impact on the 
cooling rate of the hot halo gas and hence on subsequent gas accretion 
and star formation in the galaxy.

Here, we analyse the loading factor of metals  
defined as $\beta^Z=\dot{M}^Z_{\rm eject}/(Z_{\rm g}\psi)$ (see Eq.~\ref{Eqs:SFset4}). 
The top panel of Fig.~\ref{MetalsOut} shows the metal loading factor as a function of the mass loading factor 
for galaxies at different redshifts. Galaxies at $z<2$ follow a relation which is close to 
$\beta^Z=\beta$, but which shows a flattening at $\beta \lesssim 0.5$ (i.e. in the small gas scaleheight regime). 
However, as the redshift increases, deviations become important and begin at increasingly larger $\beta$. 
At $z>6$ there is almost no correlation between $\beta^Z$ and $\beta$, with 
$\beta^Z\approx 30$ independent of $\beta$, albeit with a large dispersion. 
This behaviour is due to high-redshift galaxies having intrinsically lower metallicity gas 
from which stars form. In the low-metallicity regime, metals in bubbles coming from the swept-up gas are negligible compared to 
those coming from SNe ejecta; in the limit of $Z_{\rm g}\ll  p_{\rm SN}$ and $4\pi R^2 Z_{\rm g} \rho_{\rm d} v
_{\rm s} \ll \dot{m}^Z_{\rm inj}$, we can write the metal outflow rate due to a 
single bubble as

\begin{equation}
\dot{m}^Z_{\rm eject}=f_{\rm bo}\, p_{\rm SN}\, \psi_{\rm GMC}=f_{\rm bo}\, p_{\rm SN}\, \nu_{\rm SF}\,M_{\rm GMC}.
\end{equation}
 
\noindent The rate of metals flowing out from the galaxy in a given annulus is regulated by the number of GMCs in that annulus  
$\dot{M}^Z_{\rm eject}=f_{\rm bo}\,p_{\rm SN}\, \nu_{\rm SF}\, M_{\rm mol}$. We then calculate $\beta^Z$ per annulus in this regime 

\begin{equation}
\beta^Z=\frac{\dot{M}^Z_{\rm eject}}{Z_{\rm g}\, \nu_{\rm SF}\, M_{\rm mol}}=\frac{f_{\rm bo}\,p_{\rm SN}}{Z_{\rm g}}.
\end{equation}

\noindent Because we assume instantaneous mixing in {\tt GALFORM}, this $\beta^Z$ is representative of the 
global metal loading factor. In the limit of  $Z_{\rm g}\ll  p_{\rm SN}$, $\beta^Z$ shows no dependence 
on $h_{\rm g}$. However, the mass outflow rate has a strong dependence on $\Sigma_{\rm g}$, 
regardless of the metallicity of the ISM. This results in very little correlation between $\beta^Z$ and $\beta$ in this low-metallicity regime. 

If the ISM is already enriched with some metals, which corresponds to approximately $Z_{\rm gas}\gtrsim 0.05-0.1 Z_{\odot}$,  
 the density of the gas in the ISM also has an important 
effect on $\beta^Z$ given that the term $4\pi R^2 Z_{\rm g} \rho_{\rm d} v
_{\rm s}$ becomes comparable to or larger than the term $\dot{m}^Z_{\rm inj}$ in the evolution of single bubbles (see Eq.\ref{ener2}). In this 
case, a correlation between $\beta^Z$ and $\beta$ arises.

Although a non-linear relation between $\beta$ and $\beta^Z$ is predicted, 
we find that most galaxies in our simulation follow a relation which is close to $\beta^Z=\beta$. 
This can be seen from the distribution of $\beta$ for different redshifts in the bottom panel 
of Fig.~\ref{MetalsOut}. Quantitatively, at least $75$\% of galaxies at any redshift have $\beta > 1$ and 
at least $50$\% at $z<5$ have $\beta > 10$. This puts at least half or more of the galaxies in the regime 
where  $\beta^Z\sim \beta$. Galaxies deviating this relation are the most metal-poor ones, which 
typically correspond to those with low stellar masses.
As we show later in $\S$~\ref{Sec:Galaxies}, the inclusion of a metal loading factor with an independent 
parametrisation from the mass loading factor in {\tt GALFORM}, has a small effect on the luminosity of galaxies. However, 
if we wish to analyse in detail the gas content of galaxies and the evolution of the mass-metallicity relation, 
we would need to allow for such variations in the $\beta^Z$ parametrisation included in the model.  

\subsection{Comparison with observations and non-cosmological hydrodynamical simulations}\label{Sec:ComparisonObsHydro}

We compare our predictions for the mass loading of the wind, $\beta$,
with the values inferred from observations by \citet{Heckman00}, \citet{Martin12},
who use 
absorption features in galaxy spectra,
\citet{Newman12}, who use emission line galaxy spectra,  
\citet{Bouche12}, who use absorption lines in the lines-of-sight to background quasars (probing the
 outflow and inflow of gas), {\citet{Bolatto13}, who inferred the total outflowing mass from molecular emission, and 
\citet{Rupke13}, who simultaneously study absorption and emission lines.} 
\citet{Heckman00} and \citet{Bouche12} focus on $L^*$ galaxies at low redshift ($z\lesssim 0.1$),
 while \citet{Martin12} focus on galaxies at $z\approx 1$ and Newman et al. on galaxies at $z\approx 2$. 
{Heckman et al., Bolatto et al. and Rupke et al., do not provide stellar masses for their galaxy samples.
We therefore use the near-IR photometry available in the NASA/IPAC 
Extragalactic Database to estimate the stellar mass 
from the $K$-band luminosity. 
If only the $H$-band luminosity is given, we use the colour measurements of \citet{Boselli00}, $H-K\sim 0.25$, to convert 
to a $K$-band luminosity. We then use the median $K$-band mass-to-light ratio from \citet{Bell03b} to convert to stellar masses. 
We apply the same calculation to estimate stellar masses in the sample 
of \citet{Schwartz04}, shown in Fig.~\ref{fig:beta_obsv2}.
In the case of \citet{Bouche12}, $r$-band absolute magnitudes are given for each galaxy in the sample, so we use the 
$r$-band mass-to-light ratio from \citet{Bell03b} to convert to stellar masses. 
Finally, we adopt a correction of $0.71$ in stellar mass to convert from the adopted IMF in Bell et al., the `diet' Salpeter, to the 
\citet{Kennicutt83} IMF. Given the uncertainties in the 
scalings above, we conclude that we cannot estimate stellar masses to a factor better than $0.2$~dex and adopt 
this number as a typical error (see \citealt{Mitchell13} for a recent discussion on stellar 
mass estimate uncertainties).}

\begin{figure}
\begin{center}
\includegraphics[trim = 1mm 2mm 0.5mm 2.5mm,clip,width=0.47\textwidth]{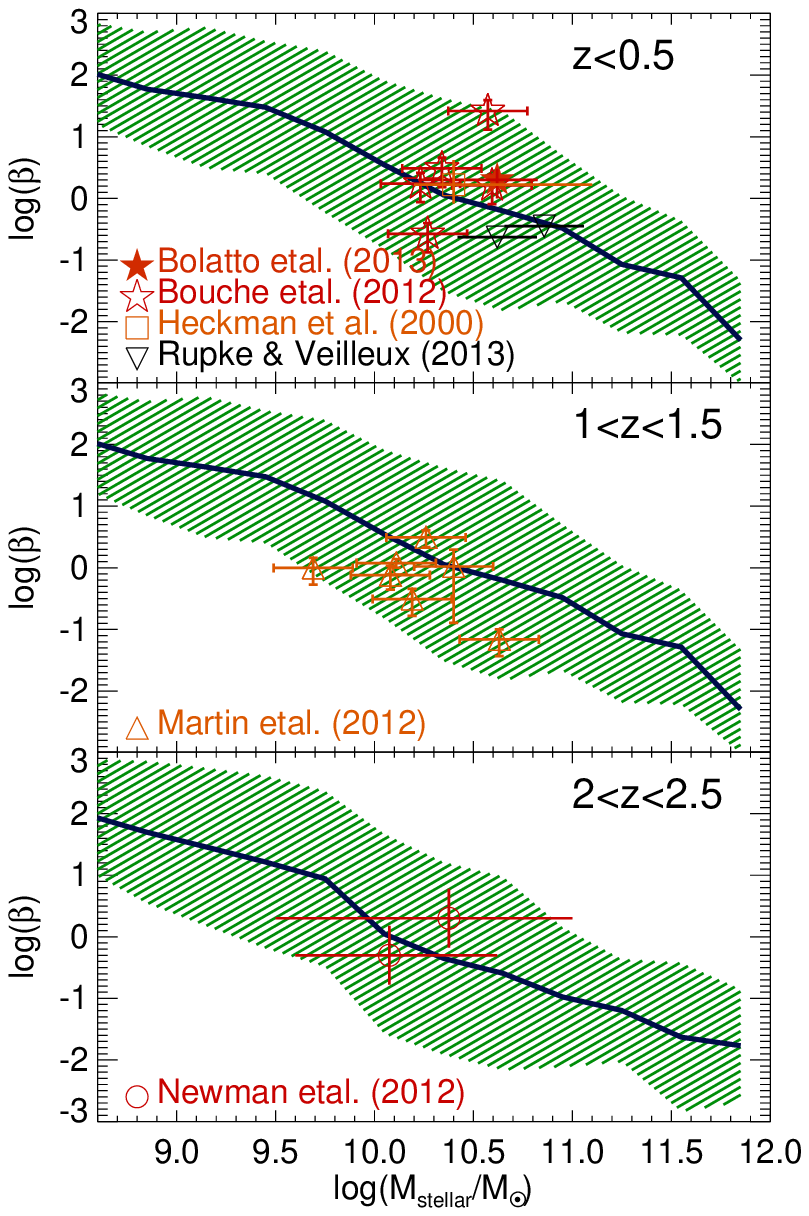}
\caption[The outflow rate to SFR ratio, as a function of
 the stellar mass.]{The mass loading, $\beta$, as a function of
 stellar mass for galaxies that have an outflow, 
in three different redshift ranges, as labelled, for
the standard set of parameters (Table~\ref{freepars}).
{Solid lines and the shaded regions indicate the median and 10 and 90\% ranges of
the distributions. 
The observationally inferred $\beta$ from 
  \citet{Heckman00}, \citet{Martin12}, \citet{Newman12}, \citet{Bouche12}, \citet{Bolatto13} and 
\citet{Rupke13} are shown using symbols, as labelled. 
The errorbars in the mass axis for Heckman et al. and Newman et al. represent 
the range of stellar masses of the galaxies in the samples and in the $y$-axis we show the range of inferred $\beta$. 
In the case of Newman et al., the two samples correspond to a low SFR sample, which has a lower median stellar mass, and a 
high SFR sample. In the cases of Bolatto et al., Bouche et al., Rupke et al. and Martin et al., 
 the error in the stellar mass and $\beta$ estimates are shown for individual galaxies. The data from Martin et al. 
plotted in the middle panel correspond to the subset of galaxies in their sample that have measured SFRs.}}
\label{fig:beta_obs}
\end{center}
\end{figure}

\begin{figure}
\begin{center}
\includegraphics[trim = 1mm 2mm 0.5mm 2.5mm,clip,width=0.47\textwidth]{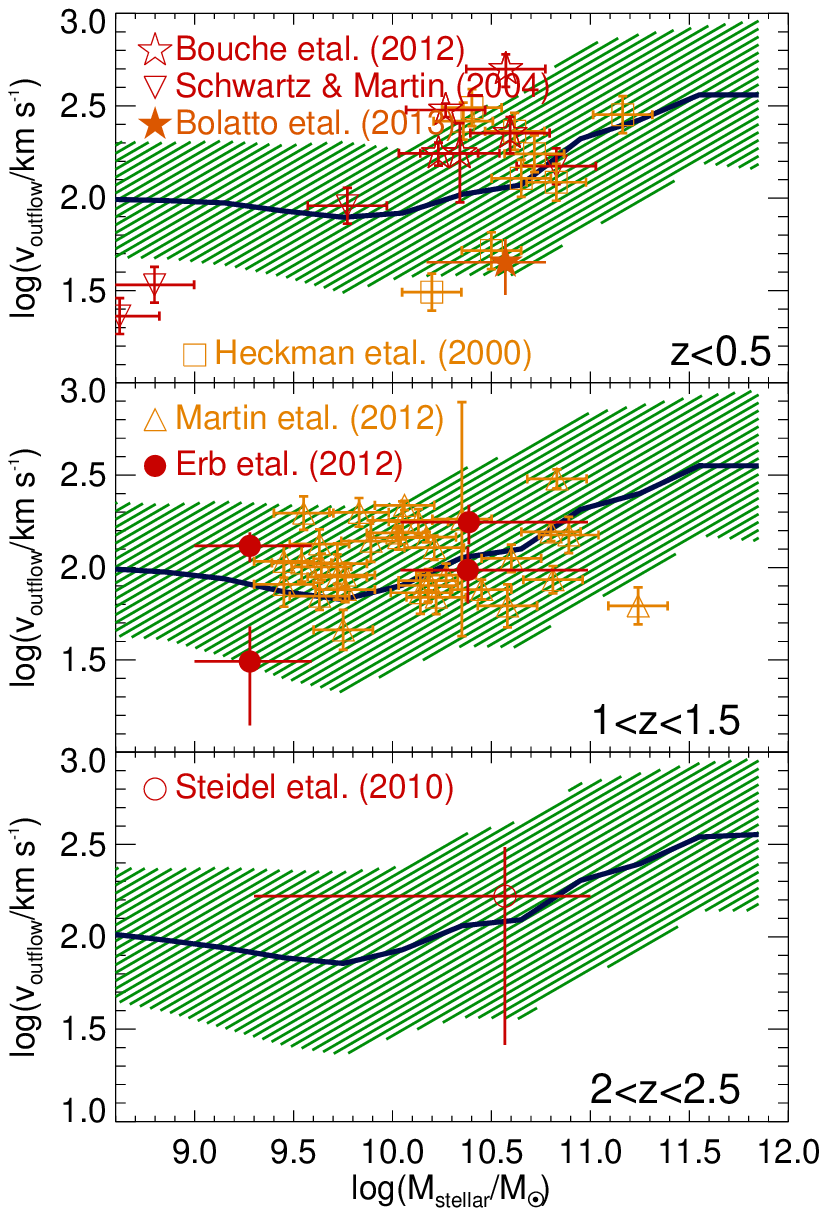}
\caption[The outflow rate to SFR ratio, as a function of
 the stellar mass.]{The mass-weighted outflow velocity, $v_{\rm outflow}$, as a function of
 stellar mass. The panels and galaxy selections are as in Fig.~\ref{fig:beta_obs}.
In the top panel, we show the observationally inferred outflow velocities of 
individual galaxies from \citet{Heckman00},
 \citet{Schwartz04}, \citet{Bolatto13} and 
 \citet{Bouche12}. 
 In the middle panel, we show the inferred outflow velocities in individual galaxies from the sample of
\citet{Martin12} and the median velocity of the galaxy samples of
\citet{Erb12}.
In the bottom panel we show the 
median outflow velocity and stellar mass of the sample of 
\citet{Steidel10}. In the case of Erb et al. and Steidel et al., the errorbars in the stellar mass axis 
correspond to the range of stellar masses in the samples, while the errorbars in the 
$y$-axis correspond to the standard deviation in the samples.
In the case of Erb et al. we plot two different velocity estimates for each stellar mass range, which 
correspond to two different iron transitions, corresponding to those giving the lowest and highest blue-shift velocities. 
Note that the number of data points in this figure from Martin et al. is larger than in Fig.~\ref{fig:beta_obs}. This is because 
 only a third of the sample had measured SFRs to provide an estimate of $\beta$.}
\label{fig:beta_obsv2}
\end{center}
\end{figure}

Fig.~\ref{fig:beta_obs} shows $\beta$ as a function
of stellar mass for our standard model (see Table~\ref{freepars}).
Symbols show the median stellar mass and the $\beta$ inferred from observational samples.
 Our model predicts $\beta$ values which are in broad agreement with those inferred from  
observations. However, there are large uncertainties  
associated with the inference of outflow rates from observations, in addition to the statistical uncertainties arising 
from the small number of objects sampled. 
The main uncertainties in the calculation of outflow rates from observations come from the
conversion between the ion and hydrogen column densities, which depends on the gas metallicity and
 ionization factor, the assumed geometry 
 (e.g. \citealt{Prochaska11}), and the still uncertain nature of absorption by low-ionisation metal
lines, in the case of absorption line studies in quasar sightlines. 
{Note that the errorbars plotted in Fig.~\ref{fig:beta_obs} do not include the systematic errors associated with the modelling 
assumptions made to derive $\beta$, and represent lower limits for the uncertainties.}

Fig.~\ref{fig:beta_obsv2} shows the mass-weighted outflow velocity 
as a function of stellar mass.  
We show the observational {estimates from \citet{Heckman00}, \citet{Schwartz04}, 
\citet{Bouche12} and \citet{Bolatto13} at $z\approx 0-0.1$, \citet{Martin12} and \citet{Erb12} at $z\approx 1$ and 
\citet{Steidel10} at $z \approx 2$. Note that for Erb et al. and Steidel et al., the errorbars 
corresponds to the standard deviation of $\beta$ in the full sample, while we plot individual errors in the rest 
of the observational samples.  
Heckman et al., Schwartz et al., Martin et al., Erb et al. and Steidel et al., 
use galaxy absorption line spectroscopy to infer an average blueshift of the 
ionised component with respect to the systemic velocity,   
Bouche et al. use MgII absorption lines in the 
lines-of-sight to background quasars to infer an outflow velocity, 
and Bolatto et al. use molecular emission lines to measure the kinematics of the 
cold gas. 
 The predicted outflow velocities are broadly consistent with those inferred from the 
observations.}
The estimates of the velocities and outflow rates from the observations 
is not straightforward,  
as the different gas phases of the outflow could have different velocities and mass loadings. { This 
becomes evident in the data points of \citet{Erb12} shown in Fig.~\ref{fig:beta_obsv2}; in a given stellar mass 
range, the two values of the outflow velocity correspond to two different iron line transitions.} 
In the case of the model, 
the plotted outflow velocities are calculated from the expansion velocities of 
bubbles at the point of break-out and are dominated by the phase that contributes the most to the outflow mass.
We predict that in many cases this corresponds to a warm or cold phase (neutral or molecular).
In the case of observations, most of the available data probe warm ionised gas and are corrected to account for the neutral 
component. Ideally, these data need to be complemented by 
deep observations at millimeter wavelengths to directly probe 
the part of the outflow that is in a cold phase. 

{ There are additional selection effects in the observations shown in Figs.~\ref{fig:beta_obs} and \ref{fig:beta_obsv2}, which 
are not taken into account in the comparison with the model. First, almost all of the 
observational samples are selected to include only highly star-forming galaxies, except for 
\citet{Bouche12}, which uses QSO absorption lines. Second, the reported outflow velocities correspond only to 
galaxies in which there was a detectable outflowing component. This biases the measurements against 
 low mass outflows. These effects need to be properly reproduced in the selection of galaxies in the model before carrying out 
a detailed comparison with the observations. 
For instance, the model predictions for the full galaxy population shown in Fig.~\ref{fig:beta_obsv2} are only marginally 
consistent with the velocities inferred by \citet{Schwartz04} for $3$ dwarf starburst galaxies. 
We calculate the median outflow velocity 
of galaxies with stellar masses 
in the range $10^{8}\,M_{\odot}-10^{9}\,M_{\odot}$ and with $\rm SFR>0.1\,M_{\odot}\,\rm yr^{-1}$, 
 corresponding to the properties of the Schwartz \& Martin sample (\citealt{Martin12}), and find $v_{\rm out}\approx 70\rm \, km\, s^{-1}$, 
with a 10 percentile of $10\rm \, km\, s^{-1}$ and a 90 percentile of $300\rm \, km\, s^{-1}$. 
The sample of Schwartz \& Martin, although not statistical, is 
broadly consistent with the predictions of the model for dwarf, star-forming galaxies.
This supports our conclusion that a careful comparison is needed.   
In a future paper we will analyse more fully the outflow mass in different 
phases and carry out a more detailed comparison with observations 
(Lagos, Baugh \& Lacey, 2013b).}

{ There are a few examples in which the different phases of the outflow are added 
to infer a total mass loading. This is the case of the starburst galaxies in \citet{Sturm11} and 
\citet{Rupke13}. Sturm et al. and Rupke et al. present estimates for the mass loading of the winds of small samples of local starbursts from 
multi-phase gas observations and derived $\beta\sim 0.1-1.1$, while in our model, we predict 
a median $\beta \approx 0.3$ for starburst galaxies with stellar masses $10^{10}<M_{\star}/M_{\odot}<10^{11}$, which overlaps 
with the stellar mass range of the observations. The predicted $\beta$ is consistent with the observations within the errorbars. 
The measured outflow velocities in the observational samples  
range from $100-800\rm \,km \,s^{-1}$, 
again consistent with the predicted 
mass-weighted velocities 
of starbursts in our model, which for the same stellar masses above, 
range between $250-1500\rm \, km \, s^{-1}$. Observationally inferred 
outflow velocities vary in a galaxy-to-galaxy basis and with the traced gas phase.}

We find that our
model agrees better with observationally inferred outflow rates compared to previous theoretical work on SNe feedback and 
 mass ejection from the ISM. 
For example, \citet{Efstathiou00} implemented a
physical model for galaxy evolution in which self-regulation was imposed: energy loss by cloud collisions is compensated by 
the energy input by SNe. Efstathiou predicted that galaxies with
$M_{\rm stellar}\approx 5\times 10^{10}\, M_{\odot}$ have a mass loading factor in winds from the ISM
of $\beta\approx 0.2$, which is a factor of more than $10$ lower than the values inferred by \citet{Martin99}
 and \citet{Bouche12}. The assumptions in the modelling of Efstathiou are different from ours.
An important difference is that we do not assume self-regulation in galaxies but instead we are able to test it. 
In addition to this, Efstathiou assumes that cooling in the interior of bubbles inflated by SNe is negligible and 
therefore SNe remnants can only contribute to the hot phase of the ISM. In our model we allow the interior of bubbles to cool 
down, which is a key process to follow, as in most of the cases cooling is efficient and bubbles enter a 
radiative phase rather quickly.

We find that our predicted outflow rates are similar to those found by \citet{Hopkins12} 
in simulations that resolve scales just below the size of GMCs and model SNe feedback by injecting
thermal energy stochastically into neighbouring particles.
However, their outflow rates correspond to the sum of several processes, such as photoevaporation and radiation pressure, 
and are not exclusively SNe driven outflows. 
They argue that in dense environments, radiation pressure dominates the overall outflow rate. 
In those environments our scheme predicts a larger contribution to the outflow rate from SNe than that predicted by Hopkins et al. 
Nonetheless, note that we indirectly assume that photoionisation takes place due to our assumption of SNe driving bubbles which expand 
against the warm medium instead of the dense gas from which stars form. 

\section{Towards a new parametrisation of the outflow rate}\label{NewBetasMod}

One of the main aims of this paper is to establish if the results of our dynamical 
model of SNe feedback can be reproduced using a simple 
parametrisation cast in terms of global galaxy properties.
In this section we use our dynamical model of SNe feedback embedded in {\tt GALFORM} to 
assess parametrisations
of the mass loading used in the literature ($\S$~\ref{Sec:vsysSNe}) and search for an improved way of reproducing 
the mass loading factor ($\S$~\ref{SubSec:NewBetaModel}).

\subsection{Dependence of the outflow rate on circular velocity}\label{Sec:vsysSNe}

\begin{figure}
\begin{center}
\includegraphics[width=0.42\textwidth]{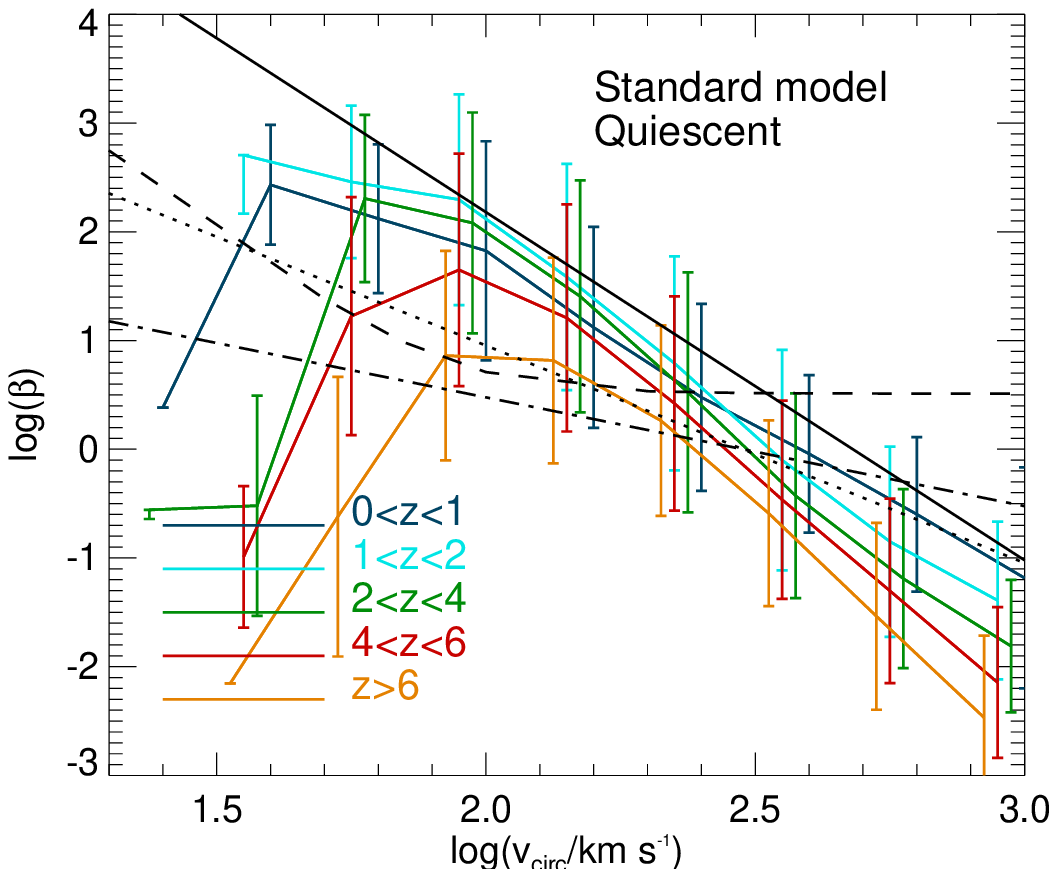}
\includegraphics[width=0.42\textwidth]{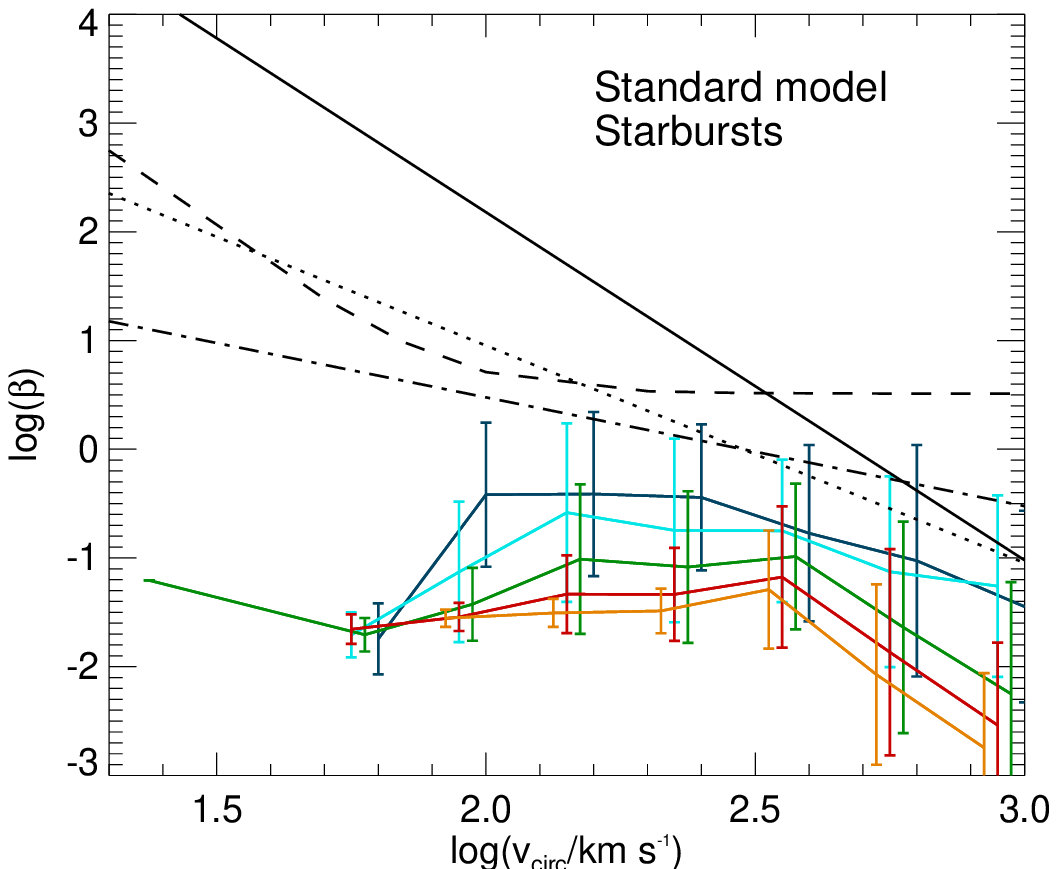}
\caption[The outflow rate to SFR ratio, as a function of the circular velocity for quiescent and starburst galaxies.]{{\it Top panel:} The 
mass loading factor, $\beta=\dot{M}_{\rm eject}/\psi$, as a function of the circular velocity of the disk for quiescent galaxies with 
$M_{\star}>10^{8}\, h^{-1} M_{\odot}$ in the model 
with the standard 
choice of parameters (see Table~\ref{freepars}).
The relation is shown for different redshift ranges, as labelled. 
 Solid lines and errorbars indicate the median and 10 and 90 percentile ranges of the relations. 
We also show the parametrizations used in a range of semi-analytic models, corresponding to 
 (i) Baugh et al. (2005; dotted line), (ii) Dutton et al. (2010; dot-dashed line), (iii) Bower et al. (2006; solid line)
and (iv) Guo et al. (2011; dashed line) (see text for details of the models).
{\it Bottom panel:} The $\beta-v_{\rm circ}$ relation in the model with 
the standard choice of ISM parameters for starburst galaxies with $M_{\star}>10^{8}\, h^{-1} M_{\odot}$ 
at different redshifts. In this case the circular velocity corresponds to that of the bulge. Lines and colours 
have the same meaning as in the top panel.}
\label{fig:beta:vsys}
\end{center}
\end{figure}

As discussed in the Introduction, a widely used approach in galaxy formation models  
is to parametrise the mass loading of the outflow solely in terms of the circular velocity, $v_{\rm circ}$, 
which is considered as a proxy for the depth of the potential well of the galaxy. 
Scalings of $\beta$ with circular velocity can be motivated by invoking momentum-conserving  
($\beta\propto v^{-1}_{\rm circ}$) or energy-conserving ($\beta\propto v^{-2}_{\rm circ}$) winds, or the power-law 
index can 
be treated as a free parameter, as in {\tt GALFORM} and most other semi-analytic models.
Our model has the power to test such assumptions by directly comparing the $\beta$ calculated 
for a given timestep with the circular velocity of the galaxy. 

Parametrisations of SNe feedback that include a direct scaling with the circular velocity of the galaxy can be 
grouped into two: those assuming a single scaling relation for both the outflow rate from the galaxy and from the halo, 
and those which separate them into two different mass loading factors, $\beta_{\rm ISM}$ for the mass loading of the galaxy and 
$\beta_{\rm halo}$ for that of the halo. 
{\tt GALFORM} is an example of the first type (see also \citealt{Lagos08} and \citealt{Cook10}). 
In the second type, we find the models of e.g. \citet{Croton06}, \citet{Monaco07}, \citet{Maccio10} and \citet{Guo11}.
For instance, \citet{Croton06} assume that the outflow rate from the galaxy scales linearly with the instantaneous
SFR, and adopt $\beta_{\rm ISM}=3.5$.
\citet{Maccio10} and \citet{Guo11} modified the form of $\beta_{\rm ISM}$ so that it makes a transition from a constant
value in high circular velocity galaxies to a form in which $\beta_{\rm ISM}$ increases as the 
circular velocity of the galaxies decreases, 
in order to better reproduce the number density of low-mass galaxies (see (iv) is the list below).
In our model we calculate $\beta_{\rm ISM}$ and compare it with the parametrisation from $4$ of the previous models.

Fig.\ref{fig:beta:vsys} shows the 
 $\beta$ predicted by the dynamical SNe feedback model after implementing it in the full galaxy formation 
simulation, plotted as a function of circular velocity for quiescent (top panel) and 
starburst galaxies (bottom panel). 
The model shown in Fig.\ref{fig:beta:vsys} corresponds 
to the standard choice of model parameters (see Table~\ref{freepars}). 
We overplot for comparison the following parametrisations for
 the mass loading from the literature:
\newline (i) $\beta=(v_{\rm circ}/300\, \rm km\, s^{-1})^{-2}$ from
\citet{Baugh05} (dotted line in Fig.~\ref{fig:beta:vsys}).
\newline (ii) $\beta=(v_{\rm circ}/300\, \rm km\, s^{-1})^{-1}$ from \citet{Dutton09} (dot-dashed line in Fig.~\ref{fig:beta:vsys}). In
the Dutton et al. model,
the normalisation velocity is calculated from the momentum injected by a single SN that ends up in the outflow,
which is $3.2\times 10^4\, M_{\odot}\, \rm km\, s^{-1}$ for a Kennicutt IMF.
\newline (iii) $\beta=(v_{\rm circ}/485\, \rm km\, s^{-1})^{-3.2}$
from  \citet{Bower06} (solid line  in Fig.~\ref{fig:beta:vsys}).
\newline (iv) $\beta=6.5\, [0.5+(v_{\rm circ}/70\, \rm km\, s^{-1})^{-3.5}]$ from \citet{Guo11} (dashed line in  in Fig.~\ref{fig:beta:vsys}),
which gives a SNe driven wind with a high mass loading even in galaxies with very high circular velocities, e.g. corresponding 
to those at the centre of clusters.

There are three key conclusions that can be drawn from Fig.\ref{fig:beta:vsys}: (i) a single power-law fit cannot describe 
the dependence of $\beta$ on $v_{\rm circ}$, (ii) there are large variations in the normalisation, but also 
in the slope of the $\beta$-$v_{\rm circ}$ relation 
with redshift, and (iii) starbursts and quiescent galaxies follow different relations. 

Regarding the shape of the $\beta$-$v_{\rm circ}$ relation, 
the top panel of Fig.\ref{fig:beta:vsys} shows that our dynamical calculations display
a trend of $\beta$ decreasing
with increasing $v_{\rm circ}$ for galaxies with $v_{\rm circ}\gtrsim 80\, \rm km\, s^{-1}$.  Below 
$v_{\rm circ}\approx 80\, \rm km\, s^{-1}$, the predicted mass loading shows a flattening or even a turnover 
followed by a positive $\beta$-$v_{\rm circ}$ relation.
The parametrisations used in the literature for the relation between $\beta$ and $v_{\rm circ}$, 
 are a poor description of the relation obtained from our physical model, which does not display a simple power-law behaviour
when plotted in this way.

\citet{Font12} discuss a phenomenological model with a saturation of the SNe feedback, which was invoked  
to reproduce the observed LF and metallicity of the Milky Way's satellites. 
Font et al. 
set a ceiling $\beta=620$ for $v_{\rm circ}<65\, \rm km\, s^{-1}$ to obtain a good match to the properties of the Milky Way's satellites. 
Our dynamical model of SNe feedback predicts a qualitatively similar behaviour to the saturated feedback scheme of Font et al.
The peak value of $\beta$ at $z=0$ is similar to the saturation value proposed by Font et al. However, we find that the peak value 
of the mass loading  
and the circular velocity at the peak occurs change with redshift. 
We also find that saturation velocity varies with the 
parameters adopted to describe the ISM and molecular clouds, spanning the range 
$v_{\rm circ,sat}\approx 70-100\, \rm km\,s^{-1}$. In our model the 
saturation velocity has no direct connection to the ratio between SNe energy and halo potential.

The redshift variation of the mass loading of the wind can be quantified  
 by fitting a
power law of the form $\beta=(v_{\rm circ}/V_{\rm hot})^{-\alpha_{\rm hot}}$ to quiescent galaxies at different redshifts
(top panel Fig.\ref{fig:beta:vsys}).
For circular velocities in the range
$v_{\rm circ}> 80 \rm km\, s^{-1}$, the dependence of $\alpha_{\rm hot}$ and $V_{\rm hot}$ on redshift 
is given by 

\begin{eqnarray}
\alpha_{\rm hot}&=& 2.7+2\, {\rm log}(1+z),              \label{Beta:Eq1}\\
V_{\rm hot}     &=& 425\, {\rm km\, s^{-1}}\,(1+z)^{-0.2}.\label{Beta:Eq2}
\end{eqnarray}

\noindent For galaxies with $v_{\rm circ}/\rm km\, s^{-1}< 80$ and for starbursts, the dependence of
$\alpha_{\rm hot}$ and $V_{\rm hot}$ on redshift is more complicated and cannot be described by simple power-law fits. 
This behaviour illustrates that the mass loading of the outflow
does not have a natural dependence on circular velocity. 

When focusing on starburst galaxies only, we find that the dependence of $\beta$ on $v_{\rm circ}$ changes dramatically 
 (see bottom panel of Fig.~\ref{fig:beta:vsys}). 
This is due to the very different conditions in the ISM in starbursts compared to
quiescent galaxies, with higher gas surface densities for a given $v_{\rm circ}$.
The turnover obtained for quiescent galaxies at $v_{\rm circ}\approx 80\, \rm km\,s^{-1}$ is also present
in starburst galaxies at $z<2$.
We find that the differences between
quiescent and starburst galaxies and the turnover at $v_{\rm circ}\approx 80\, \rm km\,s^{-1}$ can be 
explained in terms of the more fundamental relation
between $\beta$ and the gas scaleheight, 
$h_{\rm g}$.
 For the latter case, both quiescent and starburst galaxies follow nearly 
the same relation (see top panel of Fig.~\ref{fig:quiesc_burst}). This explains 
the nature of the $\beta$-$v_{\rm circ}$ relation: there 
is a correlation between $v_{\rm circ}$ and $h_{\rm g}$,  
for quiescent galaxies with $v_{\rm circ}>80\, \rm km\,s^{-1}$, but this is not present 
 at lower $v_{\rm circ}$ or in starburst galaxies. 

\subsection{A new parametrisation of the mass outflow rate}\label{SubSec:NewBetaModel}

\begin{figure*}
\begin{center}
\includegraphics[width=0.43\textwidth]{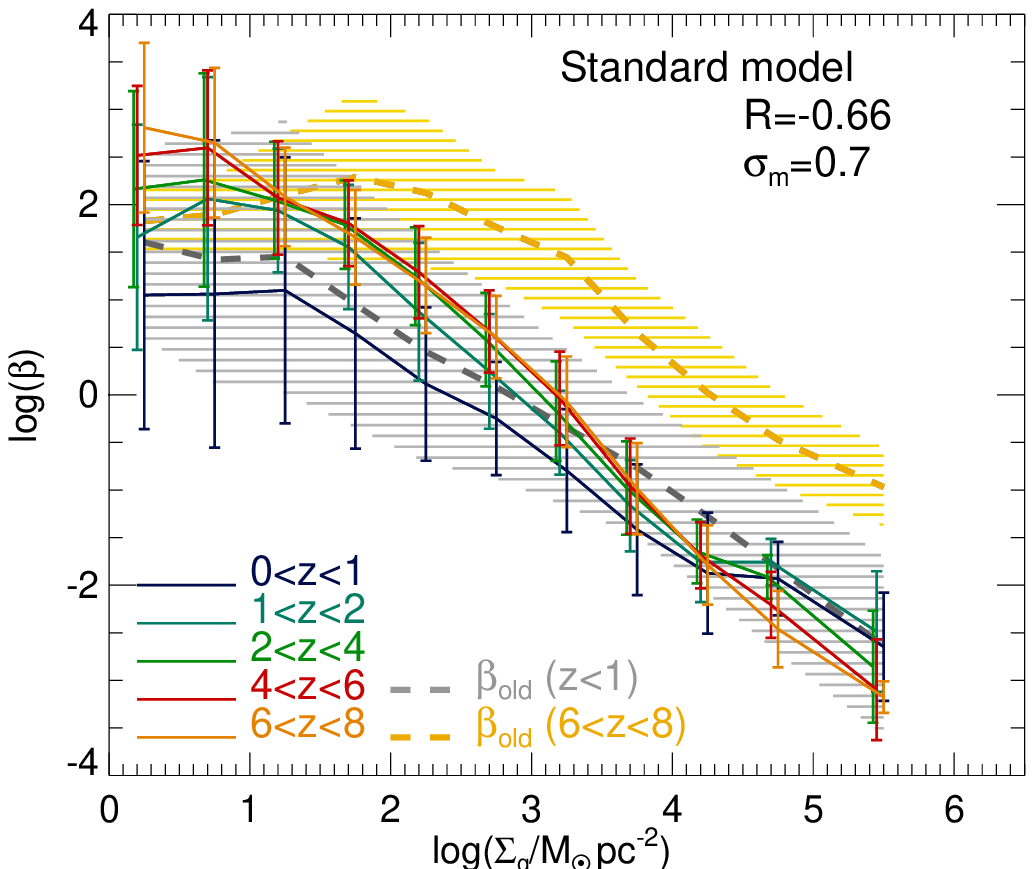}
\includegraphics[width=0.43\textwidth]{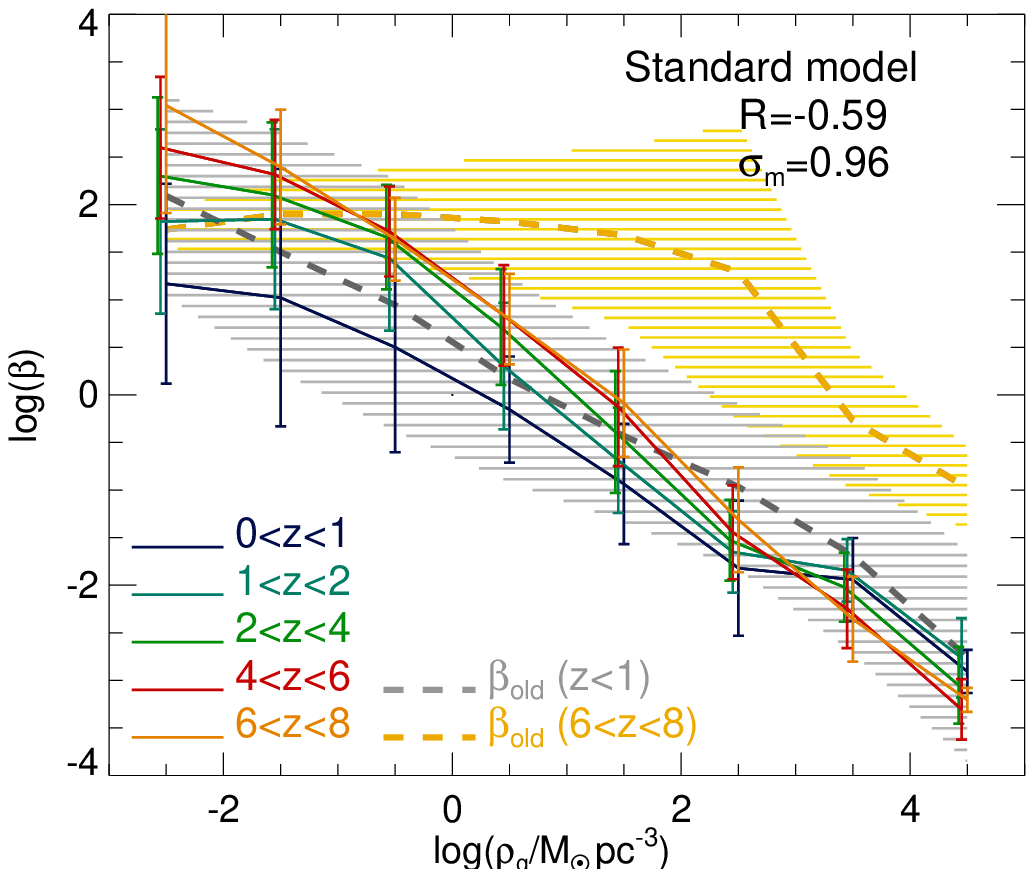}
\includegraphics[width=0.43\textwidth]{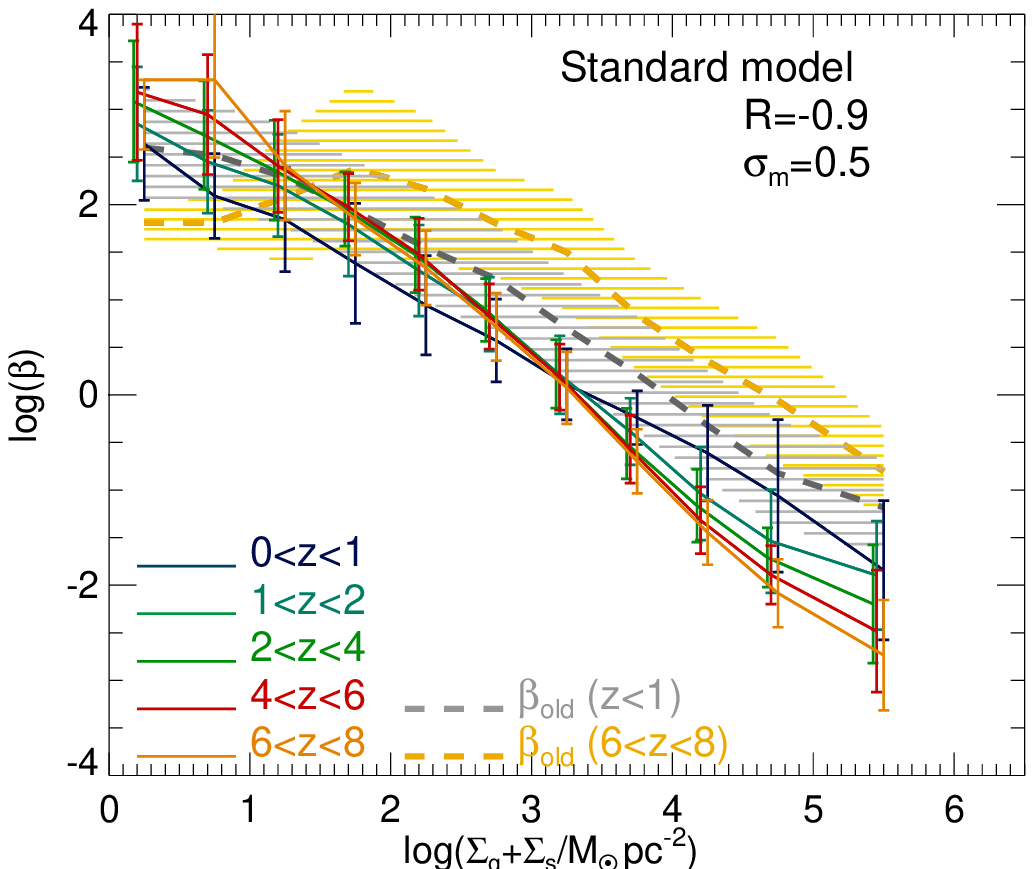}
\includegraphics[width=0.43\textwidth]{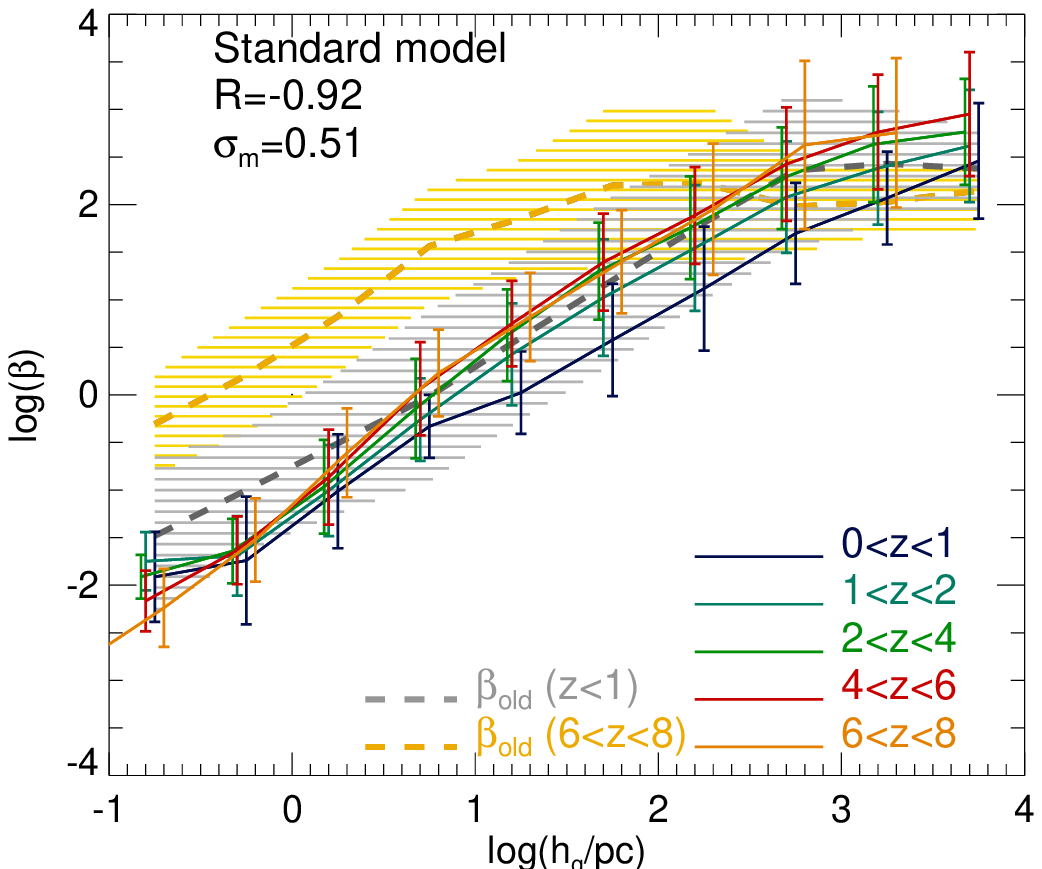}
\caption[The outflow rate to SFR ratio, as a function of 
various structural parameters of the gaseous disk.]{The 
global mass loading factor, $\beta=\dot{M}_{\rm eject}/\psi$, as a function of 
 gas surface density, $\Sigma_{\rm g}$ (top left-hand panel), 
 gas density, $\rho_{\rm g}$ (top right-hand panel),  
 gas plus stellar surface density, $\Sigma_{\rm g}+\Sigma_{\star}$ (bottom left-hand panel)
 and the gas scale height, $h_{\rm g}$ ( bottom right-hand panel), for 
galaxies with $M_{\star}>10^{8}\, h^{-1} M_{\odot}$. All quantities plotted on the 
$x$-axis are calculated at the half-mass radius of the disk in the case of quiescent SF, or the bulge 
in the case of starbursts. 
The relations are shown for different redshift ranges, as labelled, and correspond to the predictions of the 
model with the standard
choice of parameters (listed in Table~\ref{freepars}). 
 Solid lines and errorbars indicate the median and 10 and 90 percentile ranges of the relations. 
For reference, the values of the 
Pearson correlation coefficient, $R$, and the dispersion around the median, $\sigma_{\rm m}/{\rm dex}$, 
calculated for galaxies at $z<0.1$ in the new model are written on each panel.
We also show the results obtained when using the 
\citet{Bower06} choice for the outflow rate, $\beta_{\rm old}=(v_{\rm circ}/485\, {\rm km\, s^{-1}})^{-3.2}$, 
for galaxies at $z<1$ and $6<z<8$ (dashed lines) in each panel. The horizontal shading 
represents the 10 and 90 percentile ranges of the relations using the Bower et al. parametrisation.} 
\label{fig:beta1}
\end{center}
\end{figure*}

We analyse the dependence of $\beta$ on various properties of the disk in order 
to find the most natural combination of parameters to describe the mass loading. This new way of describing 
 $\beta$ can therefore be used in semi-analytic galaxy formation models and simulations. 

Fig.~\ref{fig:beta1} shows the mass loading factor, $\beta$, as a function of (i) $\Sigma_{\rm g}$, (ii) 
$\rho_{\rm g}$, (iii) $\Sigma_{\rm g}+\Sigma_{\star}$ and (iv) 
$h_{\rm g}$, 
for the standard set of parameters for GMCs and the diffuse medium (see Table~\ref{freepars}). 
Note that the third of these quantities can be written in terms of the surface density of gas and the gas 
fraction $\Sigma_{\rm g}+\Sigma_{\star}=\Sigma_{\rm g}/f_{\rm gas}$. 
All quantities above are evaluated at the half-mass radius of the disk or the bulge, $r_{\rm 50}$ (see Appendix~\ref{Profiles} 
for the definition of the profiles), and the 
predictions are shown for all galaxies, quiescent and SB, in different redshift ranges. 
We decide to study the relation between $\beta$ and these quantities due to the correlation we find between the 
mass of a single bubble at the point of break-out from the disk and the local properties $\rho_{\rm g}$, $\Sigma_{\rm g}$, 
$\Sigma_{\rm g}+\Sigma_{\star}$ and $h_{\rm g}$ (see Fig.~\ref{singlebs}).
We also show the resulting relation between $\beta$ and the quantity plotted on the $x$-axis 
if we use the old mass loading 
parametrisation (see point (iii) in list of $\S$~\ref{Sec:vsysSNe}).

We find that our results can be approximately described by the following fits  
\begin{eqnarray}
% Second equation
\beta    & =& \left[\frac{\Sigma_{\rm g}(r_{\rm 50})}{1.6\times 10^3 \, M_{\odot}\, {\rm pc}^{-2}}\right]^{-0.6}\label{beta_new_pars0}\\
\beta    & =& \left[\frac{\rho_{\rm g}(r_{\rm 50})}{14\, M_{\odot}\, {\rm pc}^{-3}}\right]^{-0.5}\\
\beta    & = & \left[\frac{\Sigma_{\rm g}(r_{\rm 50})+\Sigma_{\star}(r_{\rm 50})}{2.6\times 10^3\, M_{\odot}\, {\rm pc}^{-2}}\right]^{-1}\\
\beta    & =& \left[\frac{h_{\rm g}(r_{\rm 50})}{8\, {\rm pc}}\right]^{1.1}.
\label{beta_new_pars}
\end{eqnarray}

We quantify how good the correlation is by using two statistics, the Pearson correlation coefficient, $R$, and an 
estimate of the dispersion around the median, $\sigma_{\rm m}$. For each $x$-axis bin we calculate 
a dispersion, $\sigma_{\rm x}$, corresponding 
to the ratio between the 
sum of the square of the deviations around the median in the $y$-axis and the number of objects in the bin. 
We then calculate $\sigma_{\rm m}$, which corresponds to the square root 
of the median value of the distribution of $\sigma_{\rm x}$.
 We calculate
$\sigma_{\rm m}$ in the log-log plane, in units of dex. Note that $R$ and $\sigma_{\rm m}$ are independent statistics which can 
be used to assess how good the correlation is between two quantities. The values for both quantities for galaxies
at $z<0.1$ are written in each panel of Fig.~\ref{fig:beta1}.

\begin{figure}
\begin{center}
\includegraphics[width=0.45\textwidth]{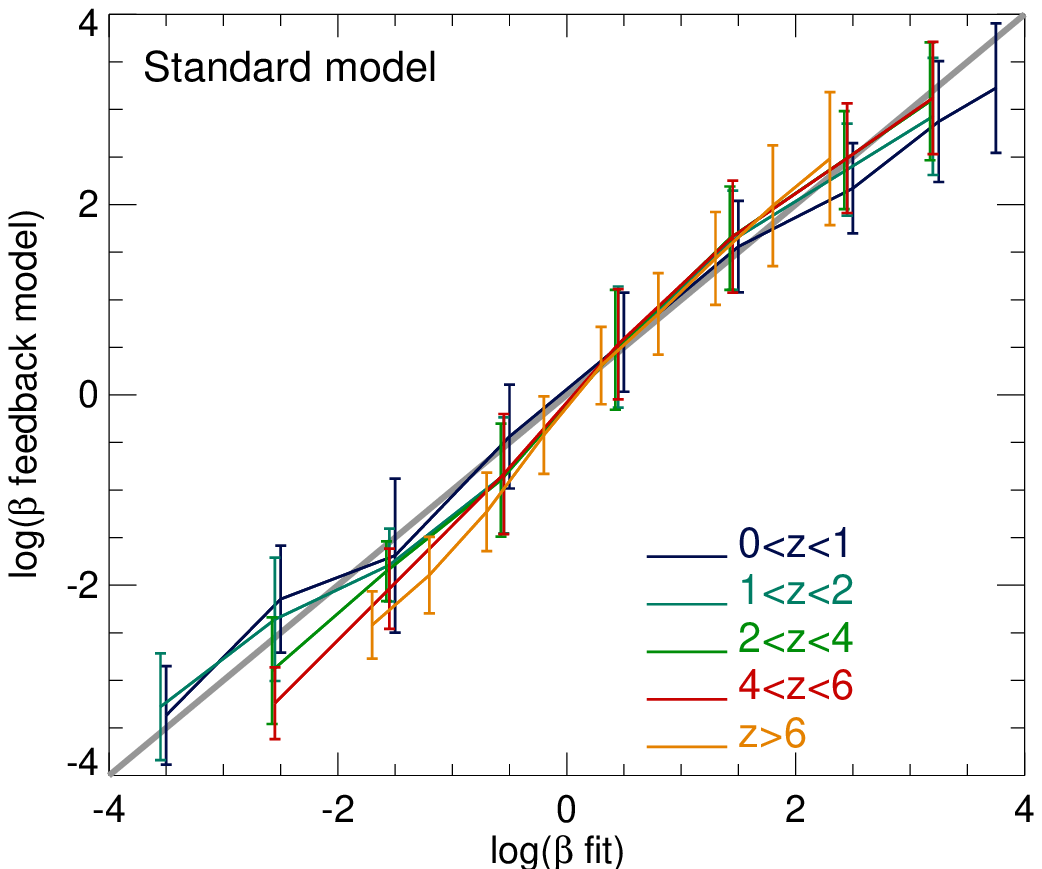}
\caption[The outflow rate to SFR ratio, as a function of 
various structural parameters of the gaseous disk.]{ 
The predicted mass loading from the full model ($y$-axis) plotted against the fit given by 
Eq.~\ref{beta_new_pars_2}, expressed in terms of the gas scaleheight and gas fraction,  
for galaxies at different redshifts, as labelled.}
\label{fig:betacomp}
\end{center}
\end{figure}

In terms of the Pearson correlation factor, $R$, and the dispersion, $\sigma_{\rm m}$ (shown in Fig.~\ref{fig:beta1}), 
the properties that best describe $\beta$ are 
$\Sigma_{\rm g}+\Sigma_{\star}$ and $h_{\rm g}$. 
Fig.~\ref{fig:beta1} shows that the normalisation and power-law index of the above relations vary with redshift, with 
high-redshift galaxies following a steeper relation than low-redshift galaxies. This trend can be understood as being due to high-redshift 
galaxies having larger gas fractions compared to lower-redshift 
galaxies. Galaxies with a high gas fraction typically have a 
molecule-dominated ISM, and these are predicted to 
follow a steeper relation between $\beta$ and $h_{\rm g}$ than those with an atomic-dominated ISM, which are typically gas poor 
(see $\S$~\ref{Sub:Analytic} for an analytic derivation of such a trend).
We find that the redshift trend can be removed by adding an extra dependence on the gas fraction to the expressions for $\beta$, 

\begin{eqnarray}
\beta   & =& \left[\frac{\Sigma_{\rm g}(r_{\rm 50})}{1600\,M_{\odot}\, {\rm pc}^{-2}}\right]^{-0.6}\left[\frac{f_{\rm gas}}{0.12}\right]^{0.8}\label{beta_new_pars_2.0}\\
\beta   & =& \left[\frac{h_{\rm g}(r_{\rm 50})}{15\, {\rm pc}}\right]^{1.1}\left[\frac{f_{\rm gas}}{0.02}\right]^{0.4}, 
\label{beta_new_pars_2}
\end{eqnarray}

\noindent which both have Pearson correlation factor of $R\approx 0.97$ 
and a dispersion $\sigma_{\rm m}\approx 0.3$~dex for galaxies at $z<0.1$. This is shown in Fig.~\ref{fig:betacomp}, 
where the fit of Eq.~\ref{beta_new_pars_2} is compared with the directly calculated 
$\beta$. Most of the redshift evolution 
seen in Fig.~\ref{fig:beta1} is removed.

\begin{figure}
\begin{center}
\includegraphics[width=0.45\textwidth]{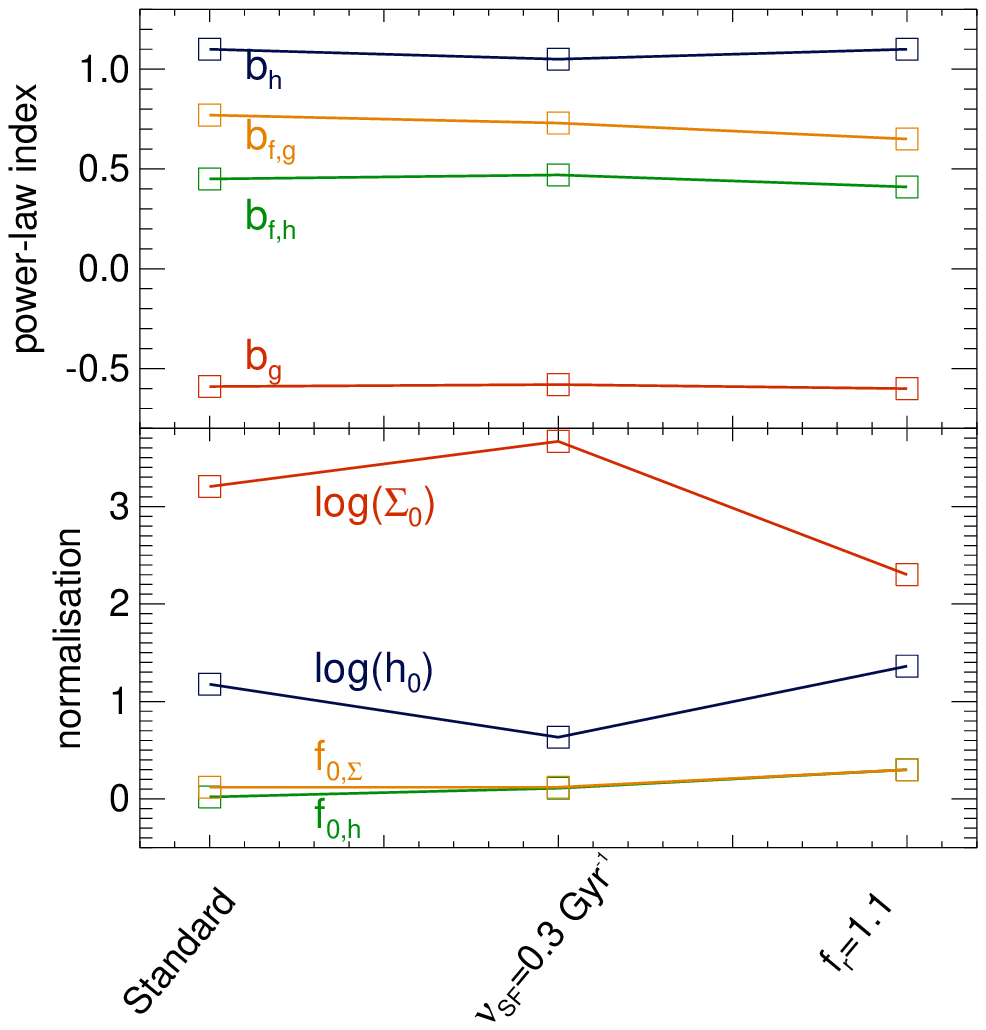}
\caption[The outflow rate to SFR ratio, as a function of 
various structural parameters of the gaseous disk.]{
{\it Top panel:} Power-law slope of the relations between $\beta$, $h_{\rm g}$, $\Sigma_{\rm g}$ and $f_{\rm gas}$, 
quantified as 
$\beta=(h_{\rm g}/h_{0})^{b_{\rm h}}\, (f_{\rm gas}/f_{\rm g,h})^{b_{\rm f,h}}$ and 
$\beta=(\Sigma_{\rm g}/\Sigma_{0})^{b_{\rm g}}\, (f_{\rm gas}/f_{\rm g,\Sigma})^{b_{\rm f,g}}$. 
Lines are as labelled in the panel.
The results for the model with the standard set of parameters and 
those predicting the highest and lowest $\beta$ are shown as symbols.
The parameters of the fit correspond to fitting the relations above in subsamples 
of galaxies at $z<8$ with different gas fractions. 
{\it Bottom panel:} Normalizations of the relations above for the three models of the top panel.
Lines are as labelled in the panel. The plot shows that the power-law slopes are not affected by 
changes in the parameters describing the ISM and SF but that only the normalisations of the relations 
change.} 
\label{fig:beta1x}
\end{center}
\end{figure}

Eqs.~\ref{beta_new_pars_2.0} and \ref{beta_new_pars_2} are also useful to characterise the mass loading $\beta$ obtained 
in the model when varying the parameters used in the ISM modelling (Table~\ref{freepars}).
This is shown in Fig.~\ref{fig:beta1x}, in which the power-law indices and normalisations for the 
relations, defined as $\beta=(\Sigma_{\rm g}/\Sigma_0)^{b_{\rm g}} (f_{\rm gas}/f_{0,\Sigma})^{b_{\rm f,g}}$  and 
$\beta=(h_{\rm g}/h_0)^{b_{\rm h}} (f_{\rm gas}/f_{\rm 0,h})^{b_{\rm f,h}}$, are shown 
for $3$ different choices of ISM model parameters.
The model using  $f_{\rm r}=1.1$ corresponds to the weakest feedback model and that with  
$\nu_{\rm SF}=0.3\, {\rm Gyr}^{-1}$ to the strongest feedback model.
The three choices of model parameters produce very little variation in the power-law indices 
of the above relations (top panel of Fig.~\ref{fig:beta1x}). 
Variations are observed in the normalisations of the relations and 
represent different feedback strengths (bottom panel of Fig.~\ref{fig:beta1x}).
This means that if we were to include the 
parametric form given by Eqs.~\ref{beta_new_pars_2.0} and \ref{beta_new_pars_2} 
in the semi-analytic model, we would need to vary the zero-point of these relations 
to reproduce the results for different parameters for the diffuse ISM and GMCs. 
Eqs.~\ref{beta_new_pars_2.0} and \ref{beta_new_pars_2} describe our results for 
the mass loading $\beta$ in galaxies at any redshift, within the range tested (i.e 
$z<10$ and $M_{\star}+M_{\rm gas,ISM}>10^8\, h^{-1}\,M_{\odot}$) with very little dependence on 
redshift or stellar mass. 

The old parametrisation (shown by the dashed lines in Fig.~\ref{fig:beta1}) results in 
a trend of $\beta$ decreasing with the properties plotted on the $x$-axis, 
given the correlation already discussed between $v_{\rm circ}$ and these variables.
However, $\beta_{\rm old}$ differs from the mass loading $\beta$ for galaxies with low surface densities of gas 
by up to a factor of $\approx 5$ in either direction, and overestimates $\beta$ at the high surface density regime 
by up to a factor of $\approx 100$, depending on the redshift. 
In Fig.~\ref{fig:beta1} $\beta_{\rm old}$ varies with redshift much more strongly than the new parametrisations, 
and therefore overestimates the SNe feedback in high-redshift galaxies. 
This reflects the importance of the analysis performed in this paper and
 the need for a revision of such parametrisations. The largest differences between the predicted $\beta$ and 
$\beta_{\rm old}$ are obtained at high-redshifts.

The difference between SBs and quiescent galaxies apparent in the
$\beta-v_{\rm circ}$ plane in Fig.~\ref{fig:beta:vsys} is greatly reduced
 in the $\beta-h_{\rm g}$ 
plane (see the top panel of Fig.~\ref{fig:quiesc_burst}). This is because SB galaxies of a given
 $v_{\rm circ}$ have much higher densities in stars and gas than their quiescent counterparts.
Although the relation is noisier due to the lower numbers of SBs in the model output 
compared to quiescent galaxies, the $\beta-h_{\rm g}$ relation
is very similar in slope and normalisation to that for quiescent galaxies. 
This suggests that the dependence of mass loading is fundamental and captures the relevant
physics determining $\beta$.

\section{The impact of the new outflow mass loading on galaxy formation}\label{Sec:Galaxies}

In this section we consider 
the impact of our dynamical model 
of SNe feedback on galaxy properties and compare with the predictions of the model which 
uses the old parametrisation. We first estimate the error associated with 
using the parametric form defined in Eq.~\ref{beta_new_pars_2} instead  
of performing  the full calculation carried out in this paper. 
Second, we analyse the net effect of our dynamical modelling on galaxy properties by focusing on two statistical properties 
of galaxies: (i) the evolution of the LF in the $K$- and $V$-bands,
and (ii) the evolution of the global SFR density. An analysis of a complete set of galaxy properties 
will be presented in a future paper (Lagos, Lacey \& Baugh, in prep.). { Note that the experiment carried out in this section attempts
to identify general trends in the LF and SFR density due to the new SNe feedback model rather than predicting 
exact normalisations of both quantities. The reasons for this are firstly, that this model does not include a self-consistent 
treatment of the re-incorporation of the gas that has escaped the galaxy, but instead uses the parametrisation described in 
$\S$~\ref{Sec:SFequations}, and secondly, the parameters associated with the AGN feedback treatment have not 
been modified to recover the agreement with the observations at the bright-end of the LF.}

\begin{figure}
\begin{center}
\includegraphics[trim = 1mm 4mm 2mm 1.5mm,clip,width=0.435\textwidth]{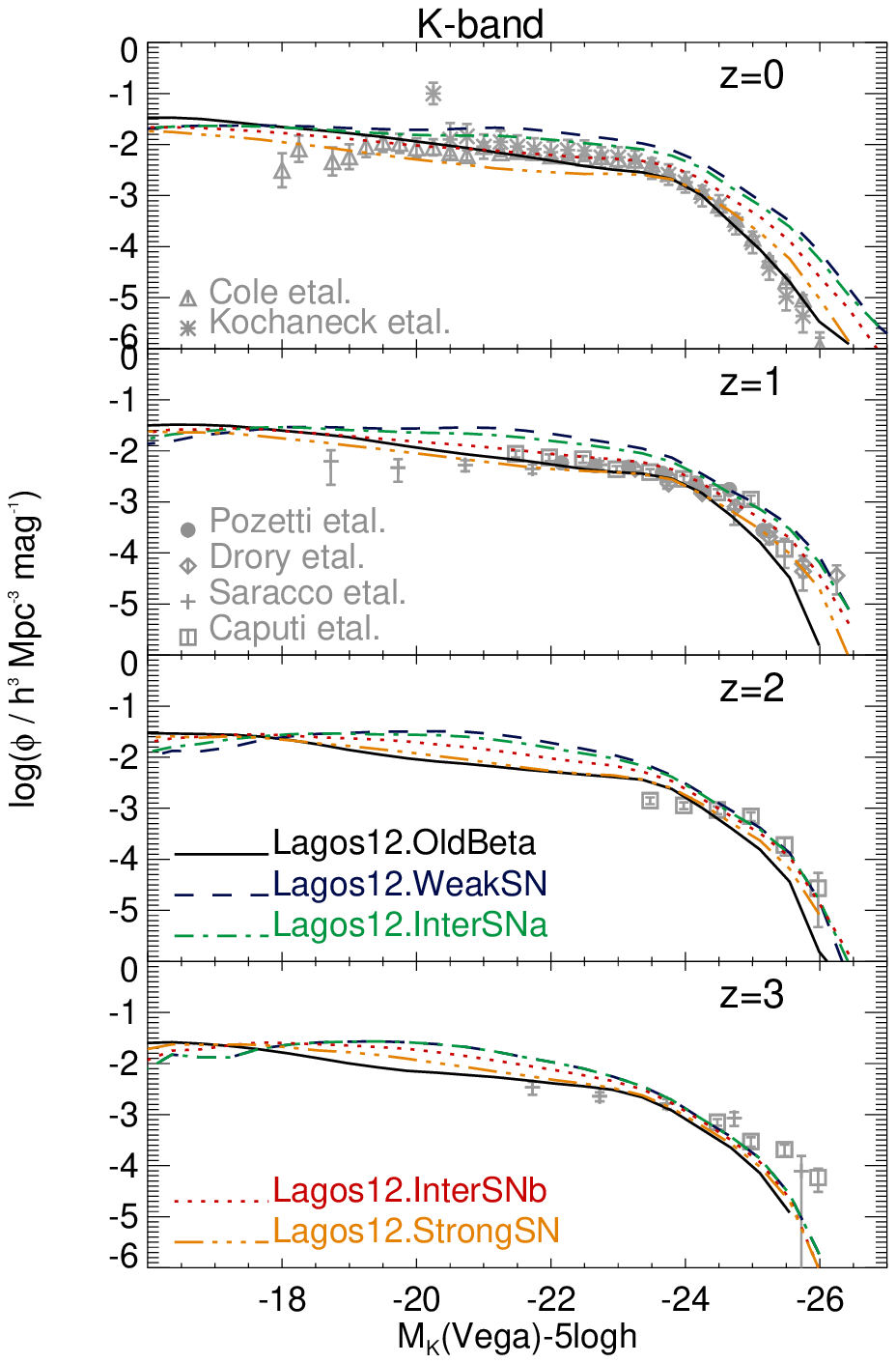}
\caption[Rest-frame $K$-band galaxy luminosity function for the Lagos12 model 
with the old and new SNe prescription.] {Rest-frame $K$-band galaxy luminosity function 
  for the Lagos12 model 
with the old and the new SNe prescriptions (see Table~\ref{Models}), at various redshifts, as labelled. 
Observational results from \citet{Pozzetti03}, \citet{Drory03}, \citet{Saracco06}
and \citet{Caputi06} are shown as grey symbols, identified by the key
in the two top panels. Note that the models have not been retuned to fit the observed LF.}
\label{Kbandboth}
\end{center}
\end{figure}

\begin{figure}
\begin{center}
\includegraphics[trim = 1mm 4mm 2mm 1.5mm,clip,width=0.435\textwidth]{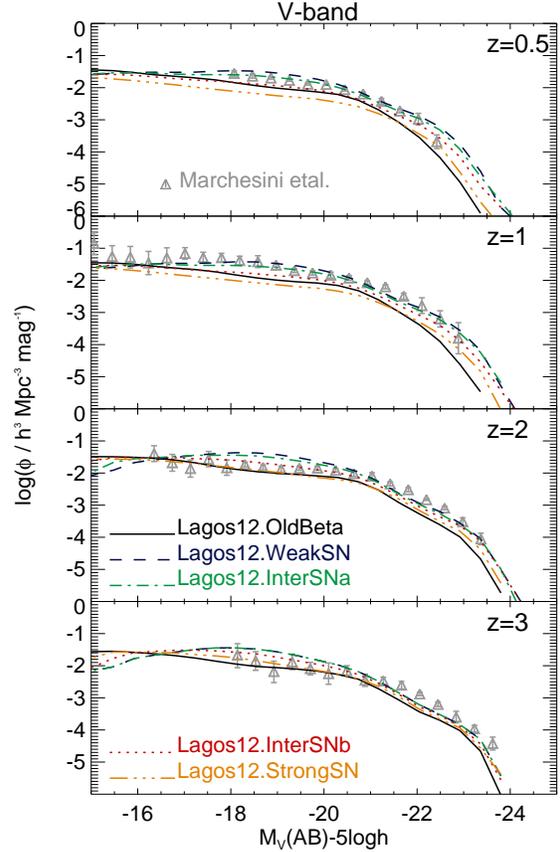}
\caption[Rest-frame $V$-band galaxy luminosity function for the Lagos12 model 
with the old and new SNe prescription.] {Rest-frame $V$-band galaxy luminosity function 
  for the Lagos12 model with the old and the new SNe prescriptions (see Table~\ref{Models}), 
at various redshifts, as labelled. 
Observational results from \citet{Marchesini12} are shown as grey symbols. 
Note that the models have not been retuned to fit the observed LF.}
\label{Vbandboth}
\end{center}
\end{figure}

We ran the full dynamical model in which $\beta$ is calculated self-consistently,  
and compare with the model using the prescription from Eq.~\ref{beta_new_pars_2} to calculate 
$\beta$, under the simplifying assumption of $\beta^Z=\beta$.
We compared the luminosity functions predicted by both procedures in the bands $900-1200$\AA, 
$b_J$, $V$, $K$ and $8\mu$m. At $z=0$, the largest differences are obtained in the far-UV band, but are at most $\approx 25$\%. 
The other bands show differences in the range $5-20$\%. 
However, at $z=6$ these differences can be as large as $80$\%. 
The reason for the larger differences at high redshifts is that we currently do not allow for variations in the parametrisation of 
 $\beta^Z$ with respect to $\beta$, like those shown in Fig.~\ref{MetalsOut}. Such variations have
only a minor effect at $z=0$, but they have an affect in $z\gtrsim 4$ galaxies, where larger differences between 
$\beta$ and $\beta^Z$ are predicted by the dynamical model. 
The main drivers of the differences seen in the luminosity functions are 
differences in the cold gas mass and mass in metals in the ISM. The stellar mass and hot gas mass functions are 
 similar to within $<40$\% at redshifts $z=0-6$. In the redshift range shown in Figs.~\ref{Kbandboth} and \ref{Vbandboth}, variations 
between the self-consistent calculation and the calculation using the $\beta$ parametrisation are not significant. 
We calculate the best parametrisations using the form of Eq.~\ref{beta_new_pars_2} for the different ISM parameter choices and 
present in Table~\ref{Models} the results for four choices of parameters spanning the full range of feedback strength\footnote{ Note that 
the weak SNe feedback model of Table~\ref{Models} gives mass loading factors that are about $3$ times lower than the standard choice of 
parameters, which is also representative of the predicted $\beta$ in the case of the extreme ISM conditions 
analysed in $\S$~\ref{Sec:xtmISM}.}. 
We find that using the prescription for $\beta$ given in  
Eq.~\ref{beta_new_pars_2} gives reliable results that closely follow the behaviour of the 
full dynamical model at $z<4$, but significantly speeds up the calculation. 

In order to analyse the effect of the new dynamical model of SNe feedback on galaxy properties, 
we focus on the \citet{Lagos12} model and vary the SNe feedback prescription. 
We compare the four alternative models listed in Table~\ref{Models}.

\begin{table}
\begin{center}
\caption{Models shown in Figs.~\ref{Kbandboth}, \ref{Vbandboth} and \ref{SFRtot}. 
The first row gives the old parametrisation used to describe the outflow. The next four rows 
 show alternative models using the new $\beta$ parametrisation of Eq.~\ref{beta_new_pars_2}. Each parametrisation represents different parameter 
choices for the full SNe feedback dynamical model, 
which is indicated in the parenthesis. The parametrisation used for each model is shown in the second column.}\label{Models}
\begin{tabular}{l l}
\\[3pt]
\hline
\small{Model} & $\beta$ parametrisation \\
\hline
Lagos12.OldBeta  & $\left(\frac{V_{\rm circ}}{485\, {\rm km\, s^{-1}}} \right)^{-3.2}$ \\
Lagos12.WeakSN ($f_{\rm r}=1.1$)& $\left(\frac{h_{\rm g}}{23\, {\rm pc}}\right)^{1.1}\left(\frac{f_{\rm gas}}{0.3}\right)^{0.4}$\\
Lagos12.InterSNa ($\tau_{\rm life,GMC}=0.03\, {\rm Gyr}$) &  $\left(\frac{h_{\rm g}}{17\, {\rm pc}}\right)^{1.1}\left(\frac{f_{\rm gas}}{0.1}\right)^{0.4}$\\
Lagos12.InterSNb (Std.) &  $\left(\frac{h_{\rm g}}{15\, {\rm pc}}\right)^{1.1}\left(\frac{f_{\rm gas}}{0.02}\right)^{0.4}$\\
Lagos12.StrongSN ($\nu_{\rm SF}=0.3\, {\rm Gyr}^{-1}$)&  $\left(\frac{h_{\rm g}}{4\, {\rm pc}}\right)^{1.1}\left(\frac{f_{\rm gas}}{0.3}\right)^{0.4}$\\
\hline
\end{tabular}
\end{center}
\end{table}

\begin{figure}
\begin{center}
\includegraphics[width=0.45\textwidth]{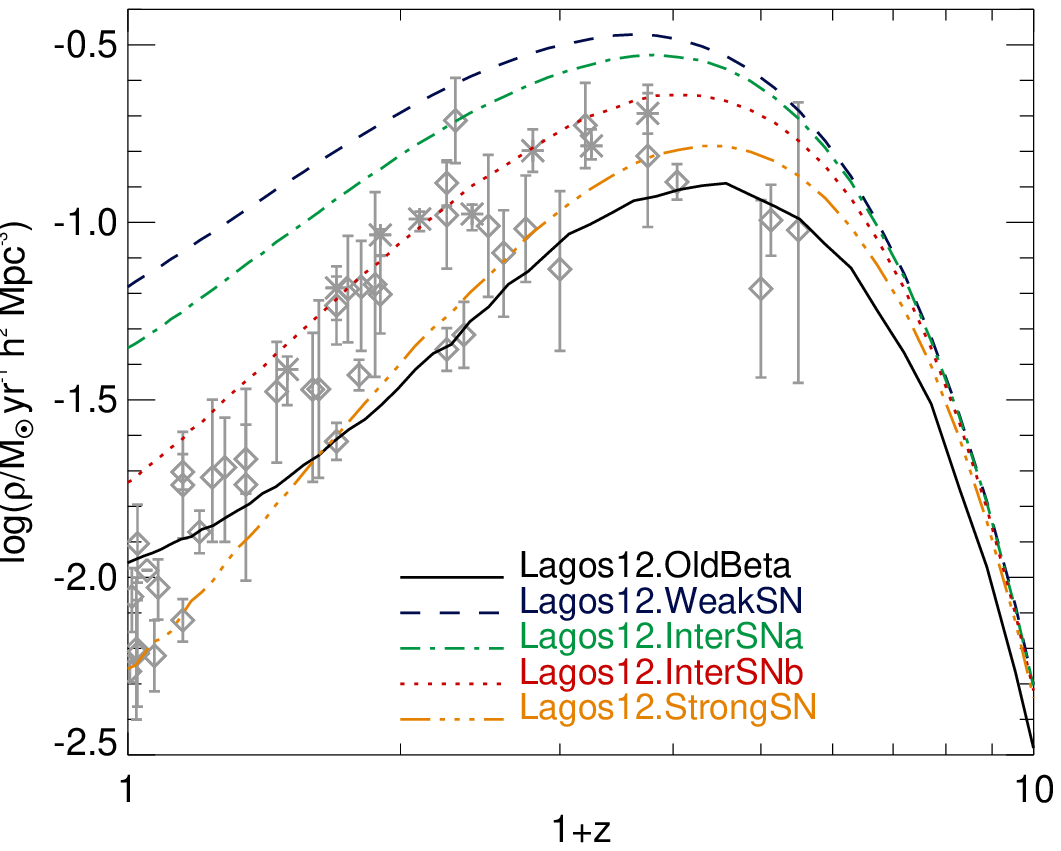}
\caption[The evolution of the cosmic star formation
rate per unit volume for the Lagos12 model 
with the old and new SNe prescription.] {The evolution of the cosmic star formation
rate per unit volume for the Lagos12 model 
with the old and the new SNe prescriptions which give rise to different strengths 
of SN feedback (see Table~\ref{Models}), as labelled. 
The observational estimates of \nocite{Karim11} Karim et al. (2011; asterisks) and the data compilation of \nocite{Hopkins04} 
Hopkins et al. (2004; diamonds) are also shown. 
\citet{Hopkins04} assumes a Salpeter IMF and \citet{Karim11} a Chabrier IMF. Therefore, SFRs have been scaled 
to a Kennicutt IMF (scaled down by a factor of $2$ in the Salpeter case and down by a factor $1.12$ in the Chabrier case).}
\label{SFRtot}
\end{center}
\end{figure}

Fig.~\ref{Kbandboth} shows the $K$-band LF at various redshifts for the $5$ 
models listed in Table~\ref{Models}. 
 We remind the reader we are not trying to fit observations here, but rather we are 
trying to see the effect the modelling of feedback has on galaxy properties 
starting from a model which uses a completely different way of calculating $\beta$. 
The most interesting feature in Fig.~\ref{Kbandboth} is that all the models that use the new 
feedback model developed in this paper give a shallower faint-end slope at $z<2.5$, regardless of the ISM model 
parameters, but produce a higher overall normalisation 
for the LF. 
The model with the strongest feedback (Lagos12.StrongSN) shows a faint end that is similar to the original model. 
There is a trend of a shallower faint end 
with weaker SN feedback models, although this trend changes with band and redshift. 
It is also clear that the models predict very weak evolution of the slope of the faint end. 
The shallower faint end slope predicted by our new feedback scheme suggests that the 
problem of the predicted steep faint end of the LF and low-mass end of the stellar mass function  
could be largely overcome by using the new parametrisation of the mass loading 
(Eq.~\ref{beta_new_pars_2}).
The physical reason behind the shallower faint end slopes obtained by using the new $\beta$ parametrisation is that 
faint galaxies typically have large $h_{\rm g}$ and therefore can reach very large values of $\beta$. 
These faint galaxies do not necessarily correspond to those with the smallest $v_{\rm circ}$, and therefore 
in these galaxies, the new parametrisation drives larger $\beta$ than that obtained with the $v_{\rm circ}$ parametrisation.  

The bright-end of the $K$-band LF predicted by the models using the new feedback prescription is higher in all the cases 
compared to the original model. This is due to the lower $\beta$ predicted by the dynamical SN feedback model compared to the parametrisation 
adopted in the Lagos12.OldBeta model. This, in addition to the unchanged gas reincorporation timescale, leads to 
more bright galaxies. 
In paper II we will model the expansion of bubbles in the halo to remove this process as a free parameter.  
We will analyse in more detail the 
effect of SN feedback on the bright end of the LF. 

Fig.~\ref{Vbandboth} is equivalent to Fig.~\ref{Kbandboth} but 
shows the $V$-band LF for $z>0.5$. The behaviour of the models in this band is 
broadly the same as in the near-IR: the new feedback scheme, regardless of the strength of the SN feedback, predicts a shallower 
faint end of the LF up to $z\approx 1.5$. However, above that redshift, the strength of the SN feedback plays an important role in determining 
whether  
the faint end is shallower or steeper than predicted by the original model. 
The slope of the faint end in the $V$-band LF varies more strongly with redshift and in a complex way 
compared to the variations seen in the $K$-band LF.

Interestingly, the different SNe feedback models of Table~\ref{Models} converge to  
similar LFs in both the $K$ and $V$ bands at $z\gtrsim 3$ but evolve differently towards $z=0$. This 
is because these models predict galaxies with different star formation histories. 
Fig.~\ref{SFRtot} shows the global SFR density evolution predicted by each of the models of Table~\ref{Models}. 
 The models using the new SN feedback scheme predict that the global SFR peaks at slightly lower 
redshifts compared to the original model, with weaker SN feedback producing a lower redshift for the peak. { 
Note that even the model with the strongest SNe feedback produces 
larger SFR densities at $z\approx 2-4$ compared to the model using the old $\beta$ parametrisation.} 
Compared to observations,   
the model with the strongest SN feedback predicts SFR densities that are too low, while the weakest SN feedback give 
SFR densities that are too high.
It is interesting to note that the model with the strongest SN feedback results in the largest decline 
in the global SFR per unit volume, dropping by a factor of $\approx 30$ from the peak to the present 
day. 
A key physical process to analyse before ruling out any of these models 
is the reincorporation timescale of the gas after outflowing from the ISM into the hot gas reservoir of the halo.
Also, other galaxy formation parameters may have to be reset, since these were based on the old outflow model. 
In any case, the fact that the use of the new $\beta$ parametrisation predicts a shallower LF of galaxies 
points to the need to revise the physics included in galaxy formation models and simulations.

\section{Discussion and Conclusions}\label{Sec:Conclus_SNepaper}

We have presented a dynamical model of SNe feedback which tracks the evolution
of bubbles inflated by SNe into the ISM of galaxies.
Our model includes a range of processes which can affect 
the expansion of bubbles: gravity, radiative energy losses,
external pressure from the diffuse medium and temporal changes in the ambient gas.
Bubbles inflated by SNe are evolved from the adiabatic to the radiative phases
until the point of break-out from the galaxy disk or bulge, or confinement in a multi-phase ISM. 
The multi-phase model of the ISM includes a diffuse, atomic phase, a dense, molecular phase and a hot, low density phase. The latter 
corresponds to the interior of bubbles. The metal enrichment of the ISM and halo due to SNe takes place through bubbles. 
The location of star-forming regions, or GMCs, which give rise to bubbles
is connected to the radial distribution of molecular gas, which allows us to study both the global outflow rate and
the radial profile of galactic outflows. The aims of this work are
(i) to test the importance of each of the physical processes included in
the expansion of bubbles and to explore the parameter space of the modelling of 
GMCs and the ISM, (ii)
 to determine which combinations of galaxy properties the
outflow rate best correlates with and (iii) to improve upon widely used parametric forms for the outflow rate used in the
literature.

To help us assess these points, we embed our calculations in the {\tt GALFORM} semi-analytic 
model, which follows the formation and evolution
of galaxies in the framework of hierarchical structure formation. 
We take advantage of the two-phase medium description introduced into 
{\tt GALFORM} by \citet{Lagos10} and \citet{Lagos11}, to
trace star formation and star forming regions using the cold molecular component of the ISM, while
allowing bubbles to sweep up gas only from the diffuse neutral atomic component.
In the Lagos et al. model, the molecular-to-atomic mass ratio is calculated from the radial profile
of the hydrostatic pressure, and the SFR is calculated from the molecular gas radial profile
(e.g. \citealt{Blitz06}; \citealt{Leroy08}). 
The semi-analytic model provides the initial conditions needed by the dynamical model of SN feedback: the stellar and
dark matter contents, the surface density of atomic and molecular gas, the 
gas metallicity and the scalelength of each mass component.
 This modelling allows us to study the relation between
the rate at which mass escapes from the galaxy disk or bulge (outflow rate) and the properties of the disk, bulge and halo, over a 
wide dynamic range.
Previous work has
focused on hydrodynamical simulations covering a narrow dynamic range, which has been chosen 
somewhat arbitrarily (\citealt{Hopkins12}; \citealt{Creasey12}), or
which have adopted Sedov analytic solutions for the evolution of bubbles (e.g. \citealt{Efstathiou00}; \citealt{Monaco04}).
One of our goals is to complement and extend this work by using 
a more general SNe feedback model and the  
galaxy population and star formation histories produced by the semi-analytic model.

We summarise our main conclusions below:

(i) We find that the mass loading of the outflow, $\beta$, 
decreases with increasing gas surface density and increases 
with increasing gas scaleheight. On the other hand, the outflow velocity increases with increasing 
gas surface density and decreases with increasing gas scaleheight. 
 These trends are seen in both
 the global and local mass loading and velocity of the wind. 

(ii) We find that the multi-phase ISM treatment included in our model is essential for reproducing
the observed outflow rates of galaxies.
When fixing the diffuse-to-cloud mass ratio instead of calculating it from the
 hydrostatic pressure, we find variations in the predicted mass loading $\beta$ of
 up to $2$ orders of magnitude in the highest gas density regimes. This emphasizes the importance of
the multi-phase ISM included in our modelling.
By adopting different, but still plausible parameters in the modelling of GMCs and the diffuse medium, we find variations  
in $\beta$ of a factor up to $\approx 3$ and in $v_{\rm outflow}$ of a factor up to $\approx 1.7$ in either direction.
We also find that by the time bubbles escape from the ISM, they are radiative in the majority of the cases.

(iii) When comparing our predicted outflow rates and velocities with those inferred from 
observations (e.g. \citealt{Martin99}; \citealt{Bouche12}), we find good agreement.
We also find that our predictions are similar to those from the non-cosmological hydrodynamical simulations
of \citet{Hopkins12} and \citet{Creasey12}, in the regimes they were able to probe. Our work therefore
confirms the finding that the surface density of gas is an important quantity in 
determining the mass loading of the outflow.

(iv) The widely used parametric forms describing SNe feedback and
relating the mass loading $\beta$ to only the circular velocity
of the galaxy do not capture the physics setting the outflow rates from galaxies. 
For instance, we find that the trend of $\beta$ decreasing with $v_{\rm circ}$ is only valid 
for galaxies with  $v_{\rm circ}\gtrsim 80\, \rm km\, s^{-1}$. Below this threshold, $\beta$ 
flattens or decreases with decreasing $v_{\rm circ}$. 
We also find that the relation between $\beta$ and $v_{\rm circ}$ changes substantially with redshift.  
We find that tighter relations are those between
$\beta$ and the gas scaleheight and gas fraction,  $\beta\propto [h_{\rm g}(r_{50})]^{1.1} [f_{\rm gas}]^{0.4}$, and between 
$\beta$ and the surface density of gas and the gas fraction, $\beta\propto[\Sigma_{\rm g}(r_{\rm 50})]^{-0.6}[f_{\rm gas}]^{0.8}$.
Changing the
parameters in the model of GMCs and the diffuse medium can change the normalisation of these relations,  
but does not alter the power-law index.  
We find that starburst and quiescent galaxies follow similar relations, with starbursts being slightly offset to lower $\beta$ 
compared to quiescent galaxies. The outflow velocities can also vary between starbursts and quiescent galaxies depending on the adopted
 star formation law. A more rapid conversion from gas to stars drives larger velocities due to the higher energy and momentum 
injection rate from SNe. 

(v) We study the effect of the dynamical model of SN feedback developed here on galaxy properties and 
test the inclusion of the new parametrisation of $\beta$ (see (iv) above). We find that 
the faint end of the near-infrared LF becomes shallower in the model using the new feedback scheme compared to the old model. 
We find that this shallowing of the faint end takes place regardless of the parameters assumed 
to describe the diffuse ISM and GMCs, with a trend of weaker SN feedback predicting a shallower faint end of the LF.

Our model is subject to simplifications required to model the evolution of bubbles
in the ISM of galaxies. A critical simplification we make is to 
 fix the GMC mass.
A more sophisticated approach would be to include a
distribution of GMC masses and their
spatial distribution following a theoretical estimate of the
spatial clustering of GMCs of different masses
(\citealt{Hopkins11}). However, such a description also requires
more detailed information about the ISM. 
Instead, we test our predictions by varying the
adopted GMC mass in the range allowed by observations (see Table~\ref{freepars}), and find
variations in the normalisation of the mass loading described in (iv), 
but with little impact on the power-law indices. 

The agreement we find between our model and detailed hydrodynamical
simulations (\citealt{Hopkins12}; \citealt{Creasey12}) suggests that we 
capture the relevant physics determining the rate at which mass escapes from the ISM
of galaxies, despite the simplifications made in our modelling.
The advantage of our calculations is that a much wider range of ISM conditions can be explored 
than is feasible in the more expensive hydrodynamical simulations.
We have given
predictions for the outflow rate for a very wide range in galaxy properties and cosmic epochs.
The method developed in this paper also allows radial profiles
of the outflow rate to be obtained. The new generation of integral field spectroscopy instruments,
 such as KMOS in the Very Large Telescope (\citealt{Sharples04}) and the
Sydney-AAO Multi-object Integral field spectrograph (\citealt{Croom12}; \citealt{Fogarty12}) will make the observations
of outflows routine in local and high-redshift galaxies, and will allow us to constrain
 our model observationally.

\nocite{Bigiel11}
\nocite{Bigiel08}
\nocite{Bertone05}
\nocite{Bertone07}
\nocite{Tsai12a}
\nocite{Tsai12b}
\nocite{Williams97}
\nocite{Dobbs11}
\nocite{Rahman11}
\nocite{Schruba11}
\nocite{Swinbank10}
\nocite{Swinbank11}

\section*{Acknowledgements}

We thank the anonymous referee for the comments that improved this paper. We also thank 
Joop Schaye, Pierluigi Monaco, Andrew Benson and Ian Smail 
for carefully reading this manuscript and for their valuable suggestions. We also thank 
Peter Creasey, Richard Bower and Luiz Felippe Rodriguez for useful comments and discussions, and 
Rodrigo Tobar and Lydia Heck for technical support.
Calculations for this paper were performed on the ICC
Cosmology Machine, which is part of the DiRAC Facility jointly funded by STFC,
the Large Facilities Capital Fund of BIS, and Durham University.
The research leading to these results has received funding from a STFC Gemini studentship, 
the European Community's Seventh 
Framework Programme ($/$FP7$/$2007-2013$/$) under grant agreement no 229517 and 
 through the Marie Curie International Research Staff Exchange Scheme 
LACEGAL (PIRSES-GA-2010-269264). This research has made use 
of the NASA/IPAC Extragalactic Database (NED) which is operated by the Jet Propulsion Laboratory, 
California Institute of Technology, under contract with the National Aeronautics and 
Space Administration.

%----------------------------------------------
\bibliographystyle{mn2e_trunc8}
\bibliography{SNfeedback_v2}

\begin{thebibliography}{184}
\expandafter\ifx\csname natexlab\endcsname\relax\def\natexlab#1{#1}\fi

\bibitem[{{Arnett} {et~al.}(1989){Arnett}, {Bahcall}, {Kirshner}, \&
  {Woosley}}]{Arnett89}
{Arnett} W.~D., {Bahcall} J.~N., {Kirshner} R.~P., {Woosley} S.~E., 1989,
  \araa, 27, 629

\bibitem[{{Ballantyne} {et~al.}(2013){Ballantyne}, {Armour}, \&
  {Indergaard}}]{Ballantyne13}
{Ballantyne} D.~R., {Armour} J.~N., {Indergaard} J., 2013, ApJ in press,
  ArXiv:1301.7020

\bibitem[{{Ballesteros-Paredes} {et~al.}(1999){Ballesteros-Paredes},
  {Hartmann}, \& {V{\'a}zquez-Semadeni}}]{Ballesteros-Paredes99}
{Ballesteros-Paredes} J., {Hartmann} L., {V{\'a}zquez-Semadeni} E., 1999, \apj,
  527, 285

\bibitem[{{Banerji} {et~al.}(2011){Banerji}, {Chapman}, {Smail},
  {Alaghband-Zadeh}, {Swinbank}, {Dunlop}, {Ivison}, \& {Blain}}]{Banerji11}
{Banerji} M., {Chapman} S.~C., {Smail} I., {Alaghband-Zadeh} S., {Swinbank}
  A.~M., {Dunlop} J.~S., {Ivison} R.~J., {Blain} A.~W., 2011, \mnras, 418, 1071

\bibitem[{{Baugh}(2006)}]{Baugh06}
{Baugh} C.~M., 2006, Reports on Progress in Physics, 69, 3101

\bibitem[{{Baugh} {et~al.}(2005){Baugh}, {Lacey}, {Frenk}, {Granato}, {Silva},
  {Bressan}, {Benson}, \& {Cole}}]{Baugh05}
{Baugh} C.~M., {Lacey} C.~G., {Frenk} C.~S., {Granato} G.~L., {Silva} L.,
  {Bressan} A., {Benson} A.~J., {Cole} S., 2005, \mnras, 356, 1191

\bibitem[{{Bell} {et~al.}(2003){Bell}, {McIntosh}, {Katz}, \&
  {Weinberg}}]{Bell03b}
{Bell} E.~F., {McIntosh} D.~H., {Katz} N., {Weinberg} M.~D., 2003, \apjs, 149,
  289

\bibitem[{{Benson}(2010)}]{Benson10b}
{Benson} A.~J., 2010, \physrep, 495, 33

\bibitem[{{Benson} {et~al.}(2003){Benson}, {Bower}, {Frenk}, {Lacey}, {Baugh},
  \& {Cole}}]{Benson03}
{Benson} A.~J., {Bower} R.~G., {Frenk} C.~S., {Lacey} C.~G., {Baugh} C.~M.,
  {Cole} S., 2003, \apj, 599, 38

\bibitem[{{Bertone} {et~al.}(2007){Bertone}, {De Lucia}, \&
  {Thomas}}]{Bertone07}
{Bertone} S., {De Lucia} G., {Thomas} P.~A., 2007, \mnras, 379, 1143

\bibitem[{{Bertone} {et~al.}(2005){Bertone}, {Stoehr}, \& {White}}]{Bertone05}
{Bertone} S., {Stoehr} F., {White} S.~D.~M., 2005, \mnras, 359, 1201

\bibitem[{{Bielby} {et~al.}(2012){Bielby}, {Hudelot}, {McCracken}, {Ilbert},
  {Daddi}, {Le F{\`e}vre}, {Gonzalez-Perez}, {Kneib}, {Marmo}, {Mellier},
  {Salvato}, {Sanders}, \& {Willott}}]{Bielby11}
{Bielby} R., {Hudelot} P., {McCracken} H.~J., {Ilbert} O., {Daddi} E., {Le
  F{\`e}vre} O., {Gonzalez-Perez} V., {Kneib} J.-P. {et~al}, 2012, \aap, 545,
  A23

\bibitem[{{Bigiel} \& {Blitz}(2012)}]{Bigiel12}
{Bigiel} F., {Blitz} L., 2012, \apj, 756, 183

\bibitem[{{Bigiel} {et~al.}(2008){Bigiel}, {Leroy}, {Walter}, {Brinks}, {de
  Blok}, {Madore}, \& {Thornley}}]{Bigiel08}
{Bigiel} F., {Leroy} A., {Walter} F., {Brinks} E., {de Blok} W.~J.~G., {Madore}
  B., {Thornley} M.~D., 2008, \aj, 136, 2846

\bibitem[{{Bigiel} {et~al.}(2011){Bigiel}, {Leroy}, {Walter}, {Brinks}, {de
  Blok}, {Kramer}, {Rix}, {Schruba}, {Schuster}, {Usero}, \&
  {Wiesemeyer}}]{Bigiel11}
{Bigiel} F., {Leroy} A.~K., {Walter} F., {Brinks} E., {de Blok} W.~J.~G.,
  {Kramer} C., {Rix} H.~W., {Schruba} A. {et~al}, 2011, \apjl, 730, L13+

\bibitem[{{Binney} \& {Tremaine}(2008)}]{Binney08}
{Binney} J., {Tremaine} S., 2008, {Galactic Dynamics: Second Edition}.
  Princeton University Press

\bibitem[{{Blitz} {et~al.}(2007){Blitz}, {Fukui}, {Kawamura}, {Leroy},
  {Mizuno}, \& {Rosolowsky}}]{Blitz07}
{Blitz} L., {Fukui} Y., {Kawamura} A., {Leroy} A., {Mizuno} N., {Rosolowsky}
  E., 2007, Protostars and Planets V, 81

\bibitem[{{Blitz} \& {Rosolowsky}(2006)}]{Blitz06}
{Blitz} L., {Rosolowsky} E., 2006, \apj, 650, 933

\bibitem[{{Blitz} \& {Shu}(1980)}]{Blitz80}
{Blitz} L., {Shu} F.~H., 1980, \apj, 238, 148

\bibitem[{{Bolatto} {et~al.}(2013){Bolatto}, {Warren}, {Leroy}, {Walter},
  {Veilleux}, {Ostriker}, {Ott}, {Zwaan}, {Fisher}, {Weiss}, {Rosolowsky}, \&
  {Hodge}}]{Bolatto13}
{Bolatto} A.~D., {Warren} S.~S., {Leroy} A.~K., {Walter} F., {Veilleux} S.,
  {Ostriker} E.~C., {Ott} J., {Zwaan} M. {et~al}, 2013, ArXiv e-prints

\bibitem[{{Boselli} {et~al.}(2000){Boselli}, {Gavazzi}, {Franzetti}, {Pierini},
  \& {Scodeggio}}]{Boselli00}
{Boselli} A., {Gavazzi} G., {Franzetti} P., {Pierini} D., {Scodeggio} M., 2000,
  \aaps, 142, 73

\bibitem[{{Bouch{\'e}} {et~al.}(2012){Bouch{\'e}}, {Hohensee}, {Vargas},
  {Kacprzak}, {Martin}, {Cooke}, \& {Churchill}}]{Bouche12}
{Bouch{\'e}} N., {Hohensee} W., {Vargas} R., {Kacprzak} G.~G., {Martin} C.~L.,
  {Cooke} J., {Churchill} C.~W., 2012, \mnras, 3207

\bibitem[{{Bower} {et~al.}(2012){Bower}, {Benson}, \& {Crain}}]{Bower12}
{Bower} R.~G., {Benson} A.~J., {Crain} R.~A., 2012, \mnras, 2860

\bibitem[{{Bower} {et~al.}(2006){Bower}, {Benson}, {Malbon}, {Helly}, {Frenk},
  {Baugh}, {Cole}, \& {Lacey}}]{Bower06}
{Bower} R.~G., {Benson} A.~J., {Malbon} R., {Helly} J.~C., {Frenk} C.~S.,
  {Baugh} C.~M., {Cole} S., {Lacey} C.~G., 2006, \mnras, 370, 645

\bibitem[{{Caputi} {et~al.}(2011){Caputi}, {Cirasuolo}, {Dunlop}, {McLure},
  {Farrah}, \& {Almaini}}]{Caputi11}
{Caputi} K.~I., {Cirasuolo} M., {Dunlop} J.~S., {McLure} R.~J., {Farrah} D.,
  {Almaini} O., 2011, \mnras, 413, 162

\bibitem[{{Caputi} {et~al.}(2006){Caputi}, {McLure}, {Dunlop}, {Cirasuolo}, \&
  {Schael}}]{Caputi06}
{Caputi} K.~I., {McLure} R.~J., {Dunlop} J.~S., {Cirasuolo} M., {Schael} A.~M.,
  2006, \mnras, 366, 609

\bibitem[{{Chen} {et~al.}(2010){Chen}, {Tremonti}, {Heckman}, {Kauffmann},
  {Weiner}, {Brinchmann}, \& {Wang}}]{Chen10}
{Chen} Y.-M., {Tremonti} C.~A., {Heckman} T.~M., {Kauffmann} G., {Weiner}
  B.~J., {Brinchmann} J., {Wang} J., 2010, \aj, 140, 445

\bibitem[{{Chieze}(1987)}]{Chieze87}
{Chieze} J.~P., 1987, \aap, 171, 225

\bibitem[{{Cole} {et~al.}(2000){Cole}, {Lacey}, {Baugh}, \& {Frenk}}]{Cole00}
{Cole} S., {Lacey} C.~G., {Baugh} C.~M., {Frenk} C.~S., 2000, \mnras, 319, 168

\bibitem[{{Combes} {et~al.}(2011){Combes}, {Garc{\'{\i}}a-Burillo}, {Braine},
  {Schinnerer}, {Walter}, \& {Colina}}]{Combes11}
{Combes} F., {Garc{\'{\i}}a-Burillo} S., {Braine} J., {Schinnerer} E., {Walter}
  F., {Colina} L., 2011, \aap, 528, 124

\bibitem[{{Cook} {et~al.}(2010){Cook}, {Evoli}, {Barausse}, {Granato}, \&
  {Lapi}}]{Cook10}
{Cook} M., {Evoli} C., {Barausse} E., {Granato} G.~L., {Lapi} A., 2010, \mnras,
  402, 941

\bibitem[{{Crain} {et~al.}(2009){Crain}, {Theuns}, {Dalla Vecchia}, {Eke},
  {Frenk}, {Jenkins}, {Kay}, {Peacock}, {Pearce}, {Schaye}, {Springel},
  {Thomas}, {White}, \& {Wiersma}}]{Crain09}
{Crain} R.~A., {Theuns} T., {Dalla Vecchia} C., {Eke} V.~R., {Frenk} C.~S.,
  {Jenkins} A., {Kay} S.~T., {Peacock} J.~A. {et~al}, 2009, \mnras, 399, 1773

\bibitem[{{Creasey} {et~al.}(2013){Creasey}, {Theuns}, \& {Bower}}]{Creasey12}
{Creasey} P., {Theuns} T., {Bower} R.~G., 2013, \mnras, 429, 1922

\bibitem[{{Crocker} {et~al.}(2011){Crocker}, {Bureau}, {Young}, \&
  {Combes}}]{Crocker11}
{Crocker} A.~F., {Bureau} M., {Young} L.~M., {Combes} F., 2011, \mnras, 410,
  1197

\bibitem[{{Croom} {et~al.}(2012){Croom}, {Lawrence}, {Bland-Hawthorn},
  {Bryant}, {Fogarty}, {Richards}, {Goodwin}, {Farrell}, {Miziarski}, {Heald},
  {Jones}, {Lee}, {Colless}, {Brough}, {Hopkins}, {Bauer}, {Birchall}, {Ellis},
  {Horton}, {Leon-Saval}, {Lewis}, {L{\'o}pez-S{\'a}nchez}, {Min}, {Trinh}, \&
  {Trowland}}]{Croom12}
{Croom} S.~M., {Lawrence} J.~S., {Bland-Hawthorn} J., {Bryant} J.~J., {Fogarty}
  L., {Richards} S., {Goodwin} M., {Farrell} T. {et~al}, 2012, \mnras, 421, 872

\bibitem[{{Croton} {et~al.}(2006){Croton}, {Springel}, {White}, {De Lucia},
  {Frenk}, {Gao}, {Jenkins}, {Kauffmann}, {Navarro}, \& {Yoshida}}]{Croton06}
{Croton} D.~J., {Springel} V., {White} S.~D.~M., {De Lucia} G., {Frenk} C.~S.,
  {Gao} L., {Jenkins} A., {Kauffmann} G. {et~al}, 2006, \mnras, 365, 11

\bibitem[{{Dale} {et~al.}(2012){Dale}, {Ercolano}, \& {Bonnell}}]{Dale12}
{Dale} J.~E., {Ercolano} B., {Bonnell} I.~A., 2012, \mnras, 424, 377

\bibitem[{{Dalla Vecchia} \& {Schaye}(2008)}]{DallaVecchia08}
{Dalla Vecchia} C., {Schaye} J., 2008, \mnras, 387, 1431

\bibitem[{{Dav{\'e}} {et~al.}(2011){Dav{\'e}}, {Oppenheimer}, \&
  {Finlator}}]{Dave11}
{Dav{\'e}} R., {Oppenheimer} B.~D., {Finlator} K., 2011, \mnras, 415, 11

\bibitem[{{Davis} {et~al.}(2012){Davis}, {Alatalo}, {Bureau}, {Cappellari},
  {Scott}, {Young}, {Blitz}, {Crocker}, {Bayet}, {Bois}, {Bournaud}, {Davies},
  {de Zeeuw}, {Duc}, {Emsellem}, {Khochfar}, {Krajnovi{\'c}}, {Kuntschner},
  {Lablanche}, {McDermid}, {Morganti}, {Naab}, {Oosterloo}, {Sarzi}, {Serra},
  \& {Weijmans}}]{Davis12}
{Davis} T.~A., {Alatalo} K., {Bureau} M., {Cappellari} M., {Scott} N., {Young}
  L.~M., {Blitz} L., {Crocker} A. {et~al}, 2012, \mnras, 286

\bibitem[{{Davis} {et~al.}(2011){Davis}, {Bureau}, {Young}, {Alatalo}, {Blitz},
  {Cappellari}, {Scott}, {Bois}, {Bournaud}, {Davies}, {de Zeeuw}, {Emsellem},
  {Khochfar}, {Krajnovi{\'c}}, {Kuntschner}, {Lablanche}, {McDermid},
  {Morganti}, {Naab}, {Oosterloo}, {Sarzi}, {Serra}, \& {Weijmans}}]{Davis11}
{Davis} T.~A., {Bureau} M., {Young} L.~M., {Alatalo} K., {Blitz} L.,
  {Cappellari} M., {Scott} N., {Bois} M. {et~al}, 2011, \mnras, 414, 968

\bibitem[{{de Avillez} \& {Berry}(2001)}]{deAvillez01}
{de Avillez} M.~A., {Berry} D.~L., 2001, \mnras, 328, 708

\bibitem[{{de Avillez} \& {Breitschwerdt}(2004)}]{Avillez04}
{de Avillez} M.~A., {Breitschwerdt} D., 2004, \aap, 425, 899

\bibitem[{{de Vaucouleurs}(1953)}]{deVaucouleurs53}
{de Vaucouleurs} G., 1953, \mnras, 113, 134

\bibitem[{{Dehnen}(1993)}]{Dehnen93}
{Dehnen} W., 1993, \mnras, 265, 250

\bibitem[{{Dekel} \& {Silk}(1986)}]{Dekel86}
{Dekel} A., {Silk} J., 1986, \apj, 303, 39

\bibitem[{{Dobbs} {et~al.}(2011){Dobbs}, {Burkert}, \& {Pringle}}]{Dobbs11}
{Dobbs} C.~L., {Burkert} A., {Pringle} J.~E., 2011, \mnras, 417, 1318

\bibitem[{{Downes} \& {Solomon}(1998)}]{Downes98}
{Downes} D., {Solomon} P.~M., 1998, \apj, 507, 615

\bibitem[{{Drory} {et~al.}(2004){Drory}, {Bender}, {Feulner}, {Hopp},
  {Maraston}, {Snigula}, \& {Hill}}]{Drory03}
{Drory} N., {Bender} R., {Feulner} G., {Hopp} U., {Maraston} C., {Snigula} J.,
  {Hill} G.~J., 2004, \apj, 608, 742

\bibitem[{{Drory} {et~al.}(2005){Drory}, {Salvato}, {Gabasch}, {Bender},
  {Hopp}, {Feulner}, \& {Pannella}}]{Drory05}
{Drory} N., {Salvato} M., {Gabasch} A., {Bender} R., {Hopp} U., {Feulner} G.,
  {Pannella} M., 2005, \apjl, 619, L131

\bibitem[{{Dutton} {et~al.}(2010){Dutton}, {van den Bosch}, \&
  {Dekel}}]{Dutton09}
{Dutton} A.~A., {van den Bosch} F.~C., {Dekel} A., 2010, \mnras, 405, 1690

\bibitem[{{Efstathiou}(2000)}]{Efstathiou00}
{Efstathiou} G., 2000, \mnras, 317, 697

\bibitem[{{Elmegreen}(1989)}]{Elmegreen89}
{Elmegreen} B.~G., 1989, \apj, 338, 178

\bibitem[{{Elmegreen}(1999)}]{Elmegreen99}
---, 1999, \apj, 527, 266

\bibitem[{{Engargiola} {et~al.}(2003){Engargiola}, {Plambeck}, {Rosolowsky}, \&
  {Blitz}}]{Engargiola03}
{Engargiola} G., {Plambeck} R.~L., {Rosolowsky} E., {Blitz} L., 2003, \apjs,
  149, 343

\bibitem[{{Erb} {et~al.}(2012){Erb}, {Quider}, {Henry}, \& {Martin}}]{Erb12}
{Erb} D.~K., {Quider} A.~M., {Henry} A.~L., {Martin} C.~L., 2012, \apj, 759, 26

\bibitem[{{Ferri{\`e}re}(2001)}]{Ferriere01}
{Ferri{\`e}re} K.~M., 2001, Reviews of Modern Physics, 73, 1031

\bibitem[{{Fogarty} {et~al.}(2012){Fogarty}, {Bland-Hawthorn}, {Croom},
  {Green}, {Bryant}, {Lawrence}, {Richards}, {Allen}, {Bauer}, {Birchall},
  {Brough}, {Colless}, {Ellis}, {Farrell}, {Goodwin}, {Heald}, {Hopkins},
  {Horton}, {Jones}, {Lee}, {Lewis}, {L{\'o}pez-S{\'a}nchez}, {Miziarski},
  {Trowland}, {Leon-Saval}, {Min}, {Trinh}, {Cecil}, {Veilleux}, \&
  {Kreimeyer}}]{Fogarty12}
{Fogarty} L.~M.~R., {Bland-Hawthorn} J., {Croom} S.~M., {Green} A.~W., {Bryant}
  J.~J., {Lawrence} J.~S., {Richards} S., {Allen} J.~T. {et~al}, 2012, \apj,
  761, 169

\bibitem[{{Font} {et~al.}(2011){Font}, {Benson}, {Bower}, {Frenk}, {Cooper},
  {De Lucia}, {Helly}, {Helmi}, {Li}, {McCarthy}, {Navarro}, {Springel},
  {Starkenburg}, {Wang}, \& {White}}]{Font12}
{Font} A.~S., {Benson} A.~J., {Bower} R.~G., {Frenk} C.~S., {Cooper} A., {De
  Lucia} G., {Helly} J.~C., {Helmi} A. {et~al}, 2011, \mnras, 417, 1260

\bibitem[{{Fu} {et~al.}(2010){Fu}, {Guo}, {Kauffmann}, \& {Krumholz}}]{Fu10}
{Fu} J., {Guo} Q., {Kauffmann} G., {Krumholz} M.~R., 2010, \mnras, 409, 515

\bibitem[{{Fujita} {et~al.}(2009){Fujita}, {Martin}, {Mac Low}, {New}, \&
  {Weaver}}]{Fujita09}
{Fujita} A., {Martin} C.~L., {Mac Low} M.-M., {New} K.~C.~B., {Weaver} R.,
  2009, \apj, 698, 693

\bibitem[{{Fukugita} {et~al.}(1998){Fukugita}, {Hogan}, \&
  {Peebles}}]{Fukugita98}
{Fukugita} M., {Hogan} C.~J., {Peebles} P.~J.~E., 1998, \apj, 503, 518

\bibitem[{{Genzel} {et~al.}(2008){Genzel}, {Burkert}, {Bouch{\'e}}, {Cresci},
  {F{\"o}rster Schreiber}, {Shapley}, {Shapiro}, {Tacconi}, {Buschkamp},
  {Cimatti}, {Daddi}, {Davies}, {Eisenhauer}, {Erb}, {Genel}, {Gerhard},
  {Hicks}, {Lutz}, {Naab}, {Ott}, {Rabien}, {Renzini}, {Steidel}, {Sternberg},
  \& {Lilly}}]{Genzel08}
{Genzel} R., {Burkert} A., {Bouch{\'e}} N., {Cresci} G., {F{\"o}rster
  Schreiber} N.~M., {Shapley} A., {Shapiro} K., {Tacconi} L.~J. {et~al}, 2008,
  \apj, 687, 59

\bibitem[{{Genzel} {et~al.}(2010){Genzel}, {Tacconi}, {Gracia-Carpio},
  {Sternberg}, {Cooper}, {Shapiro}, {Bolatto}, {Bouch{\'e}}, {Bournaud},
  {Burkert}, {Combes}, {Comerford}, {Cox}, {Davis}, {Schreiber},
  {Garcia-Burillo}, {Lutz}, {Naab}, {Neri}, {Omont}, {Shapley}, \&
  {Weiner}}]{Genzel10}
{Genzel} R., {Tacconi} L.~J., {Gracia-Carpio} J., {Sternberg} A., {Cooper}
  M.~C., {Shapiro} K., {Bolatto} A., {Bouch{\'e}} N. {et~al}, 2010, \mnras,
  407, 2091

\bibitem[{{Glazebrook}(2013)}]{Glazebrook13}
{Glazebrook} K., 2013, ArXiv e-prints

\bibitem[{{Gonzalez-Perez} {et~al.}(2012){Gonzalez-Perez}, {Lacey}, {Baugh},
  {Frenk}, \& {Wilkins}}]{Gonzalez-Perez12}
{Gonzalez-Perez} V., {Lacey} C.~G., {Baugh} C.~M., {Frenk} C.~S., {Wilkins}
  S.~M., 2012, MNRAS submitted (ArXiv:1209.2152)

\bibitem[{{Granato} {et~al.}(2000){Granato}, {Lacey}, {Silva}, {Bressan},
  {Baugh}, {Cole}, \& {Frenk}}]{Granato00}
{Granato} G.~L., {Lacey} C.~G., {Silva} L., {Bressan} A., {Baugh} C.~M., {Cole}
  S., {Frenk} C.~S., 2000, \apj, 542, 710

\bibitem[{{Green}(2009)}]{Green09}
{Green} D.~A., 2009, Bulletin of the Astronomical Society of India, 37, 45

\bibitem[{{Guo} {et~al.}(2013){Guo}, {White}, {Angulo}, {Henriques}, {Lemson},
  {Boylan-Kolchin}, {Thomas}, \& {Short}}]{Guo12}
{Guo} Q., {White} S., {Angulo} R.~E., {Henriques} B., {Lemson} G.,
  {Boylan-Kolchin} M., {Thomas} P., {Short} C., 2013, \mnras, 428, 1351

\bibitem[{{Guo} {et~al.}(2011){Guo}, {White}, {Boylan-Kolchin}, {De Lucia},
  {Kauffmann}, {Lemson}, {Li}, {Springel}, \& {Weinmann}}]{Guo11}
{Guo} Q., {White} S., {Boylan-Kolchin} M., {De Lucia} G., {Kauffmann} G.,
  {Lemson} G., {Li} C., {Springel} V. {et~al}, 2011, \mnras, 413, 101

\bibitem[{{Haffner} {et~al.}(2009){Haffner}, {Dettmar}, {Beckman}, {Wood},
  {Slavin}, {Giammanco}, {Madsen}, {Zurita}, \& {Reynolds}}]{Haffner09}
{Haffner} L.~M., {Dettmar} R.-J., {Beckman} J.~E., {Wood} K., {Slavin} J.~D.,
  {Giammanco} C., {Madsen} G.~J., {Zurita} A. {et~al}, 2009, Reviews of Modern
  Physics, 81, 969

\bibitem[{{Hatton} {et~al.}(2003){Hatton}, {Devriendt}, {Ninin}, {Bouchet},
  {Guiderdoni}, \& {Vibert}}]{Hatton03}
{Hatton} S., {Devriendt} J.~E.~G., {Ninin} S., {Bouchet} F.~R., {Guiderdoni}
  B., {Vibert} D., 2003, \mnras, 343, 75

\bibitem[{{Heckman} {et~al.}(2000){Heckman}, {Lehnert}, {Strickland}, \&
  {Armus}}]{Heckman00}
{Heckman} T.~M., {Lehnert} M.~D., {Strickland} D.~K., {Armus} L., 2000, \apjs,
  129, 493

\bibitem[{{Heiles}(1979)}]{Heiles79}
{Heiles} C., 1979, \apj, 229, 533

\bibitem[{{Heitsch} {et~al.}(2005){Heitsch}, {Burkert}, {Hartmann}, {Slyz}, \&
  {Devriendt}}]{Heitsch05}
{Heitsch} F., {Burkert} A., {Hartmann} L.~W., {Slyz} A.~D., {Devriendt}
  J.~E.~G., 2005, \apjl, 633, L113

\bibitem[{{Henriques} {et~al.}(2013){Henriques}, {White}, {Thomas}, {Angulo},
  {Guo}, {Lemson}, \& {Springel}}]{Henriques13}
{Henriques} B.~M.~B., {White} S.~D.~M., {Thomas} P.~A., {Angulo} R.~E., {Guo}
  Q., {Lemson} G., {Springel} V., 2013, \mnras, 431, 3373

\bibitem[{{Hopkins}(2004)}]{Hopkins04}
{Hopkins} A.~M., 2004, \apj, 615, 209

\bibitem[{{Hopkins}(2012)}]{Hopkins11}
{Hopkins} P.~F., 2012, \mnras, 423, 2016

\bibitem[{{Hopkins} {et~al.}(2012){Hopkins}, {Quataert}, \&
  {Murray}}]{Hopkins12}
{Hopkins} P.~F., {Quataert} E., {Murray} N., 2012, \mnras, 421, 3522

\bibitem[{{Karim} {et~al.}(2011){Karim}, {Schinnerer},
  {Mart{\'{\i}}nez-Sansigre}, {Sargent}, {van der Wel}, {Rix}, {Ilbert},
  {Smol{\v c}i{\'c}}, {Carilli}, {Pannella}, {Koekemoer}, {Bell}, \&
  {Salvato}}]{Karim11}
{Karim} A., {Schinnerer} E., {Mart{\'{\i}}nez-Sansigre} A., {Sargent} M.~T.,
  {van der Wel} A., {Rix} H.-W., {Ilbert} O., {Smol{\v c}i{\'c}} V. {et~al},
  2011, \apj, 730, 61

\bibitem[{{Kennicutt}(1983)}]{Kennicutt83}
{Kennicutt} Jr. R.~C., 1983, \apj, 272, 54

\bibitem[{{Kennicutt}(1998)}]{Kennicutt98}
---, 1998, \apj, 498, 541

\bibitem[{{Kim} {et~al.}(2011){Kim}, {Baugh}, {Benson}, {Cole}, {Frenk},
  {Lacey}, {Power}, \& {Schneider}}]{Kim10}
{Kim} H.-S., {Baugh} C.~M., {Benson} A.~J., {Cole} S., {Frenk} C.~S., {Lacey}
  C.~G., {Power} C., {Schneider} M., 2011, \mnras, 414, 2367

\bibitem[{{Kornei} {et~al.}(2012){Kornei}, {Shapley}, {Martin}, {Coil}, {Lotz},
  {Schiminovich}, {Bundy}, \& {Noeske}}]{Kornei12}
{Kornei} K.~A., {Shapley} A.~E., {Martin} C.~L., {Coil} A.~L., {Lotz} J.~M.,
  {Schiminovich} D., {Bundy} K., {Noeske} K.~G., 2012, \apj, 758, 135

\bibitem[{{Koyama} \& {Ostriker}(2009)}]{Koyama09}
{Koyama} H., {Ostriker} E.~C., 2009, \apj, 693, 1346

\bibitem[{{Kregel} {et~al.}(2002){Kregel}, {van der Kruit}, \& {de
  Grijs}}]{Kregel02}
{Kregel} M., {van der Kruit} P.~C., {de Grijs} R., 2002, \mnras, 334, 646

\bibitem[{{Krumholz} \& {McKee}(2005)}]{Krumholz05}
{Krumholz} M.~R., {McKee} C.~F., 2005, \apj, 630, 250

\bibitem[{{Krumholz} {et~al.}(2009){Krumholz}, {McKee}, \&
  {Tumlinson}}]{Krumholz09}
{Krumholz} M.~R., {McKee} C.~F., {Tumlinson} J., 2009, \apj, 699, 850

\bibitem[{{Lacey} {et~al.}(2011){Lacey}, {Baugh}, {Frenk}, \&
  {Benson}}]{Lacey11}
{Lacey} C.~G., {Baugh} C.~M., {Frenk} C.~S., {Benson} A.~J., 2011, \mnras, 45

\bibitem[{{Lacey} {et~al.}(2008){Lacey}, {Baugh}, {Frenk}, {Silva}, {Granato},
  \& {Bressan}}]{Lacey08}
{Lacey} C.~G., {Baugh} C.~M., {Frenk} C.~S., {Silva} L., {Granato} G.~L.,
  {Bressan} A., 2008, \mnras, 385, 1155

\bibitem[{{Lagos} {et~al.}(2011{\natexlab{a}}){Lagos}, {Baugh}, {Lacey},
  {Benson}, {Kim}, \& {Power}}]{Lagos11}
{Lagos} C.~D.~P., {Baugh} C.~M., {Lacey} C.~G., {Benson} A.~J., {Kim} H.-S.,
  {Power} C., 2011{\natexlab{a}}, \mnras, 418, 1649

\bibitem[{{Lagos} {et~al.}(2012){Lagos}, {Bayet}, {Baugh}, {Lacey}, {Bell},
  {Fanidakis}, \& {Geach}}]{Lagos12}
{Lagos} C.~d.~P., {Bayet} E., {Baugh} C.~M., {Lacey} C.~G., {Bell} T.~A.,
  {Fanidakis} N., {Geach} J.~E., 2012, \mnras, 426, 2142

\bibitem[{{Lagos} {et~al.}(2008){Lagos}, {Cora}, \& {Padilla}}]{Lagos08}
{Lagos} C.~D.~P., {Cora} S.~A., {Padilla} N.~D., 2008, \mnras, 388, 587

\bibitem[{{Lagos} {et~al.}(2011{\natexlab{b}}){Lagos}, {Lacey}, {Baugh},
  {Bower}, \& {Benson}}]{Lagos10}
{Lagos} C.~D.~P., {Lacey} C.~G., {Baugh} C.~M., {Bower} R.~G., {Benson} A.~J.,
  2011{\natexlab{b}}, \mnras, 416, 1566

\bibitem[{{Larson}(1974)}]{Larson74}
{Larson} R.~B., 1974, \mnras, 169, 229

\bibitem[{{Law} {et~al.}(2007){Law}, {Steidel}, {Erb}, {Larkin}, {Pettini},
  {Shapley}, \& {Wright}}]{Law07}
{Law} D.~R., {Steidel} C.~C., {Erb} D.~K., {Larkin} J.~E., {Pettini} M.,
  {Shapley} A.~E., {Wright} S.~A., 2007, \apj, 669, 929

\bibitem[{{Leroy} {et~al.}(2008){Leroy}, {Walter}, {Brinks}, {Bigiel}, {de
  Blok}, {Madore}, \& {Thornley}}]{Leroy08}
{Leroy} A.~K., {Walter} F., {Brinks} E., {Bigiel} F., {de Blok} W.~J.~G.,
  {Madore} B., {Thornley} M.~D., 2008, \aj, 136, 2782

\bibitem[{{Li} \& {White}(2009)}]{Li09}
{Li} C., {White} S.~D.~M., 2009, \mnras, 398, 2177

\bibitem[{{Liu} {et~al.}(2010){Liu}, {Yang}, {Mo}, {van den Bosch}, \&
  {Springel}}]{Liu10}
{Liu} L., {Yang} X., {Mo} H.~J., {van den Bosch} F.~C., {Springel} V., 2010,
  \apj, 712, 734

\bibitem[{{Livermore} {et~al.}(2012){Livermore}, {Jones}, {Richard}, {Bower},
  {Ellis}, {Swinbank}, {Rigby}, {Smail}, {Arribas}, {Rodriguez Zaurin},
  {Colina}, {Ebeling}, \& {Crain}}]{Livermore12}
{Livermore} R.~C., {Jones} T., {Richard} J., {Bower} R.~G., {Ellis} R.~S.,
  {Swinbank} A.~M., {Rigby} J.~R., {Smail} I. {et~al}, 2012, \mnras, 427, 688

\bibitem[{{Mac Low} {et~al.}(1989){Mac Low}, {McCray}, \& {Norman}}]{MacLow89}
{Mac Low} M.-M., {McCray} R., {Norman} M.~L., 1989, \apj, 337, 141

\bibitem[{{Macci{\`o}} {et~al.}(2010){Macci{\`o}}, {Kang}, {Fontanot},
  {Somerville}, {Koposov}, \& {Monaco}}]{Maccio10}
{Macci{\`o}} A.~V., {Kang} X., {Fontanot} F., {Somerville} R.~S., {Koposov} S.,
  {Monaco} P., 2010, \mnras, 402, 1995

\bibitem[{{Maciejewski} {et~al.}(1996){Maciejewski}, {Murphy}, {Lockman}, \&
  {Savage}}]{Maciejewski96}
{Maciejewski} W., {Murphy} E.~M., {Lockman} F.~J., {Savage} B.~D., 1996, \apj,
  469, 238

\bibitem[{{MacLow} \& {McCray}(1988)}]{MacLow88}
{MacLow} M.-M., {McCray} R., 1988, \apj, 324, 776

\bibitem[{{Maloney}(1988)}]{Maloney88}
{Maloney} P., 1988, \apj, 334, 761

\bibitem[{{Mandelbaum} {et~al.}(2006){Mandelbaum}, {Seljak}, {Kauffmann},
  {Hirata}, \& {Brinkmann}}]{Mandelbaum06}
{Mandelbaum} R., {Seljak} U., {Kauffmann} G., {Hirata} C.~M., {Brinkmann} J.,
  2006, \mnras, 368, 715

\bibitem[{{Marchesini} {et~al.}(2012){Marchesini}, {Stefanon}, {Brammer}, \&
  {Whitaker}}]{Marchesini12}
{Marchesini} D., {Stefanon} M., {Brammer} G.~B., {Whitaker} K.~E., 2012, \apj,
  748, 126

\bibitem[{{Marchesini} {et~al.}(2009){Marchesini}, {van Dokkum}, {F{\"o}rster
  Schreiber}, {Franx}, {Labb{\'e}}, \& {Wuyts}}]{Marchesini09}
{Marchesini} D., {van Dokkum} P.~G., {F{\"o}rster Schreiber} N.~M., {Franx} M.,
  {Labb{\'e}} I., {Wuyts} S., 2009, \apj, 701, 1765

\bibitem[{{Marigo}(2001)}]{Marigo01}
{Marigo} P., 2001, \aap, 370, 194

\bibitem[{{Martin}(1999)}]{Martin99}
{Martin} C.~L., 1999, \apj, 513, 156

\bibitem[{{Martin}(2005)}]{Martin05}
---, 2005, \apj, 621, 227

\bibitem[{{Martin} {et~al.}(2012){Martin}, {Shapley}, {Coil}, {Kornei},
  {Bundy}, {Weiner}, {Noeske}, \& {Schiminovich}}]{Martin12}
{Martin} C.~L., {Shapley} A.~E., {Coil} A.~L., {Kornei} K.~A., {Bundy} K.,
  {Weiner} B.~J., {Noeske} K.~G., {Schiminovich} D., 2012, \apj, 760, 127

\bibitem[{{McKee} \& {Cowie}(1975)}]{McKee75}
{McKee} C.~F., {Cowie} L.~L., 1975, \apj, 195, 715

\bibitem[{{McKee} \& {Holliman}(1999)}]{McKee99}
{McKee} C.~F., {Holliman} II J.~H., 1999, \apj, 522, 313

\bibitem[{{McKee} \& {Ostriker}(2007)}]{McKee07}
{McKee} C.~F., {Ostriker} E.~C., 2007, \araa, 45, 565

\bibitem[{{Merson} {et~al.}(2013){Merson}, {Baugh}, {Helly}, {Gonzalez-Perez},
  {Cole}, {Bielby}, {Norberg}, {Frenk}, {Benson}, {Bower}, {Lacey}, \&
  {Lagos}}]{Merson12}
{Merson} A.~I., {Baugh} C.~M., {Helly} J.~C., {Gonzalez-Perez} V., {Cole} S.,
  {Bielby} R., {Norberg} P., {Frenk} C.~S. {et~al}, 2013, \mnras, 429, 556

\bibitem[{{Mitchell} {et~al.}(2013){Mitchell}, {Lacey}, {Baugh}, \&
  {Cole}}]{Mitchell13}
{Mitchell} P.~D., {Lacey} C.~G., {Baugh} C.~M., {Cole} S., 2013, ArXiv e-prints

\bibitem[{{Monaco}(2004{\natexlab{a}})}]{Monaco04b}
{Monaco} P., 2004{\natexlab{a}}, \mnras, 354, 151

\bibitem[{{Monaco}(2004{\natexlab{b}})}]{Monaco04}
---, 2004{\natexlab{b}}, \mnras, 352, 181

\bibitem[{{Monaco} {et~al.}(2007){Monaco}, {Fontanot}, \& {Taffoni}}]{Monaco07}
{Monaco} P., {Fontanot} F., {Taffoni} G., 2007, \mnras, 375, 1189

\bibitem[{{Murray} {et~al.}(2005){Murray}, {Quataert}, \&
  {Thompson}}]{Murray05}
{Murray} N., {Quataert} E., {Thompson} T.~A., 2005, \apj, 618, 569

\bibitem[{{Narayanan} {et~al.}(2008){Narayanan}, {Cox}, {Kelly}, {Dav{\'e}},
  {Hernquist}, {Di Matteo}, {Hopkins}, {Kulesa}, {Robertson}, \&
  {Walker}}]{Narayanan08}
{Narayanan} D., {Cox} T.~J., {Kelly} B., {Dav{\'e}} R., {Hernquist} L., {Di
  Matteo} T., {Hopkins} P.~F., {Kulesa} C. {et~al}, 2008, \apjs, 176, 331

\bibitem[{{Navarro} {et~al.}(1997){Navarro}, {Frenk}, \& {White}}]{Navarro97}
{Navarro} J.~F., {Frenk} C.~S., {White} S.~D.~M., 1997, \apj, 490, 493

\bibitem[{{Newman} {et~al.}(2012){Newman}, {Genzel}, {F{\"o}rster-Schreiber},
  {Shapiro Griffin}, {Mancini}, {Lilly}, {Renzini}, {Bouch{\'e}}, {Burkert},
  {Buschkamp}, {Carollo}, {Cresci}, {Davies}, {Eisenhauer}, {Genel}, {Hicks},
  {Kurk}, {Lutz}, {Naab}, {Peng}, {Sternberg}, {Tacconi}, {Vergani}, {Wuyts},
  \& {Zamorani}}]{Newman12}
{Newman} S.~F., {Genzel} R., {F{\"o}rster-Schreiber} N.~M., {Shapiro Griffin}
  K., {Mancini} C., {Lilly} S.~J., {Renzini} A., {Bouch{\'e}} N. {et~al}, 2012,
  \apj, 761, 43

\bibitem[{{Oka} {et~al.}(2001){Oka}, {Hasegawa}, {Sato}, {Tsuboi}, {Miyazaki},
  \& {Sugimoto}}]{Oka01}
{Oka} T., {Hasegawa} T., {Sato} F., {Tsuboi} M., {Miyazaki} A., {Sugimoto} M.,
  2001, \apj, 562, 348

\bibitem[{{Ostriker} \& {McKee}(1988)}]{Ostriker88}
{Ostriker} J.~P., {McKee} C.~F., 1988, Reviews of Modern Physics, 60, 1

\bibitem[{{Pelupessy} \& {Papadopoulos}(2009)}]{Pelupessy09}
{Pelupessy} F.~I., {Papadopoulos} P.~P., 2009, \apj, 707, 954

\bibitem[{{P{\'e}rez-Gonz{\'a}lez} {et~al.}(2008){P{\'e}rez-Gonz{\'a}lez},
  {Rieke}, {Villar}, {Barro}, {Blaylock}, {Egami}, {Gallego}, {Gil de Paz},
  {Pascual}, {Zamorano}, \& {Donley}}]{Perez-Gonzalez08}
{P{\'e}rez-Gonz{\'a}lez} P.~G., {Rieke} G.~H., {Villar} V., {Barro} G.,
  {Blaylock} M., {Egami} E., {Gallego} J., {Gil de Paz} A. {et~al}, 2008, \apj,
  675, 234

\bibitem[{{Pidopryhora} {et~al.}(2007){Pidopryhora}, {Lockman}, \&
  {Shields}}]{Pidopryhora07}
{Pidopryhora} Y., {Lockman} F.~J., {Shields} J.~C., 2007, \apj, 656, 928

\bibitem[{{Portinari} {et~al.}(1998){Portinari}, {Chiosi}, \&
  {Bressan}}]{Portinari98}
{Portinari} L., {Chiosi} C., {Bressan} A., 1998, \aap, 334, 505

\bibitem[{{Power} {et~al.}(2010){Power}, {Baugh}, \& {Lacey}}]{Power10}
{Power} C., {Baugh} C.~M., {Lacey} C.~G., 2010, \mnras, 406, 43

\bibitem[{{Pozzetti} {et~al.}(2003){Pozzetti}, {Cimatti}, {Zamorani}, {Daddi},
  {Menci}, {Fontana}, {Renzini}, {Mignoli}, {Poli}, {Saracco}, {Broadhurst},
  {Cristiani}, {D'Odorico}, {Giallongo}, \& {Gilmozzi}}]{Pozzetti03}
{Pozzetti} L., {Cimatti} A., {Zamorani} G., {Daddi} E., {Menci} N., {Fontana}
  A., {Renzini} A., {Mignoli} M. {et~al}, 2003, \aap, 402, 837

\bibitem[{{Prochaska} {et~al.}(2011){Prochaska}, {Kasen}, \&
  {Rubin}}]{Prochaska11}
{Prochaska} J.~X., {Kasen} D., {Rubin} K., 2011, \apj, 734, 24

\bibitem[{{Puech} {et~al.}(2007){Puech}, {Hammer}, {Lehnert}, \&
  {Flores}}]{Puech07}
{Puech} M., {Hammer} F., {Lehnert} M.~D., {Flores} H., 2007, \aap, 466, 83

\bibitem[{{Putman} {et~al.}(2012){Putman}, {Peek}, \& {Joung}}]{Putman12}
{Putman} M.~E., {Peek} J.~E.~G., {Joung} M.~R., 2012, \araa, 50, 491

\bibitem[{{Rahman} {et~al.}(2012){Rahman}, {Bolatto}, {Xue}, {Wong}, {Leroy},
  {Walter}, {Bigiel}, {Rosolowsky}, {Fisher}, {Vogel}, {Blitz}, {West}, \&
  {Ott}}]{Rahman11}
{Rahman} N., {Bolatto} A.~D., {Xue} R., {Wong} T., {Leroy} A.~K., {Walter} F.,
  {Bigiel} F., {Rosolowsky} E. {et~al}, 2012, \apj, 745, 183

\bibitem[{{Rees} \& {Ostriker}(1977)}]{Rees77}
{Rees} M.~J., {Ostriker} J.~P., 1977, \mnras, 179, 541

\bibitem[{{Rosolowsky} \& {Blitz}(2005)}]{Rosolowsky05}
{Rosolowsky} E., {Blitz} L., 2005, \apj, 623, 826

\bibitem[{{Rubin} {et~al.}(2010){Rubin}, {Weiner}, {Koo}, {Martin},
  {Prochaska}, {Coil}, \& {Newman}}]{Rubin10}
{Rubin} K.~H.~R., {Weiner} B.~J., {Koo} D.~C., {Martin} C.~L., {Prochaska}
  J.~X., {Coil} A.~L., {Newman} J.~A., 2010, \apj, 719, 1503

\bibitem[{{Rupke} {et~al.}(2005){Rupke}, {Veilleux}, \& {Sanders}}]{Rupke05}
{Rupke} D.~S., {Veilleux} S., {Sanders} D.~B., 2005, \apjs, 160, 87

\bibitem[{{Rupke} \& {Veilleux}(2013)}]{Rupke13}
{Rupke} D.~S.~N., {Veilleux} S., 2013, \apj, 768, 75

\bibitem[{{Samui} {et~al.}(2008){Samui}, {Subramanian}, \&
  {Srianand}}]{Samui08}
{Samui} S., {Subramanian} K., {Srianand} R., 2008, \mnras, 385, 783

\bibitem[{{Saracco} {et~al.}(2006){Saracco}, {Fiano}, {Chincarini}, {Vanzella},
  {Longhetti}, {Cristiani}, {Fontana}, {Giallongo}, \& {Nonino}}]{Saracco06}
{Saracco} P., {Fiano} A., {Chincarini} G., {Vanzella} E., {Longhetti} M.,
  {Cristiani} S., {Fontana} A., {Giallongo} E. {et~al}, 2006, \mnras, 367, 349

\bibitem[{{Sato} {et~al.}(2009){Sato}, {Martin}, {Noeske}, {Koo}, \&
  {Lotz}}]{Sato09}
{Sato} T., {Martin} C.~L., {Noeske} K.~G., {Koo} D.~C., {Lotz} J.~M., 2009,
  \apj, 696, 214

\bibitem[{{Scannapieco} {et~al.}(2006){Scannapieco}, {Tissera}, {White}, \&
  {Springel}}]{Scannapieco06}
{Scannapieco} C., {Tissera} P.~B., {White} S.~D.~M., {Springel} V., 2006,
  \mnras, 371, 1125

\bibitem[{{Scannapieco} {et~al.}(2012){Scannapieco}, {Wadepuhl}, {Parry},
  {Navarro}, {Jenkins}, {Springel}, {Teyssier}, {Carlson}, {Couchman}, {Crain},
  {Vecchia}, {Frenk}, {Kobayashi}, {Monaco}, {Murante}, {Okamoto}, {Quinn},
  {Schaye}, {Stinson}, {Theuns}, {Wadsley}, {White}, \&
  {Woods}}]{Scannapieco12}
{Scannapieco} C., {Wadepuhl} M., {Parry} O.~H., {Navarro} J.~F., {Jenkins} A.,
  {Springel} V., {Teyssier} R., {Carlson} E. {et~al}, 2012, \mnras, 423, 1726

\bibitem[{{Schaye}(2004)}]{Schaye04}
{Schaye} J., 2004, \apj, 609, 667

\bibitem[{{Schaye} {et~al.}(2010){Schaye}, {Dalla Vecchia}, {Booth}, {Wiersma},
  {Theuns}, {Haas}, {Bertone}, {Duffy}, {McCarthy}, \& {van de
  Voort}}]{Schaye10}
{Schaye} J., {Dalla Vecchia} C., {Booth} C.~M., {Wiersma} R.~P.~C., {Theuns}
  T., {Haas} M.~R., {Bertone} S., {Duffy} A.~R. {et~al}, 2010, \mnras, 402,
  1536

\bibitem[{{Schruba} {et~al.}(2011){Schruba}, {Leroy}, {Walter}, {Bigiel},
  {Brinks}, {de Blok}, {Dumas}, {Kramer}, {Rosolowsky}, {Sandstrom},
  {Schuster}, {Usero}, {Weiss}, \& {Wiesemeyer}}]{Schruba11}
{Schruba} A., {Leroy} A.~K., {Walter} F., {Bigiel} F., {Brinks} E., {de Blok}
  W.~J.~G., {Dumas} G., {Kramer} C. {et~al}, 2011, \aj, 142, 37

\bibitem[{{Schwartz} \& {Martin}(2004)}]{Schwartz04}
{Schwartz} C.~M., {Martin} C.~L., 2004, \apj, 610, 201

\bibitem[{{Schwartz} {et~al.}(2006){Schwartz}, {Martin}, {Chandar},
  {Leitherer}, {Heckman}, \& {Oey}}]{Schwartz06}
{Schwartz} C.~M., {Martin} C.~L., {Chandar} R., {Leitherer} C., {Heckman}
  T.~M., {Oey} M.~S., 2006, \apj, 646, 858

\bibitem[{{Serra} {et~al.}(2012){Serra}, {Oosterloo}, {Morganti}, {Alatalo},
  {Blitz}, {Bois}, {Bournaud}, {Bureau}, {Cappellari}, {Crocker}, {Davies},
  {Davis}, {de Zeeuw}, {Duc}, {Emsellem}, {Khochfar}, {Krajnovi{\'c}},
  {Kuntschner}, {Lablanche}, {McDermid}, {Naab}, {Sarzi}, {Scott}, {Trager},
  {Weijmans}, \& {Young}}]{Serra12}
{Serra} P., {Oosterloo} T., {Morganti} R., {Alatalo} K., {Blitz} L., {Bois} M.,
  {Bournaud} F., {Bureau} M. {et~al}, 2012, \mnras, 2823

\bibitem[{{Shapley} {et~al.}(2003){Shapley}, {Steidel}, {Pettini}, \&
  {Adelberger}}]{Shapley03}
{Shapley} A.~E., {Steidel} C.~C., {Pettini} M., {Adelberger} K.~L., 2003, \apj,
  588, 65

\bibitem[{{Sharples} {et~al.}(2004){Sharples}, {Bender}, {Lehnert}, {Ramsay
  Howat}, {Bremer}, {Davies}, {Genzel}, {Hofmann}, {Ivison}, {Saglia}, \&
  {Thatte}}]{Sharples04}
{Sharples} R.~M., {Bender} R., {Lehnert} M.~D., {Ramsay Howat} S.~K., {Bremer}
  M.~N., {Davies} R.~L., {Genzel} R., {Hofmann} R., {Ivison} R.~J., {Saglia}
  R., {Thatte} N.~A., 2004, in Society of Photo-Optical Instrumentation
  Engineers (SPIE) Conference Series, Vol. 5492, Society of Photo-Optical
  Instrumentation Engineers (SPIE) Conference Series, {Moorwood} A.~F.~M.,
  {Iye} M., eds., pp. 1179--1186

\bibitem[{{Shetty} \& {Ostriker}(2012)}]{Shetty12}
{Shetty} R., {Ostriker} E.~C., 2012, \apj, 754, 2

\bibitem[{{Shu} {et~al.}(2005){Shu}, {Mo}, \& {Shu-DeMao}}]{Shu05}
{Shu} C.-G., {Mo} H.-J., {Shu-DeMao}, 2005, Chinese Journal of Ast. and
  Astrophysics, 5, 327

\bibitem[{{Silk}(1997)}]{Silk97}
{Silk} J., 1997, \apj, 481, 703

\bibitem[{{Silk}(2003)}]{Silk03}
---, 2003, \mnras, 343, 249

\bibitem[{{Solomon} {et~al.}(1997){Solomon}, {Downes}, {Radford}, \&
  {Barrett}}]{Solomon97}
{Solomon} P.~M., {Downes} D., {Radford} S.~J.~E., {Barrett} J.~W., 1997, \apj,
  478, 144

\bibitem[{{Solomon} {et~al.}(1987){Solomon}, {Rivolo}, {Barrett}, \&
  {Yahil}}]{Solomon87}
{Solomon} P.~M., {Rivolo} A.~R., {Barrett} J., {Yahil} A., 1987, \apj, 319, 730

\bibitem[{{Springel} \& {Hernquist}(2003)}]{Springel03}
{Springel} V., {Hernquist} L., 2003, \mnras, 339, 289

\bibitem[{{Springel} {et~al.}(2005){Springel}, {White}, {Jenkins}, {Frenk},
  {Yoshida}, {Gao}, {Navarro}, {Thacker}, {Croton}, {Helly}, {Peacock}, {Cole},
  {Thomas}, {Couchman}, {Evrard}, {Colberg}, \& {Pearce}}]{Springel05}
{Springel} V., {White} S.~D.~M., {Jenkins} A., {Frenk} C.~S., {Yoshida} N.,
  {Gao} L., {Navarro} J., {Thacker} R. {et~al}, 2005, \nat, 435, 629

\bibitem[{{Springel} {et~al.}(2001){Springel}, {White}, {Tormen}, \&
  {Kauffmann}}]{Springel01}
{Springel} V., {White} S.~D.~M., {Tormen} G., {Kauffmann} G., 2001, \mnras,
  328, 726

\bibitem[{{Steidel} {et~al.}(2010){Steidel}, {Erb}, {Shapley}, {Pettini},
  {Reddy}, {Bogosavljevi{\'c}}, {Rudie}, \& {Rakic}}]{Steidel10}
{Steidel} C.~C., {Erb} D.~K., {Shapley} A.~E., {Pettini} M., {Reddy} N.,
  {Bogosavljevi{\'c}} M., {Rudie} G.~C., {Rakic} O., 2010, \apj, 717, 289

\bibitem[{{Stringer} {et~al.}(2012){Stringer}, {Bower}, {Cole}, {Frenk}, \&
  {Theuns}}]{Stringer12}
{Stringer} M.~J., {Bower} R.~G., {Cole} S., {Frenk} C.~S., {Theuns} T., 2012,
  \mnras, 423, 1596

\bibitem[{{Sturm} {et~al.}(2011){Sturm}, {Gonz{\'a}lez-Alfonso}, {Veilleux},
  {Fischer}, {Graci{\'a}-Carpio}, {Hailey-Dunsheath}, {Contursi}, {Poglitsch},
  {Sternberg}, {Davies}, {Genzel}, {Lutz}, {Tacconi}, {Verma}, {Maiolino}, \&
  {de Jong}}]{Sturm11}
{Sturm} E., {Gonz{\'a}lez-Alfonso} E., {Veilleux} S., {Fischer} J.,
  {Graci{\'a}-Carpio} J., {Hailey-Dunsheath} S., {Contursi} A., {Poglitsch} A.
  {et~al}, 2011, \apjl, 733, L16

\bibitem[{{Sutherland} \& {Dopita}(1993)}]{Sutherland93}
{Sutherland} R.~S., {Dopita} M.~A., 1993, \apjs, 88, 253

\bibitem[{{Swinbank} {et~al.}(2011){Swinbank}, {Papadopoulos}, {Cox}, {Krips},
  {Ivison}, {Smail}, {Thomson}, {Neri}, {Richard}, \& {Ebeling}}]{Swinbank11}
{Swinbank} A.~M., {Papadopoulos} P.~P., {Cox} P., {Krips} M., {Ivison} R.~J.,
  {Smail} I., {Thomson} A.~P., {Neri} R. {et~al}, 2011, \apj, 742, 11

\bibitem[{{Swinbank} {et~al.}(2010){Swinbank}, {Smail}, {Longmore}, {Harris},
  {Baker}, {De Breuck}, {Richard}, {Edge}, {Ivison}, {Blundell}, {Coppin},
  {Cox}, {Gurwell}, {Hainline}, {Krips}, {Lundgren}, {Neri}, {Siana},
  {Siringo}, {Stark}, {Wilner}, \& {Younger}}]{Swinbank10}
{Swinbank} A.~M., {Smail} I., {Longmore} S., {Harris} A.~I., {Baker} A.~J., {De
  Breuck} C., {Richard} J., {Edge} A.~C. {et~al}, 2010, \nat, 464, 733

\bibitem[{{Tsai} {et~al.}(2012){Tsai}, {Matsushita}, {Kong}, {Matsumoto}, \&
  {Kohno}}]{Tsai12a}
{Tsai} A.-L., {Matsushita} S., {Kong} A.~K.~H., {Matsumoto} H., {Kohno} K.,
  2012, \apj, 752, 38

\bibitem[{{Tsai} {et~al.}(2009){Tsai}, {Matsushita}, {Nakanishi}, {Kohno},
  {Kawabe}, {Inui}, {Matsumoto}, {Tsuru}, {Peck}, \& {Tarchi}}]{Tsai12b}
{Tsai} A.-L., {Matsushita} S., {Nakanishi} K., {Kohno} K., {Kawabe} R., {Inui}
  T., {Matsumoto} H., {Tsuru} T.~G. {et~al}, 2009, \pasj, 61, 237

\bibitem[{{van der Kruit} \& {Freeman}(2011)}]{VanDerKruit11}
{van der Kruit} P.~C., {Freeman} K.~C., 2011, \araa, 49, 301

\bibitem[{{V{\'a}zquez-Semadeni} {et~al.}(2006){V{\'a}zquez-Semadeni}, {Ryu},
  {Passot}, {Gonz{\'a}lez}, \& {Gazol}}]{Vazquez-Semadeni06}
{V{\'a}zquez-Semadeni} E., {Ryu} D., {Passot} T., {Gonz{\'a}lez} R.~F., {Gazol}
  A., 2006, \apj, 643, 245

\bibitem[{{Veilleux} {et~al.}(2005){Veilleux}, {Cecil}, \&
  {Bland-Hawthorn}}]{Veilleux05}
{Veilleux} S., {Cecil} G., {Bland-Hawthorn} J., 2005, \araa, 43, 769

\bibitem[{{Wada} {et~al.}(2002){Wada}, {Meurer}, \& {Norman}}]{Wada02}
{Wada} K., {Meurer} G., {Norman} C.~A., 2002, \apj, 577, 197

\bibitem[{{Walch} {et~al.}(2012){Walch}, {Whitworth}, {Bisbas}, {W{\"u}nsch},
  \& {Hubber}}]{Walch12}
{Walch} S.~K., {Whitworth} A.~P., {Bisbas} T., {W{\"u}nsch} R., {Hubber} D.,
  2012, \mnras, 427, 625

\bibitem[{{Weaver} {et~al.}(1977){Weaver}, {McCray}, {Castor}, {Shapiro}, \&
  {Moore}}]{Weaver77}
{Weaver} R., {McCray} R., {Castor} J., {Shapiro} P., {Moore} R., 1977, \apj,
  218, 377

\bibitem[{{Weiner} {et~al.}(2009){Weiner}, {Coil}, {Prochaska}, {Newman},
  {Cooper}, {Bundy}, {Conselice}, {Dutton}, {Faber}, {Koo}, {Lotz}, {Rieke}, \&
  {Rubin}}]{Weiner09}
{Weiner} B.~J., {Coil} A.~L., {Prochaska} J.~X., {Newman} J.~A., {Cooper}
  M.~C., {Bundy} K., {Conselice} C.~J., {Dutton} A.~A. {et~al}, 2009, \apj,
  692, 187

\bibitem[{{Weinmann} {et~al.}(2012){Weinmann}, {Pasquali}, {Oppenheimer},
  {Finlator}, {Mendel}, {Crain}, \& {Macci{\`o}}}]{Weinmann12}
{Weinmann} S.~M., {Pasquali} A., {Oppenheimer} B.~D., {Finlator} K., {Mendel}
  J.~T., {Crain} R.~A., {Macci{\`o}} A.~V., 2012, \mnras, 426, 2797

\bibitem[{{White} \& {Frenk}(1991)}]{White91}
{White} S.~D.~M., {Frenk} C.~S., 1991, \apj, 379, 52

\bibitem[{{White} \& {Rees}(1978)}]{White78}
{White} S.~D.~M., {Rees} M.~J., 1978, \mnras, 183, 341

\bibitem[{{Williams} \& {McKee}(1997)}]{Williams97}
{Williams} J.~P., {McKee} C.~F., 1997, \apj, 476, 166

\bibitem[{{Wong} \& {Blitz}(2002)}]{Wong02}
{Wong} T., {Blitz} L., 2002, \apj, 569, 157

\bibitem[{{Woosley} \& {Weaver}(1995)}]{Woosley95}
{Woosley} S.~E., {Weaver} T.~A., 1995, \apjs, 101, 181

\end{thebibliography}
%---------------------------------------------------------------------

\label{lastpage}
\appendix
\section[]{The recycle fraction and yield of different stellar populations}\label{App:YieldRecycle}

The number of SNe per solar mass of stars formed, $\eta_{\rm SN}$, is calculated from the IMF, $\phi(m)\propto {\rm d}N(m)/{\rm d}m$, as, 

\begin{equation}
\eta_{\rm SN}=\int_{m_{\rm SN}}^{m_{\rm max}}\, \phi(m)\, {\rm d} m, 
\end{equation} 

\noindent where %$m_{\rm min}=0.1\, M_{\odot}$, 
$m_{\rm SN}=8\, M_{\odot}$ and $m_{\rm max}=120\, M_{\odot}$. 
For the \citet{Kennicutt83} IMF adopted here, $\eta_{\rm SN}=9.4\times10^{-3} M^{-1}_{\odot}$ (in the 
case of a Salpeter IMF, $\eta_{\rm SN}=7.3\times10^{-3} M^{-1}_{\odot}$).
In $\S$~\ref{Sec:Ener}, we define the mass injection rate from SNe depending on the recycled fraction 
of massive stars, $R_{\rm SN}$. This recycled fraction also depends on the IMF as,  

\begin{equation}
R_{\rm SN}=\int_{m_{\rm SN}}^{m_{\rm max}}\, (m-m_{\rm remn})\phi(m)\, {\rm d} m, 
\label{Eq:ejec}
\end{equation} 

\noindent where $m_{\rm rem}$ is the remnant mass. Similarly, we define the yield from SNe 
as 

\begin{equation}
p_{\rm SN} =\int_{m_{\rm SN}}^{m_{\rm max}}\, m_{\rm i}(m)\phi(m) {\rm d} m, 
\label{Eq:yield}
\end{equation} 

\noindent where $m_{\rm i}(m)$ is the mass of metals produced by stars of initial mass $m$.
We use the stellar evolution models of 
\citet{Marigo01} and \citet{Portinari98} to calculate the ejected mass from 
intermediate and massive stars, respectively. For a Kennicutt IMF, we obtain $R_{\rm SN}=0.14$ and 
$p_{\rm SN}=0.018$.

\section[]{Radial profiles of the stellar and dark matter components and the midplane pressure}\label{Profiles}

An important driver in the evolution of bubbles is the gravitational attraction exerted by 
the stellar and dark matter components. We describe here how we calculate the mass 
enclosed by a sphere of radius $R$ located at a distance $d$ from the centre of the galaxy. 
We perform our calculations of bubble evolution in shells in the disk, which defines $d$ (see $\S$\ref{Sec:GMCs}). 

The total stellar plus dark matter mass within a sphere of radius $R$ 
displaced by $d$ from the centre of the galaxy corresponds to 

\begin{equation}
M_{\rm t}(R,d)=M_{\star}(R,d)+M_{\rm DM}(R,d),
\label{MassTot}
\end{equation}

\noindent where $M_{\star}(R,d)=M_{\star,\rm disk}(R,d)+M_{\star,\rm bulge}(R,d)$ is the 
total stellar mass, $M_{\star,\rm disk}(R,d)$ and $M_{\star,\rm bulge}(R,d)$ represent 
the mass in the disk and the bulge, respectively,  
and $M_{\rm DM}(R,d)$ the mass in DM, in all cases enclosed in $R$. 
 We describe below
 how we calculate the variables of Eq.~\ref{MassTot}.\\
 
\noindent {\it Disk radial profile.} We assume disks are well described by a 
radial exponential profile with a scale radius $r_{\rm s}$, which is related to the 
half-mass radius as $r_{\rm 50,disk}=1.67\, r_{\rm s}$ \citep{Binney08}. We define  
the stellar surface density of the disk 
at a distance $d$ from the centre as 

\begin{equation}
\Sigma_{\star, \rm disk}(d)=\frac{M^{\rm t}_{\star,\rm disk}}{2\pi\, r^2_{\rm s}} e^{-d/r_{\rm s}}.
\end{equation}

\noindent Here, $M^{\rm t}_{\star,\rm disk}$ is the total stellar mass in the disk. 
If the relevant sphere of radius $R$ is at a distance $d$ from the centre, then the 
stellar mass in the midplane of the disk exerting the gravitational attraction on  the bubble is 
approximately

\begin{eqnarray}
M_{\star, \rm disk}(R,d)&\approx& \frac{4\pi R^3}{3} \, \frac{\Sigma_{\star, \rm disk}(d)}{2\, h_{\star}}. 
\end{eqnarray}

\noindent Here, $h_{\star}$ is the scale height of the stars, which we estimate from the scale radius of the disk 
following the empirical results of \citet{Kregel02}, $r_{\rm s}/h_{\star}=7.3$. 

\noindent {\it Bulge radial profile.} The potential well of a galactic bulge, $\Phi(r)$, can be well 
described by a Dehnen profile \citep{Dehnen93} with  
$\gamma_{\rm b}=3/2$ which closely resembles a \citet{deVaucouleurs53} $r^{1/4}$ profile,

\begin{equation}
\Phi(r)=\frac{G\,M^{\rm t}_{\star,\rm bulge}}{r_0}\, \frac{1}{2-\gamma_{\rm b}}\left[1-\left(\frac{r}{r+r_0}\right)^{2-\gamma_{\rm b}}\right], 
\end{equation}

\noindent where $r_{0}$ is the scale radius and $M^{\rm t}_{\star,\rm bulge}$ is the total stellar mass 
of the bulge. The scale radius relates to  
 the half-mass radius of the bulge, $r_{50,\rm b}$ as 

\begin{equation}
r_{\rm 50, b}= r_0\,(2^{1/(3-\gamma_{\rm b})}-1)^{-1}.
\end{equation}

 In this definition of potential well, the 
volume density profile of stars is, 

\begin{equation}
\rho_{\star,\rm bulge}(r)=\frac{(3-\gamma_{\rm b})}{4\pi}\, \frac{M^{\rm t}_{\star,\rm bulge}\, r_{0}}{r^{\gamma_{\rm b}}(r+r_0)^{4-\gamma_{\rm b}}}.
\label{Eq:rhobulge}
\end{equation}

\noindent Although the stars in the bulge follow a De Vacouleurs profile, the gas is assumed 
to be better characterised by an exponential profile, as has been observed in early-type galaxies 
(e.g. \citealt{Crocker11}; \citealt{Davis11}; \citealt{Serra12}). This means that the same geometry 
adopted for the case of disks applies here: bubbles expand in a coordinate system 
displaced by $d$ in the x-axis. However, the difference with the case of the disk is that 
here the stellar profile has spherical symmetry. 
With this in mind, we approximate the stellar mass enclosed by a bubble of radius R  
displaced by $d$ from the centre as, 
 
\begin{eqnarray}
M_{\star, \rm bulge}(R,d)&\approx&\frac{4\pi R^3}{3}\, \rho_{\star,\rm bulge}(d),
\label{Eq:mbulge}
\end{eqnarray}

We use the equations above to calculate the $M_{\star}(R,d)$ that goes into Eqs.~\ref{ener1}-\ref{ener2}, 
\ref{press1}-\ref{press2} and \ref{mom1}-\ref{mom2}.\\

\noindent{\it Dark matter radial profile.} Here we assume that DM halos are well 
described by a NFW profile \citep{Navarro97}. We follow the description of \citet{Cole00}, where 
 halos contract in response to the presence of baryons. The galaxy disk, bulge 
and DM halo adjust to each other adiabatically. 

The volume mass density of DM is described in a NFW profile as 

\begin{equation}
\rho_{\rm DM}(r)=\frac{\delta_{\rm c}\, \rho_{\rm c}}{(r/r_{\rm s})(1+r/r_{\rm s})^2},
\label{Eq:rhodm}
\end{equation} 

\noindent where $r_{\rm s}$ is the DM scale radius, $\delta_{\rm c}$ is the characteristic 
(dimensionless) density and $\rho_{\rm c}$ is the critical density of the universe.   
As before, the mass enclosed within a sphere of radius $R$ displaced by $d$ from the centre of the potential well,
 
\begin{eqnarray}
M_{\rm DM}(R,d)&\approx&\frac{4\pi R^3}{3} \, \rho_{\rm DM}(d),
\label{Eq:mdm} 
\end{eqnarray}

\noindent assuming $\rho_{\rm DM}(d)$ is approximately constant within the bubble. 

Note that Eqs.~\ref{Eq:mbulge} and \ref{Eq:mdm} are accurate 
in the regime where $d/R \gg 1$. 
In this paper we neglect the effect of tidal forces on bubbles, which arise from 
the asymmetric gravitational field, which distort their shape. This would 
affect the size of bubbles perpendicular to the gaseous disk and therefore 
the break-out of bubbles. 

\subsection{The midplane hydrostatic pressure of disk galaxies and the gas scaleheight}\label{App:BR}

Under the assumptions of local isothermal stellar and gas layers, and
$\sigma_{\star}>\sigma_{\rm gas}$, the midplane hydrostatic pressure in disks,
$P_{\rm ext}$, can be approximated to within
$10$\% by \citep{Elmegreen89}

\begin{equation}
P_{\rm ext}(r)\approx \frac{\pi}{2} G \Sigma_{\rm gas}(r) \left[ \Sigma_{\rm gas}(r) +
\left( \frac{\sigma_{\rm d}}{\sigma_{\star}(r)}\right)\Sigma_{\star}(r)\right], \label{PressureDef}
\end{equation}
where $\Sigma_{\rm gas}$ and $\Sigma_{\star}$ are
the surface densities of gas and stars at $r$, respectively, and
$\sigma_{\rm g}$ and $\sigma_{\star}$ give the vertical velocity
dispersion of the gas and stars. We assume a constant gas velocity
dispersion, $\sigma_{\rm d}=10\, \rm km\,\rm s^{-1}$ \citep{Leroy08}.
By assuming that $\Sigma_{\star}\gg\Sigma_{\rm gas}$,
$\sigma_{\star}(r)=\sqrt{\pi \rm G\, \rm h_{\star}\Sigma_{\star}({\it r})}$,
where $h_{\star}$ is the stellar scale height.
This
approximation could break down for very high redshift galaxies, whose
disks are gas dominated. In such cases, we assume a floor of
$\sigma_{\star}\ge\sigma_{\rm g}$. 

{In the case of the gas scaleheight, we simply assume vertical equilibrium, where 
the gravitational force is balanced by the pressure of the gas, $P=\sigma^2_{\rm d}\,\rho_{\rm g}$, where 
$\rho_{\rm g}=\Sigma_{\rm g}/2\,h_{\rm g}$ and $\Sigma_{\rm g}$ is the gas surface density (molecular 
plus atomic gas). Using Eq.~\ref{PressureDef} to define the pressure on the midplane of the disk 
due to the gravitational force, we can write,}

\begin{equation}
h_{\rm g}(r)\approx \frac{\sigma^{2}_{\rm d}}{\pi\,G \left[ \Sigma_{\rm gas}(r) +
\left( \frac{\sigma_{\rm d}}{\sigma_{\star}(r)}\right)\Sigma_{\star}(r)\right]}. \label{PressureDef2}
\end{equation}

\section[]{Calculation of swept-up, confinement and break-out mass rates}\label{App:MassRates}

The contribution from bubbles to the rate of change of the mass and metallicity in the ISM and hot halo gas, 
depends on their evolution. In this appendix, we briefly describe how we calculate the overall 
contribution from bubbles in different evolutionary stages included in the set 
of Eqs.~\ref{Eqs:SFset1}-\ref{Eqs:SFset2}.\\

{\it The swept-up mass.} Each galaxy has generations of bubbles whose evolution depends on the time 
they started their expansion and their spatial distribution in the galaxy. Each galaxy has its 
star formation history (SFH) sampled in a fine grid in time that goes down to the current time, $t_{\rm c}$.
Each time interval, ${\rm d}t^{\prime}$, in the SFH of a galaxy 
has associated a new generation of $N_{\rm GMC,i,t^{\prime}}$ set of bubbles in the annulus $i$ of the galaxy disk. 
Each of these bubbles 
 have swept-up a mass $m_{\rm sw}(r_{\rm i},t^{\prime})$ from the diffuse medium and have a 
total mass $m_{\rm b}(r_{\rm i},t^{\prime})$ at $t^{\prime}$.
The number of annuli used to solve the equations of bubbles expansion ($\S 4.1$) is $N_{\rm r}$.
The overall rate of swept-up mass is 

\begin{eqnarray}
  \dot{M}_{\rm sw,ISM}(t_{\rm c}) &= &\int^{t_{\rm c}}_0 \sum^{{\rm i}=N_{\rm r}}_{\rm i=1} N_{\rm GMC,i,t^{\prime}}\,\dot{m}_{\rm sw}(r_{\rm i},t^{\prime})\,(1-H_{r,h})\nonumber\\
 & & (1-H_{v,\sigma})\,{\rm d}t^{\prime}.
\label{set:dragism}
\end{eqnarray}

\noindent Here, $H_{r,h}$ and $H_{v,\sigma}$ are step functions defined in terms of the 
radius of bubbles, $R_{\rm s}$, the gas scaleheight, $h_{\rm g}$, the expansion speed of bubbles, $v_{\rm s}$, 
and the velocity dispersion of the warm gas phase of the ISM, $\sigma_{\rm d}$, as, 
$H_{r,h}=H[f_{\rm r}h_{\rm g}(r_{\rm i},t^{\prime})-R_{\rm b}(r_{\rm i},t^{\prime})]$ and 
$H_{v,\sigma}=H[\sigma_{\rm d}-v_{\rm s}(r_{\rm i},t^{\prime})]$.
The quantities $h_{\rm g}$, $R_{\rm b}$ and $v_{\rm s}$ depend on time and annulus. Eq.~\ref{set:dragism} 
implies that all bubbles contribute to the swept-up mass rate unless they 
have been confined or broken-out from the ISM in previous times. 
Bubbles at different evolutionary stages can coexist in 
an annulus.\\

{\it Confined bubbles.} Confined bubbles contribute positively to $\dot{M}_{\rm g,ISM}$. The confinement of bubbles 
depends on whether the expansion velocity of bubbles reach or exceed the velocity dispersion of the warm phase 
in the ISM, $\sigma_{\rm d}$.
The rate of mass transferred to the ISM by confinement is

 \begin{eqnarray}
  \dot{M}_{\rm conf,ISM}(t_{\rm c}) &= &\int^{t_{\rm c}}_0 \sum^{{\rm i}=N_{\rm r}}_{\rm i=1} N_{\rm GMC,i,t^{\prime}}\,\dot{m}_{\rm b}(r_{\rm i},t^{\prime})\,H_{v,\sigma}\,{\rm d}t^{\prime}.
\label{set:conf}
\end{eqnarray}

{\it Break-out of bubbles.} The break-out of bubbles from the ISM contributes positively to the 
ISM gas due to the fraction of gas mass in the bubbles 
that stays in the ISM, $(1-f_{\rm bo})$.
The condition for break-out 
is that the radius of the bubbles reaches a factor $f_{\rm r}$ of the gas scaleheight, 
$R_{\rm b}\ge f_{\rm r}h_{\rm g}$. The 
rate of break-out gas mass in the ISM is 

\begin{eqnarray}
 \dot{M}_{\rm bo,ISM}(t_{\rm c}) &= &\int^{t_{\rm c}}_0 \sum^{{\rm i}=N_{\rm r}}_{\rm i=1} N_{\rm GMC,i,t^{\prime}}\,\dot{m}_{\rm b}(r_{\rm i},t^{\prime})\,H_{r,h}\,{\rm d}t^{\prime}.
\label{set:boISM}
\end{eqnarray}

\end{document}